\documentclass[iop,article,numberedappendix]{emulateapj}
\usepackage{natbib}
\usepackage{amsmath}
\usepackage{verbatim}
\usepackage[]{graphicx}
\usepackage{enumerate}
\usepackage{color}
\usepackage[pdftex]{hyperref}

\newcommand{\bymv}{$1.06\pm0.12$\smallskip}
\newcommand{\btmv}{$0.35\pm0.05$}
\newcommand{\byxmv}{$1.36\pm0.06$\smallskip}
\newcommand{\isymv}{$(25\pm9)\%$\smallskip}
\newcommand{\isyyxv}{$(21\pm7)\%$\smallskip}
\newcommand{\byyxv}{$0.84\pm0.07$\smallskip}
\newcommand{\isymlogv}{$0.11\pm0.04$\smallskip}
\newcommand{\isyyxlogv}{$0.09\pm0.03$\smallskip}
\newcommand{\bym}{$\beta_1^{y|m}$}
\newcommand{\btm}{$\beta_1^{t|m}$}
\newcommand{\aym}{$\beta_0^{y|m}$}
\newcommand{\sigym}{$\sigma_{y|m}$}
\newcommand{\bymg}{$\beta_1^{y|m_g}$}
\newcommand{\byxm}{$\beta_1^{y_x|m}$}
\newcommand{\bmyx}{$\beta_1^{m|y_x}$}
\newcommand{\byyx}{$\beta_1^{y|y_x}$}
\newcommand{\olm}{$\theta_{l|m}$}
\newcommand{\otm}{$\theta_{t|m}$}
\newcommand{\oym}{$\theta_{y|m}$}
\newcommand{\fgas}{$f_{\rm gas}$}
\newcommand{\fbig}{$f_{\rm gas,500}$}
\newcommand{\fsmall}{$f_{\rm gas,2500}$}
\newcommand{\Mtot}{$M_{\rm tot}$}
\newcommand{\Msmall}{$M_{2500}$}
\newcommand{\Mbig}{$M_{500}$}
\newcommand{\Mgas}{$M_{\rm gas}$}
\newcommand{\Mgassmall}{$M_{\rm gas,2500}$}
\newcommand{\Ysz}{$Y_{\rm SZ}$}
\newcommand{\Yx}{$Y_{\rm X}$}
\newcommand{\Ysmall}{$Y_{2500}$}
\newcommand{\Ybig}{$Y_{500}$}

\newcommand{\Tx}{$T_{\rm X}$}
\newcommand{\Tbig}{$kT$}
\newcommand{\Lx}{$L_{\rm x}$}
\newcommand{\Lbig}{$L_{500}$}
\newcommand{\rbig}{$r_{500}$}
\newcommand{\rsmall}{$r_{2500}$}
\newcommand{\wbig}{$w_{500}$}
\newcommand{\Msun}{$M_{\odot}$}
\newcommand{\chisqr}{$\chi^2$}
\newcommand*{\myalign}[2]{\multicolumn{1}{#1}{#2}}
\begin{document}
\title{Galaxy Cluster Scaling Relations between Bolocam Sunyaev-Zel'dovich Effect and \emph{Chandra} X-ray Measurements}
\author{
N. G. Czakon\altaffilmark{1,2},
J.~Sayers\altaffilmark{3}, 
A.~Mantz\altaffilmark{4}, 
S. R. Golwala\altaffilmark{3},
T.P. Downes\altaffilmark{3},
P.M. Koch\altaffilmark{1},
K.-Y. Lin\altaffilmark{1},
S. M. Molnar\altaffilmark{5},
L. A.~Moustakas\altaffilmark{6},
T. Mroczkowski\altaffilmark{7,8},
E.~Pierpaoli\altaffilmark{9},
J. A. Shitanishi\altaffilmark{9},
S. Siegel\altaffilmark{3},
K.~Umetsu\altaffilmark{1}
}
\altaffiltext{1}
{Institute of Astronomy and Astrophysics, Academia Sinica, P.O. Box 23-141, Taipei 10617, Taiwan}
\altaffiltext{2}
             {czakon@asiaa.sinica.edu.tw}
\altaffiltext{3}
             {Division of Physics, Math, and Astronomy, California Institute of Technology, 1200 East California Blvd, Pasadena, CA 91125}
\altaffiltext{4}
		{Kavli Institute for Cosmological Physics, University of Chicago, 5640 South Ellis Avenue, Chicago, IL 60637}
\altaffiltext{5}
		{Department of Physics, National Taiwan University, 106, Taipei, Taiwan}
\altaffiltext{6}
             {Jet Propulsion Laboratory,4800 Oak Grove Drive, Pasadena, CA 91109}
\altaffiltext{7}
             {Naval Research Laboratory, 1204 S. Washington St. 629, Alexandria, VA 22314}
\altaffiltext{8}
             {National Research Council Fellow}
\altaffiltext{9}
             {Department of Physics and Astronomy, University of Southern California, 3620 McClintock Avenue, Los Angeles, CA 90089}
\begin{abstract}
We present scaling relations between the integrated Sunyaev-Zel'dovich Effect (SZE) signal, \Ysz, its X-ray analogue, \Yx$\equiv$\Mgas\Tx, and total mass, \Mtot, for the 45 galaxy clusters in the Bolocam X-ray-SZ (BOXSZ) sample. All parameters are integrated within \rsmall. 
\Ysmall~values are measured using SZE data collected with Bolocam, operating at 140 GHz at the Caltech Submillimeter Observatory (CSO).
The temperature, \Tx, and mass, \Mgassmall, of the intracluster medium are determined using X-ray data collected with \emph{Chandra}, and \Mtot~is derived from \Mgas{ }assuming a constant gas mass fraction.
Our analysis accounts for several potential sources of bias, including: selection effects, contamination from radio point sources, and the loss of SZE signal due to noise filtering and beam-smoothing effects.
We measure the \Ysmall--\Yx~scaling to have a power-law index of \byyxv, and a fractional intrinsic scatter in \Ysmall~of \isyyxv~at fixed \Yx, both of which are consistent with previous analyses.
We also measure the scaling between \Ysmall\ and \Msmall, finding a power-law index of \bymv~and a fractional intrinsic scatter in \Ysmall~at fixed mass of \isymv.
While recent SZE scaling relations using X-ray mass proxies have found power-law indices consistent with the self-similar prediction of 5/3, our measurement stands apart by differing from the self-similar prediction by approximately 5$\sigma$.
Given the good agreement between the measured \Ysmall--\Yx~scalings, much of this discrepancy appears to be caused by differences in the calibration of the X-ray mass proxies adopted for each particular analysis.
\end{abstract}
\section{Introduction}
The mass distribution in the universe is an essential prediction for any cosmological model and must be observationally tested. 
Galaxy clusters offer a window to study this mass distribution because, with masses ranging from approximately $10^{13}$ to $10^{15}$\Msun, they are the largest gravitationally bound objects in the universe.
Furthermore, galaxy clusters are natural probes of dark energy as their growth progressively slows and eventually freezes out in the presence of accelerated cosmic expansion \citep{Voit2005,Vikhlinin2009c,Mantz2010a,Allen2011,Benson2013,Planck2013SZEcosmo}.

The deep gravitational potential wells of galaxy clusters accrete large amounts of baryonic matter that is compressively heated to $10^7$--$10^8$ Kelvin,  forming a highly ionized intracluster medium (ICM, \citealt{Sarazin1988}). 
This ICM produces the two observables used in this analysis: X-ray emission (primarily from thermal bremsstrahlung) and the distortion of the cosmic microwave background radiation (CMB) via Compton scattering off of the ICM, known as the Sunyaev-Zel'dovich effect (SZE, \citealt{Sunyaev1972}).
Simulations indicate that simple self-similar scaling relations assuming hydrostatic equilibrium (HSE) provide a reasonably good, but not perfect, description linking the physical properties of galaxy clusters with observables. \citep{Bertschinger1985,Kaiser1986,Kravtsov2012,Angulo2012}.
Observationally, deviations from self-similarity have been identified in the scaling between X-ray luminosity, temperature, and cluster mass (e.g., \citealt{Edge1991,Henry1991,WhiteJonesForman1997,Reiprich2002, Arnaud2005, Stanek2006, Maughan2006}). These deviations might arise from a variety of factors, such as cluster morphology, departures from HSE, and physical processes that include but are not limited to: radiative cooling and star formation (CSF) and active galactic nucleus (AGN) feedback. How these features affect the measured scaling relations has been investigated in simulations \citep{Nagai2006,Nagai2007HSE,Nagai2007CSF,Fabjan2011,Battaglia2013,Sembolini2013}.

While X-ray observations have long been used to constrain the thermal properties of the ICM, SZE measurements have now emerged as an additional observational tool for studying the ICM.
Because the SZE produces a fractional shift in the energy of CMB photons, it does not dim with redshift
and is therefore a promising probe to study cosmology in the epoch where, according to the standard cosmological model, dark energy begins to affect cosmic expansion \citep{Carlstrom2002}. Several astronomical surveys have recently produced SZE-selected cluster catalogs \citep{Vanderlinde2010, Marriage2011, Reichardt2013,Planck2011ESZ, Planck2013SZECat} and have used these to constrain cosmological parameters with a precision comparable to those from X-ray cluster surveys (e.g., \citealt{Benson2013, Reichardt2012,Hasselfield2013,Planck2013SZEcosmo}).

Significant systematic uncertainty remains as to the exact mass scaling of the SZE signal, which limits the impact of cosmological studies using SZE-selected clusters. 
Large efforts have been directed at both simulation \citep{Sehgal2010, Vanderlinde2010,Sembolini2013} and observational programs \citep{Andersson2011, Benson2013, Planck2013SZEcosmo}
to remedy this situation, but an approximate 10--20\% calibration uncertainty still limits recent cosmological results. 
For example, \citet{Benson2013} anticipate the need for an absolute mass-observable scaling uncertainty of less than 5\% (with less than a 6\% uncertainty in the redshift evolution of this scaling) in order to obtain measurement-noise-limited rather than calibration-limited constraints on the dark energy equation of state for the South Pole Telescope (SPT) 2500 $\rm deg^2$ cluster cosmology analysis.

In addition to large SZE surveys, smaller field-of-view SZE instruments have observed large samples of previously known clusters.
These instruments thereby provide additional data outside of the survey areas of the dedicated survey instruments (notably in the Northern Hemisphere), in part to further improve the SZE-observable/mass calibration.
Some examples of SZE results derived from such instruments are: the Atacama Pathfinder Experiment-SZ (APEX-SZ) \citep{Nord2009}, the Arcminute Microkelvin Imager (AMI) \citep{AMI2012}, the SZ Array (SZA) \citep{Reese2012}, the Array for Microwave Background Anisotropy (AMiBA) \citep{Huang2010}, and Bolocam \citep{SayersMorphology}.
There have also been a handful of SZE-observable/mass scaling relations derived from pointed observations of previously known clusters (e.g., \citealt{Bonamente2008, Marrone2009, Marrone2012,Plagge2010,Bender2014}).
In addition, some groups have combined resolved SZE data with optical and/or X-ray data sets to obtain joint-observable total cluster mass estimates for single clusters (e.g., \citealt{Nord2009, Basu2010, Morandi2012, SerenoA1689}), and such measurements are likely to become more common given the rapidly improving quality of SZE data.

In the present analysis, we compare the integrated SZE signal measured with Bolocam to \emph{Chandra} X-ray-determined cluster masses. 
The methodology for measuring cluster mass from X-ray observations has been an increasingly active area of research since \emph{Chandra} and XMM-\emph{Newton} launched in 1999. X-ray analyses offer abundant, low-scatter mass proxies, thereby providing an ideal tool to estimate the masses of the BOXSZ sample.
X-ray-derived masses have already been used in several large cosmological analyses, for example, by \citet[][hereafter V09]{Vikhlinin2009b} and \citet[][hereafter M10]{Mantz2010b}.

This manuscript is arranged as follows.
Section~\ref{sec:BOXSZ} introduces the BOXSZ cluster sample.
In Section~\ref{sec:xray}, we give a brief overview of the X-ray data reduction and the adopted methodology for mass estimation.
Section~\ref{sec:boloy} reviews the relevant physics of the Sunyaev-Zel'dovich effect as it pertains to this work and gives a more extensive overview of the SZE data reduction and noise characterization.
In Section~\ref{sec:SR}, we introduce our formalism for fitting scaling relations and give an overview of the simulation-derived biases in the determined parameters due to selection effects.
Finally, in Section~\ref{sec:sr_results}, we present the results of the BOXSZ scaling relations, which are compared with those of other groups, and explore key differences in our analysis that might explain the discrepencies between the results of different groups.

Several appendices provide more detail on our methods and results.
Appendices~\ref{sec:sig_offset}, \ref{sec:r500_v_r2500}, \ref{sec:ftest}, and \ref{sec:sf} explain technical aspects of our analysis.
In Appendix~\ref{sec:fgas}, we give a detailed comparison between our mass proxy and the mass proxies used in similar SZE scaling relation studies, and we describe how an alternative parameterization of our mass proxy would affect our results.
The maps for all of the clusters in our sample are given at the end of the manuscript in Appendix~\ref{sec:thumbnails}.

For this analysis, we adhere to the convention of measuring cluster properties within a radius, $r_{\rm\Delta}$, within which the mean cluster density is $\Delta$ times the critical density of the universe at the redshift of the cluster, $\rho_{\rm{c}}(z)$. 
We assume a $\Lambda\rm CDM$ cosmology, 
$H_{\rm{0}} = 70\ \rm km \ s^{-1}Mpc^{-1}$, $\Omega_{\rm M}= 0.3$, and $\Omega_\Lambda=0.7$. 
The redshift evolution of the Hubble parameter with respect to its present value is taken to be $H(z) = H_0E(z)$ with $E(z)=\sqrt{\Omega_{\rm{M}}(1+z)^3+\Omega_\Lambda}$.
\section{The Bolocam X-Ray SZ (BOXSZ) Sample}
\label{sec:BOXSZ}
\setlength{\tabcolsep}{.3em}{
\begin{deluxetable*}{lrrccccc} \tabletypesize{\scriptsize} \tablecaption{Observational Information for the BOXSZ cluster sample.} \tablewidth{0pt} \tablehead{ Name&\myalign{c}{RA}&\myalign{c}{DEC}&SZE S/N&SZE RMS&~~SZE $\rm t_{int}$&CLASH&WtG\\ &\myalign{c}{(J2000)}&\myalign{c}{(J2000)}&&($\mu \rm K_{CMB}$-arcmin)&~~(hours)}
\startdata 
Abell~2204&16:32:47.2&+05:34:33&22.3&18.5&12.7&\nodata&\checkmark\\[3pt] 
Abell~383&02:48:03.3&-03:31:46&\phn{9.6}&18.9&24.3&\checkmark&\checkmark\\[3pt] 
Abell~209&01:31:53.1&-13:36:48&13.9&22.3&17.8&\checkmark&\checkmark\\[3pt] 
Abell~963&10:17:03.6&+39:02:52&\phn{8.3}&35.7&11.0&\nodata&\checkmark\\[3pt]
Abell~1423&11:57:17.4&+33:36:40&\phn{5.8}&31.7&11.5&\checkmark&\nodata\\[3pt] 
Abell~2261&17:22:27.0&+32:07:58&10.2&15.9&17.5&\checkmark&\checkmark\\[3pt] 
Abell~2219&16:40:20.3&+46:42:30&11.1&39.6&\phn{6.3}&\nodata&\checkmark\\[3pt] 
Abell~267&01:52:42.2&+01:00:30&\phn{9.6}&23.0&20.7&\nodata&\nodata\\[3pt]
RX~J2129.6+0005&21:29:39.7&+00:05:18&\phn{8.0}&23.7&16.0&\checkmark&\checkmark\\[3pt] 
Abell~1835&14:01:01.9&+02:52:40&15.7&16.2&14.0&\nodata&\checkmark\\[3pt] 
Abell~697&08:42:57.6&+36:21:57&22.6&17.4&14.3&\nodata&\nodata\\[3pt] 
Abell~611&08:00:56.8&+36:03:26&10.8&25.0&18.7&\checkmark&\checkmark\\[3pt]
MS~2137&21:40:15.1&-23:39:40&\phn{6.5}&27.3&12.8&\checkmark&\checkmark\\[3pt] 
Abell~S1063&22:48:44.8&-44:31:45&10.2&48.6&\phn{5.5}&\checkmark&\nodata\\[3pt] 
MACS~J1931.8-2634&19:31:49.6&-26:34:34&10.1&28.7&\phn{7.5}&\checkmark&\nodata\\[3pt] 
MACS~J1115.8+0129&11:15:51.9&+01:29:55&10.9&22.8&15.7&\checkmark&\checkmark\\[3pt]
MACS~J1532.8+3021&15:32:53.8&+30:20:59&\phn{8.0}&22.3&14.8&\checkmark&\checkmark\\[3pt] 
Abell~370&02:39:53.2&-01:34:38&12.8&28.9&11.8&\nodata&\checkmark\\[3pt] 
MACS~J1720.2+3536&17:20:16.7&+35:36:23&10.6&23.5&16.8&\checkmark&\checkmark\\[3pt] 
ZWCL~0024+17&00:26:35.8&+17:09:41&\phn{3.3}&26.6&\phn{8.3}&\nodata&\nodata\\[3pt]
MACS~J2211.7-0349&22:11:45.9&-03:49:42&14.7&38.6&\phn{6.5}&\nodata&\checkmark\\[3pt] 
MACS~J0429.6-0253&04:29:36.0&-02:53:06&\phn{8.9}&24.1&17.0&\checkmark&\checkmark\\[3pt] 
MACS~J0416.1-2403&04:16:08.8&-24:04:14&\phn{8.5}&29.3&\phn{7.8}&\checkmark&\nodata\\[3pt] 
MACS~J0451.9+0006&04:51:54.7&+00:06:19&\phn{8.1}&22.7&14.2&\nodata&\checkmark\\[3pt]
MACS~J1206.2-0847&12:06:12.3&-08:48:06&21.7&24.9&11.3&\checkmark&\checkmark\\[3pt] 
MACS~J0417.5-1154&04:17:34.3&-11:54:27&22.7&22.7&\phn{9.8}&\nodata&\checkmark\\[3pt] 
MACS~J0329.6-0211&03:29:41.5&-02:11:46&12.1&22.5&10.3&\checkmark&\checkmark\\[3pt] 
MACS~J1347.5-1144&13:47:30.8&-11:45:09&36.6&19.7&15.5&\checkmark&\checkmark\\[3pt]
MACS~J1311.0-0310&13:11:01.7&-03:10:40&\phn{9.6}&22.5&14.2&\checkmark&\nodata\\[3pt] 
MACS~J2214.9-1359&22:14:57.3&-14:00:11&12.6&27.3&\phn{7.2}&\nodata&\checkmark\\[3pt] 
MACS~J0257.1-2325&02:57:09.1&-23:26:04&10.1&39.0&\phn{5.0}&\nodata&\checkmark\\[3pt] 
MACS~J0911.2+1746&09:11:10.9&+17:46:31&\phn{4.8}&33.5&\phn{6.2}&\nodata&\checkmark\\[3pt]
MACS~J0454.1-0300&04:54:11.4&-03:00:51&24.3&18.2&14.5&\nodata&\checkmark\\[3pt] 
MACS~J1423.8+2404&14:23:47.9&+24:04:43&\phn{9.4}&22.3&21.7&\checkmark&\checkmark\\[3pt]
MACS~J1149.5+2223&11:49:35.4&+22:24:04&17.4&24.0&17.7&\checkmark&\checkmark\\[3pt] 
MACS~J0018.5+1626&00:18:33.4&+16:26:13&15.7&21.0&\phn{9.8}&\nodata&\checkmark\\[3pt]
MACS~J0717.5+3745&07:17:32.1&+37:45:21&21.3&29.4&12.5&\checkmark&\checkmark\\[3pt] 
MS~2053.7-0449&20:56:21.0&-04:37:49&\phn{5.1}&18.0&18.7&\nodata&\nodata\\[3pt] 
MACS~J0025.4-1222&00:25:29.9&-12:22:45&12.3&19.7&14.3&\nodata&\checkmark\\[3pt] 
MACS~J2129.4-0741&21:29:25.7&-07:41:31&15.2&21.3&13.2&\checkmark&\checkmark\\[3pt]
MACS~J0647.7+7015&06:47:49.7&+70:14:56&14.4&22.0&11.7&\checkmark&\checkmark\\[3pt] 
MACS~J0744.8+3927&07:44:52.3&+39:27:27&13.3&20.6&16.3&\checkmark&\checkmark\\[3pt] 
MS~1054.4-0321&10:56:58.5&-03:37:34&17.4&13.9&18.3&\nodata&\nodata\\[3pt] 
CL~J0152.7&01:52:41.1&-13:58:07&10.2&23.4&\phn{9.3}&\nodata&\nodata\\[3pt]
CL~J1226.9+3332&12:26:57.9&+33:32:49&13.0&22.9&11.8&\checkmark&\nodata
\enddata \tablecomments{From left to right: the cluster catalog and ID, X-ray centroid coordinates (J2000), the peak SZE S/N in the optimally filtered images (see \citet{SayersESZ}), RMS noise level of the SZE images, and the total Bolocam integration time. The final two columns indicate whether the cluster is in the CLASH sample of \citet{Postman2012} and/or in the WtG sample of \citet{vonderLinden2014}.} \label{tab:obs_info} \end{deluxetable*}}

The Bolocam X-Ray SZ (BOXSZ) sample is a compilation of 45 clusters with existing \emph{Chandra} data observed with Bolocam at 140 GHz \citep{SayersPressure}.
Bolocam is a 144-element bolometric camera with a 58\arcsec~full-width at half maximum (FWHM) point-spread function (PSF) at the SZE-emission-weighted band center of 140 GHz \citep{Glenn1998,Haig2004}. 
The Bolocam data were collected over five years (from Fall 2006 to Spring 2012) in 14 different observing runs at the Caltech Submillimeter Observatory.
Table~\ref{tab:obs_info} includes the relevant observational information for these clusters.

Bolocam's field-of-view is well-matched to observe intermediate redshift clusters, and therefore many of the clusters in the BOXSZ sample were selected based on having a redshift between 0.3 and 0.6. 
The BOXSZ sample spans from $z=0.15$ to $z=0.9$, with a median redshift of $\langle z \rangle =0.42$.
This redshift distribution is similar to the initial ground-based SZE-selected catalogs of both the SPT, $\langle z\rangle =0.57$ \citep{Song2012}, and the Atacama Cosmology Telescope, $\langle z \rangle =0.44$ \citep{Menanteau2010}.
In contrast, the early {\it Planck} SZE catalog has a median redshift of $\langle z \rangle = 0.15$ \citep{Planck2011ESZ}, and the 2013 {\it Planck} SZE catalog has a 
median redshift of $\langle z \rangle = 0.22$ \citep{Planck2013SZECat}.
In addition to redshift, many of the clusters in the BOXSZ sample were selected based on their higher-than-average X-ray spectroscopic temperatures, \Tx, given the expected correlation between 
\Tx~and SZE brightness.
Finally, a few clusters were chosen solely due to their membership either in the CLASH \citep{Postman2012} or the MACS high-redshift \citep{Ebeling2007} catalogs, both of which are fully contained in the BOXSZ sample. 
Recently, the Weighing the Giants (WtG) team presented weak-lensing measurements for 51 X-ray selected galaxy clusters for the primary purpose of calibrating X-ray mass proxies for cosmology (\citealt{vonderLinden2014,Kelly2014,Applegate2014}; \citealt{Mantz2015}), and 33 BOXSZ clusters are in the WtG cluster sample. Although not directly relevant to the present analysis, future cluster studies will benefit from the available multi-wavelength data sets associated with these cluster samples and BOXSZ cluster membership in either the CLASH or WtG samples is indicated in Table \ref{tab:obs_info}.
Despite having a large amount of overlap with other X-ray defined cluster samples, the BOXSZ sample as a whole lacks a well-defined selection function. We explore the effects of the BOXSZ cluster selection in Appendix \ref{sec:sf}.

BOXSZ SZE data have already been used for individual cluster studies \citep{Morandi2012,Umetsu2012,ZitrinA383,Mroczkowski2012,Zemcov2012,Mauskopf2012,Medezinski2013,Sayers0717}, to characterize the contamination from radio galaxies in 140 GHz SZE measurements \citep{SayersRadio}, and to measure the average pressure profile of the sample \citep{SayersPressure}.
\section{X-ray Data and Mass Estimation} 
\label{sec:xray}
X-ray luminosity and temperature measurements for the BOXSZ clusters were either taken directly from M10 or derived from archival \emph{Chandra} data in an identical manner, as described in \citet{SayersPressure}. To estimate cluster gas masses and total masses, we follow the procedure laid out in M10, with the exception that we calculate the integrated cluster parameters within \rsmall~rather than \rbig.

In brief, gas mass profiles are non-parametrically derived from each cluster's 0.7--2.0\,keV surface brightness profile following the technique of \citet{White1997}.
In converting soft-band brightness to gas density, the best-fit global temperature is used; however, for the relevant temperatures of the BOXSZ sample, the temperature dependence of this conversion is negligible. 
For high-mass clusters, like those in the BOXSZ sample, \citet[][hereafter Allen08]{Allen2008} measure the gas mass fraction, \fgas, to be consistent with a constant value at \rsmall~for dynamically relaxed clusters with mean temperatures above 5 keV---a result that is also supported by simulations \citep[][]{Eke1998,Crain2007,Battaglia2013,Planelles2013}. 
Some observational and simulation results, e.g., \citet{Vikhlinin2009c,Pratt2009,Battaglia2012,Sembolini2013}, support a non-constant \fgas~model. In Appendix~\ref{sec:fgas}, we discuss the relevance of these measurements to the BOXSZ cluster sample, and explore the effect that non-constant \fgas~models would have on our results.
Given that 43 out of the 45 BOXSZ clusters have cluster temperatures greater than 5~keV (the other two have cluster temperatures of 4.5~keV and 4.7~keV), the constant \fgas~found by Allen08 should be valid for the BOXSZ cluster sample as well.
The gas mass profile is used to derive \rsmall~and \Msmall~by solving an implicit equation,
\begin{equation}
M_{2500} = \frac{M_{\rm{gas,2500}}}{f_{\rm{gas},2500}}
=  2500 \times \frac{4}{3} \pi \rho_{\rm{cr}}(z) r_{2500}^3,
\label{eq:M10mass}
\end{equation}
using the reference value \fgas$(r_{2500}) = 0.1104$ measured by Allen08.

As detailed in M10, our procedure incorporates systematic allowances for calibration uncertainties, projection-induced scatter in \Mgas~measurements (using expectations from simulations \citep{Nagai2007HSE}), and intrinsic scatter in \fgas{ }(Allen08, see also \citealt{Mantz2014}), with a final systematic uncertainty of 8\% on the value of \Msmall.
Note that the intrinsic scatter in \fgas{ }is not expected to differ markedly between relaxed clusters such as those used by A08 and the cluster population generally. In simulations, \citet{Battaglia2013} find a fractional intrinsic scatter at \rsmall~of $\sim$$9$\% for a representative sample of massive clusters, consistent with our estimate of systematic uncertainties.

\cite{Kravtsov2006} propose an alternative, \Ysz-like, X-ray observable, \Yx$\equiv$\Mgas$T_x$.
Several groups have used \Yx~as a mass proxy for both cosmological analysis \citep[e.g.,][]{Benson2013} and scaling relations \citep[e.g.,][]{Arnaud2010,Andersson2011,Planck2011SR}. 
Although we do not use \Yx~as a mass proxy in this work, we do fit scaling relations between \Ysmall~and \Yx~in order to provide a direct comparison between our SZE and X-ray data that is independent of mass calibration and the choice of mass proxy.

\begin{figure}
  \centering
  \epsscale{1.15}
  \plotone{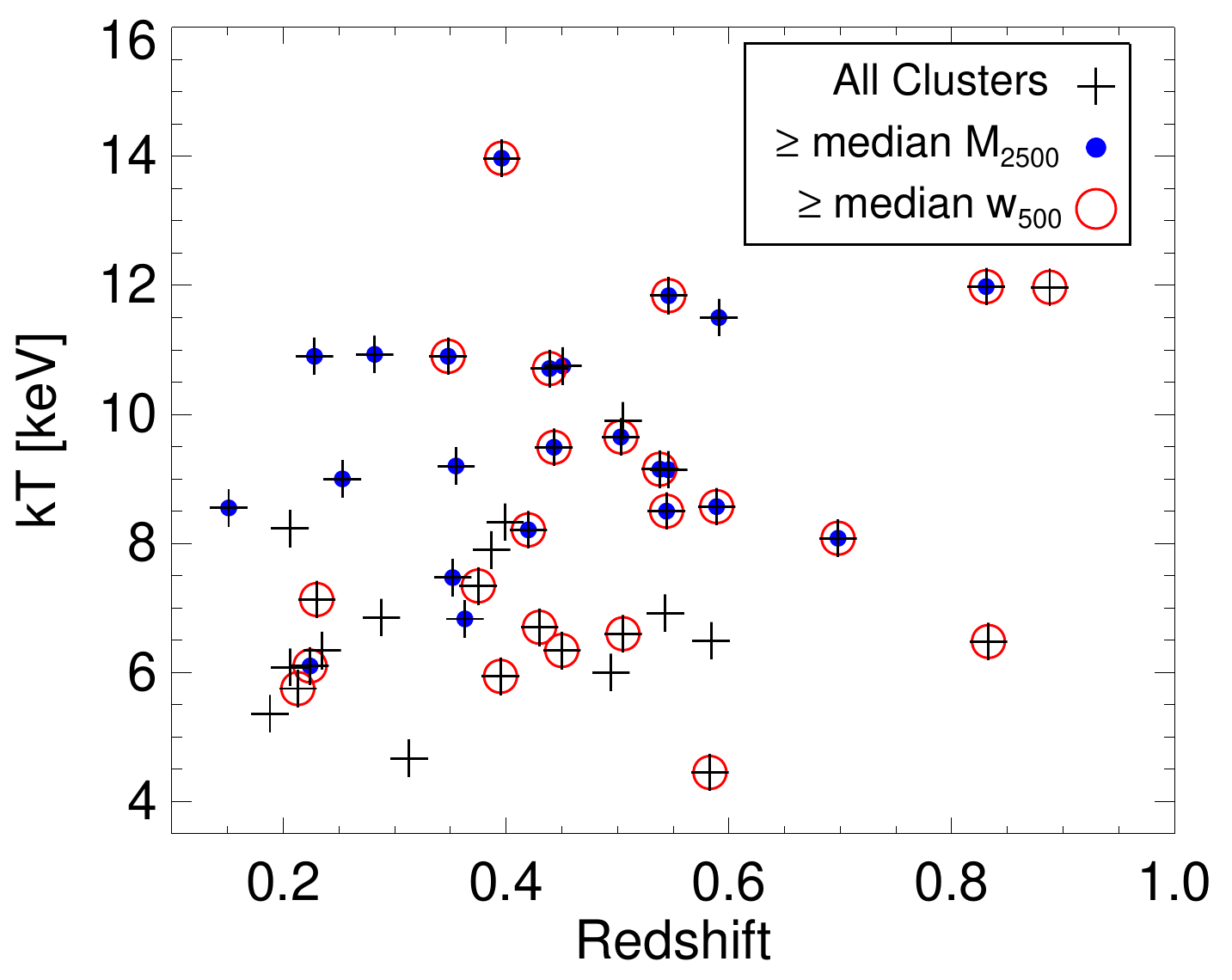}
  \caption{X-ray temperature (keV) and redshift distribution of the BOXSZ cluster sample.
Black crosses: the entire BOXSZ cluster sample.
Filled blue circles: the 23 clusters more massive than the median, $\langle M_{2500}\rangle=3.0\times10^{14}$\Msun.
We use $w_{500}$ (described in Section \ref{sec:xray}) to quantify the degree to which a cluster is dynamically disturbed.
Open red circles: the 23 clusters with $w_{500}$ values greater than the median, $\langle w_{500}\rangle=0.7\times10^{-2}$.
Of all the observables shown in this plot, the only clear correlation within the BOXSZ sample is between mass and X-ray temperature.
}
\label{fig:z_v_t}
\end{figure}

The present work uses centroid variance, \wbig, a measure of how much the body of the X-ray emission is displaced from its core~\citep{Mohr1993}, as a proxy for the dynamical state of the BOXSZ clusters. 
The \wbig~measurements were calculated based on the method of \citet{Maughan2008,Maughan2012} and are presented in \citet{SayersPressure}, where clusters with \wbig~$>0.01$ (approximately one third of our sample) are classified as disturbed. 
The temperature and redshift distributions of the BOXSZ sample, as well as subsamples based on the median values of \wbig~and \Msmall, are depicted in Figure~\ref{fig:z_v_t}.
The fractions of disturbed and cool-core clusters, the former an indicator of morphological state and the latter an indicator of entropic state, are consistent with the fractions found in samples selected on X-ray luminosity at comparable redshifts \citep[e.g.,][]{Allen2011}. 
\section{Bolocam Sunyaev-Zel'dovich Effect Data}
\label{sec:boloy}
\subsection{The Sunyaev Zel'dovich Effect}
\label{sec:ysz}
The thermal SZE spectral distortion of the CMB can be expressed as:
\begin{equation}
\Delta T_{\rm{SZE}} = f(x,T_{\rm{e}})yT_{\rm CMB},
\end{equation}
with
\begin{equation}
f(x,T_{\rm{e}}) = \left(x \frac{e^x+1}{e^x-1} - 4\right)\left(1+\delta_{\rm R}(x,T_{\rm{e}})\right).
\end{equation}
The $f(x,T_{\rm{e}})$ term contains all the spectral information and, in the low $T_{\rm e}$ limit, it is solely a function of the Boltzmann ratio of the CMB itself, $x=h\nu/k_{\rm{B}}T_{\rm CMB}$. 
Here, $h$ is the Planck constant, $k_{\rm{B}}$ is the Boltzmann constant, $T_{\rm{CMB}}$ is the CMB temperature, and $\nu$ is the photon frequency.
CMB photons receive a net blueshift via the SZE, and at approximately 219 GHz the net photon gain balances the net photon loss in occupation number, resulting in a null signal. 
Relativistic corrections to the SZE signal can be included by multiplying $f(x)$ by the frequency and electron-temperature dependent factor $(1 + \delta_{\rm{R}}(x,T_{\rm{e}}))$ \citep{Itoh1998}. We use the \Tx~values listed in Table \ref{tab:szx_info}{ }as the $T_e$ values with which to compute a single value for the relativistic correction for each cluster, which is generally $\lesssim10$\%. 
Since temperature profiles of clusters are not strictly constant, using a single temperature to compute the relativistic corrections may result in a bias.
However, even in the extreme case of strong cool-core clusters, the total variation in temperature within \rsmall~is generally less than 50\% of the average temperature.
Therefore, even if we consider one of these extreme cases, and if we further assume the limiting scenario where the bias is equal to the maximum deviation from the average temperature, then the resulting bias in the relativistically corrected SZE signal would be $\lesssim5$\%, which is small when added in quadrature to our statistical uncertainty in measuring the SZE signal (see Table \ref{tab:szx_info}).

The Compton parameter, $y$, represents the magnitude of the SZE distortion. 
This term is directly proportional to the electron pressure, $P_{\rm{e}}$, integrated along the line-of-sight:
\begin{equation}
y = (\sigma_{\rm{T}}/m_{\rm{e}}c^2)\int P_{\rm{e}}~dl.
\label{eq:little_y}
\end{equation}
The SZE signal is often expressed as a volume integral:
\begin{equation}
  \label{eq:big_y}
  Y_{\rm{SZ}}D_{\rm{A}}^2  =  \int y~dA = D_{\rm{A}}^2\int y~d\Omega,
\end{equation}
where $D_{\rm{A}}$ is the angular diameter distance of the cluster and $\Omega$ is 
the solid angle of the integration. 
$Y_{\rm{SZ}}D_{\rm{A}}^2$ is proportional to the total thermal energy of the ICM, which under the limit of HSE corresponds 
directly to the total cluster mass and motivates the use of $Y_{\rm{SZ}}D_{\rm{A}}^2$ as a mass proxy.
If the integration solid angle does not encircle the entire signal region, then the PSF may cause an apparent signal loss by transferring signal from the inner regions of the cluster to the outer regions.
We model this effect as a multiplicative parameter analogous to $\delta_{\rm{R}}$, i.e. $(1+\delta_{\rm PSF}(r_{\rm{\Delta}}))$.
The $\delta_{\rm{R}}$ and $\delta_{\rm PSF}$ correction factors are listed in Table \ref{tab:szx_info} and we discuss these corrections further in Appendix \ref{sec:r500_v_r2500}.

Together with Equation \ref{eq:little_y}, Equation \ref{eq:big_y} presents \Ysz~as a cylindrical integral of the electron pressure. 
As a result, our \Ysmall--\Msmall~scaling relation analysis uses a cylindrical \Ysz~measurement and a spherical \Mtot~measurement, with both parameters integrated within a solid angle extending to \rsmall.
Simulations and observations indicate that clusters, regardless of morphology, have similar scaled pressure profiles beyond \rsmall~\citep[see for example][]{SayersPressure}.
Therefore, the power-law index relating \Ysmall~and \Msmall~should be the same regardless of whether a spherical or cylindrical
integral is used to obtain \Ysmall.
However, given the cluster-to-cluster scatter about the average scaled pressure profile, scaling relations using cylindrical \Ysmall~may suffer larger scatter than those using spherical \Ysmall.
\subsection{Calibration, Noise Removal, and Transfer Function Deconvolution}
\label{sec:calibration}
We now highlight the main features of the Bolocam data reduction presented in \citet{SayersMorphology}.
Pointing models are constructed for each cluster using 10-minute-long observations of mm-bright point sources taken approximately once per hour during cluster observations.
These models are accurate to $\simeq5$\arcsec, and this pointing uncertainty produces an effective broadening of our point-spread function (PSF).
Specifically, an effective PSF is determined by convolving Bolocam's nominal PSF, which has a FWHM of 58\arcsec, with a two-dimensional Gaussian profile of width $\sigma = 5$\arcsec.
This broadening of our PSF due to pointing uncertainties is small, and does not have a significant impact on our derived results (especially for resolved objects like galaxy clusters).
Flux calibration is performed with nightly 20-minute observations of Uranus and Neptune together with other secondary calibrators given in \citet{Sandell1994}.
The absolute fluxes of Uranus and Neptune were determined using the models of \citet{Griffin1993}, rescaled based on recent WMAP measurements \citep{Weiland2011} as detailed in \citet{SayersFluxcal}.
The overall uncertainty on our flux calibration is 5\%.
Atmospheric brightness fluctuations are removed from the data-streams of each detector by first subtracting the response-weighted mean detector signal and then applying a 250 mHz high-pass filter.
This process removes some cluster signal and is weakly dependent on cluster shape.
As described in detail in \citet{SayersMorphology}, an iterative process is used to determine the signal transfer function separately for each cluster.
Each iteration involves processing a parametric model through the data reduction pipeline, computing a signal transfer function by comparing the output shape of this model to the input shape, fitting a parametric model to the data assuming this transfer function, and then using this parametric fit as the input to the next iteration. This process converges quickly---generally within two iterations.
The measured signal transfer function can then be applied to a model cluster profile in order to compare it with the processed Bolocam image of the cluster, or it can be used to deconvolve the signal transfer function to obtain an unbiased image of the cluster.
The processed images are 14\arcmin$\times$14\arcmin~in size, while the deconvolved images are reduced to 10\arcmin$\times$10\arcmin~in size to prevent significant amplification of the largest-scale noise during the deconvolution.
Both sets of images are included in Appendix~\ref{sec:thumbnails}.
\subsection{Noise Characterization}
\label{sec:noise}
Extracting scaling relation information from observations depends critically on an accurate characterization of the noise in the data.
This is because a misestimate of the noise will not only affect the derived uncertainty estimates, but it will also bias the determination of the best-fit scaling relation.
The Bolocam SZE cluster images contain noise from a wide range of sources: atmospheric fluctuations, instrument noise, primary CMB anisotropies, and emission from the non-uniform distribution of foreground and background galaxies.
We describe our characterization of these different sources of noise in further detail below.
There is also an uncertainty in the overall normalization of the SZE signal due to uncertainties in the absolute flux calibration.
In Section~\ref{sec:xfer_model_offset}, we discuss additional uncertainties due to the deconvolution of the signal transfer function, and in Section~\ref{sec:ysz_est} we quantify the noise in our \Ysmall~estimates that arises from our uncertainties in the overdensity radius used for integration.

For each cluster we form a set of 1000 noise realizations, which together represent our best characterization of the noise properties of the co-added Bolocam maps for that cluster.
The base for these noise estimates is created by jackknifing the approximately 50 to 100 10-minute Bolocam observations (where each observation consists of a complete sets of scans) performed on each cluster.
Specifically, we generate a jackknife map by multiplying a randomly chosen subset of half of these observations by $-1$ prior to coadding them, repeating the process 1000 times. While the resulting images contain no astronomical signal, they do retain the statistical properties of the atmospheric and instrumental noise for the ensemble of observations.

We also account for several sources of astronomical contamination.
First, using the measured angular power spectrum from SPT\citep{Keisler2011, Reichardt2012} and assuming the fluctuations are Gaussian, we generate 1000 random CMB realizations of the 140~GHz astronomical sky, adding one unique realization to each difference map.
In addition, we account for noise fluctuations due to unresolved dusty galaxies using the measured SPT power spectra from \citet{Hall2010}, again under the assumption that the fluctuations are Gaussian.
The resulting noise realizations are statistically identical to Bolocam maps of blank fields, thereby verifying that this noise model provides an adequate description of the Bolocam data.

Because bright and/or cluster-member radio galaxies are not accounted for in the SPT power spectrum, we therefore characterize and subtract them from our maps (see \citealt{SayersRadio} for a full description of this procedure).
The brightest cluster galaxy (BCG), in particular, is often a bright radio emitter, and this emission will systematically reduce the magnitude of the SZE decrement towards the cluster.
Bolocam detects a total of 6 bright radio sources in the BOXSZ maps.
These are subtracted from the cluster maps by using the Bolocam data to constrain the normalization of a point-source template centered on the coordinates determined by the 
NVSS radio survey \citep{Condon1998}.
In addition, there are NVSS-detected sources near the centers of 11 clusters in the BOXSZ sample that have extrapolated 140~GHz flux densities greater than 0.5~mJy, which is the threshold found to produce more than a 1\% bias in the SZE signal towards the cluster.
All of these sources are subtracted using the extrapolated flux density based on 1.4 GHz NVSS and 30 GHz OVRO/BIMA/SZA measurements.
The uncertainties on these subtracted point sources are accounted for in the estimated error of the measured SZE parameters by adding to each noise realization the corresponding point-source template multiplied by a random value drawn from a Gaussian distribution. 
The standard deviation of the distribution is equal to either the uncertainty on the normalization of the detected source, or based on a fixed 30\% uncertainty on the extrapolated flux density for the undetected radio sources.
\subsection{Model Fits and SZE Signal Offset Corrections}
\label{sec:xfer_model_offset}
In this analysis, we use parametric model fits for two main purposes.
First, as described in Section~\ref{sec:calibration}, we employ a particular cluster's best-fit model to determine our analysis pipeline's transfer function. 
Second, as we will describe in Section \ref{sec:ysz_est}, because the above transfer function is not well defined at zero spatial frequency, we use the model fits to constrain the deconvolved map's mean signal offset level (which we term the ``SZE signal offset"), necessary in the estimation of \Ysmall.  In this section, we describe the procedure for model fitting and offset estimation.

One of the first and most widely adopted models describing the physical properties of the ICM is the isothermal $\beta$-model \citep{Cavaliere1976}.
As higher quality X-ray data and cosmological simulations have become available, it is now clear that the $\beta$-model is insufficient in describing cluster properties at both small and large radii. 
Cosmological simulations performed by \citet[hereafter NFW]{NFW1995}, reveal a characteristic NFW dark matter profile.
Under the influence of thermal and non-thermal pressure, baryonic matter departs from faithfully mirroring the dark matter profile.
Recent work by \citet{Nagai2007CSF} and \citet[hereafter Arnaud10]{Arnaud2010} combine X-ray data at small cluster radii with simulations at large cluster radii, showing overlap in the region near \rbig. 
The characteristic profile is well-described by a generalized-NFW model (gNFW):
\begin{equation}
  p(r)=\frac{p_0}{(c_{500}~r/r_{500})^\gamma\left[1+(c_{500}~r/r_{500})^\alpha \right]^{(\beta-\gamma)/\alpha}},
  \label{eq:GNFW}
\end{equation}
where $p_0$ is the pressure normalization, $c_{500}$ is the concentration parameter which sets the radial scale, and $\alpha, \beta$, and $\gamma$ are the power-law slopes at moderate, large, and small radii. High quality SZE data, collected by the \emph{Planck}, SPT, and Bolocam instruments, have recently allowed constraints on this gNFW model using a combination of X-ray and SZE data \citep{Planck2013Pressure} and SZE data alone \citep{Plagge2010,SayersPressure}.
We follow the widely accepted practice in the literature, and use the measured gNFW power law indices of the Arnaud10 model for this analysis with $[\alpha,\beta,\gamma]=[1.05,5.49,0.31]$.
We allow $p_0$ to float in all cases and further generalize our fits to allow for ellipticity by replacing $r$ with $(1-\epsilon/2)\sqrt{r_1^2+r_2^2/(1-\epsilon)^2}$,\footnote{We choose this multiplicative prefactor so that the arithmetic mean of the major and minor axes is constant under the transformation.} where $\epsilon$ is the ellipticity and $r_1$ and $r_2$ represent the semi-major and semi-minor axes in the plane of the sky, respectively. 

The elliptical generalization of Equation \ref{eq:GNFW} is numerically integrated using Equations \ref{eq:little_y} and \ref{eq:big_y} with the additional assumption that the extent of the cluster along the line-of-sight lies between the extent of the cluster along the major and minor axes in the plane of the sky:
\begin{align}
r^2\rightarrow \ \ \ \ \ \ \ \ \ \ \ \ \ \ \ \ \ \ \ \ \ \ \ \ \ \ \ \ \ \ \ \ \ \ \ \ \ \ \ \ \ \ \ \ \ \ \ \ \ \ \ \ \ \ \ \ \ \ \ \ \nonumber \\
(1-\epsilon/2)^2\left[ r_1^2+\frac{r_2^2}{(1-\epsilon)^2} + \frac{r_3^2}{2}\left(1+\frac{1}{(1-\epsilon)^2}\right)\right].
\end{align}
That is, we assume that the cluster principal axes are in the plane of the sky and along the line-of-sight and that the semi-axis along the line-of-sight is the inverse root-mean-square average of the two semi-axes in the plane of the sky.

Our procedure for fitting the model to the data is described in detail in Section 4.3 of \citet{SayersMorphology}, and we briefly summarize it here.
First, the two-dimensional projection of the candidate model is convolved with both the Bolocam PSF and the transfer function of the Bolocam reduction pipeline.
The result is then compared to the processed map of the Bolocam data, and a \chisqr{ }value is computed based on the noise RMS of each pixel in the map (i.e., the noise covariance matrix is assumed to be diagonal).
We vary the model parameters to minimize the value of \chisqr{ }using the generalized least-squares algorithm \ttfamily MPFITFUN\normalfont\footnote{\url{http://www.physics.wisc.edu/~craigm/idl/fitting.html}}\citep{Markwardt2009}.

Due to the variety of cluster morphologies and SZE signal-to-noise within the BOXSZ sample,
the number of free parameters needed to sufficiently describe our data varies from cluster to cluster.
For all model fits, we allow $p_0$ and the model centroid to float.
We implement a statistical test, described in Appendix \ref{sec:ftest}, to determine whether to 
allow the values of $c_{500}$ and $\epsilon$ in Equation \ref{eq:GNFW} to deviate from the fiducial Arnaud10 values ($c_{500}=1.18$ and $\epsilon=0$) for individual clusters. 
This gives us four models with four different numbers of model parameters (MPs):
(1) $c_{500}$ and $\epsilon$ are fixed, 
(2) $c_{500}$ is allowed to float and $\epsilon$ is fixed,
(3) $c_{500}$ is fixed and $\epsilon$ and the position angle, $\theta$, (East of North) on the sky are allowed to float, 
and (4) $c_{500}$, $\epsilon$, and $\theta$ are allowed to float.
We will subsequently refer to these models in terms of their number of MPs: 1, 2, 3 or 4.\footnote{Since we allow the model centroid to float in RA and dec, technically, there are two additional MPs for all of these fits. For simplicity, we have chosen the numbering scheme to start with 1.} 

Once a minimal model is selected for a given cluster according to the procedure outlined in Appendix \ref{sec:ftest}, this model is used for all subsequent steps in our analysis.
The model chosen for each cluster is given in the last column of Table \ref{tab:fit_info}.
The largest fraction of the BOXSZ cluster sample, 16 clusters, are best-described using a 1-MP model, which is a spherical gNFW model with $c_{500}$ fixed to the Arnaud10/X-ray-determined value. 
The higher-order 2-, 3-, and 4-MP models are selected for 10, 12, and 7 clusters in the sample, respectively.
Therefore, approximately 42\% of the clusters in our sample prefer an elliptical over a spherical model fit, and approximately 38\% of the clusters prefer a concentration parameter that differs from X-ray-derived  value of $c_{500}=1.18$.
While the choice of model does affect the value of \Ysmall~for an individual cluster, it has little to no effect on the observed scaling relations discussed in the next section.

\begin{figure}
\centering
\includegraphics[width=0.485\textwidth]{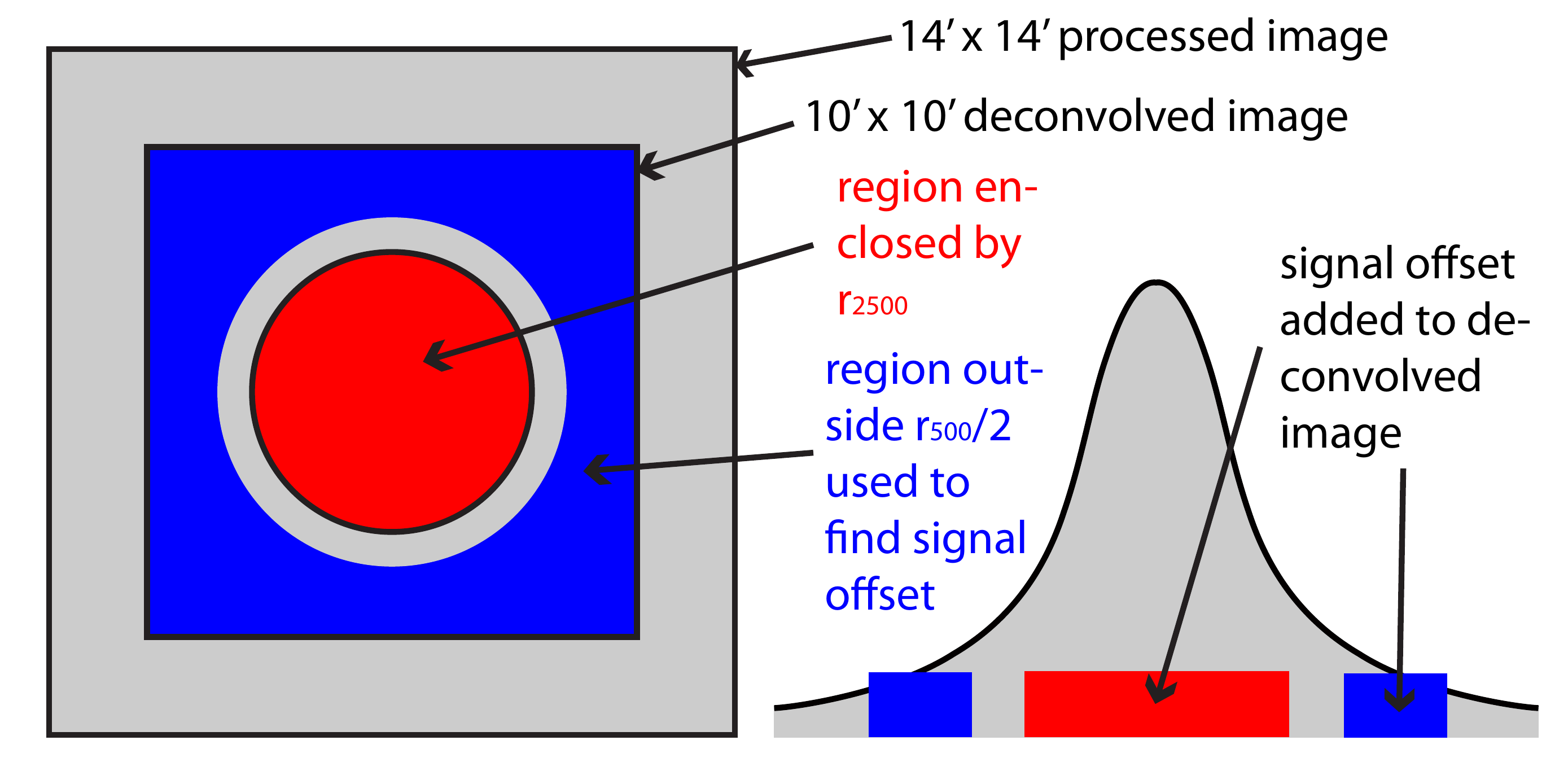}
\caption{Schematic demonstrating how the SZE signal offset for each of our deconvolved images is determined. 
The large gray box on the left shows the extent of our processed images (14\arcmin$\times$14\arcmin). Prior to deconvolving the signal transfer function, the processed image is trimmed to a 10\arcmin$\times$10\arcmin~box. 
As the transfer function for the overall signal offset of our cluster maps is not well defined, 
the SZE signal offset is determined as described in Section~\ref{sec:xfer_model_offset} and Appendix~\ref{sec:sig_offset} using the region outside of \rbig/2 (shown in dark blue).
The circular region used to compute \Ysmall~is denoted in red.
The right-hand figure depicts a 14\arcmin~one-dimensional slice through a simplified cluster illustration, showing that the cluster SZE signal is non-zero even at the edge of the image. 
The dark blue and red boxes indicate the approximate value of the SZE signal offset added to the deconvolved image and the shading corresponds to the same regions depicted in the left-hand diagram.
The SZE signal offset generally contributes $\sim$30\% of \Ysmall{ }for the lowest redshift clusters and $\sim$10\% of \Ysmall{ }for the highest redshift clusters (see Section \ref{sec:ysz_est} and Figure \ref{fig:offsety}).}
\label{fig:schematic}
\end{figure}

The minimal model required to adequately describe each cluster is then used to determine the signal offset in the deconvolved images.
In Figure \ref{fig:schematic}, we provide a schematic to aid in visualizing the following description of this process.
For each cluster, the mean signal for the deconvolved image in the region $r \ge r_{500}/2$ is set to the noise-weighted mean signal of the minimal model in the same region, and this value is called the ``SZE signal offset".\footnote{For Abell~2204, the region outside of \rbig$/2$ does not contain a sufficient number of pixels for this purpose, and we use the region outside of 4\arcmin{ }(approximately $0.45\times$\rbig) instead.}
In Section \ref{sec:ysz_est}, we quantify how the SZE signal offset affects our \Ysmall~measurements. 

In addition to $r_{500}/2$, we have explored a range of other radii to define the region used to compute the mean signal offset.
Our goal was to find a radius large enough so that the region of the image used to compute this offset is independent from the region used to determine \Ysmall, thus minimizing the model dependence of the \Ysmall~estimates.
However, at larger radii, the measurement noise on the mean signal increases quickly because the number of map pixels included in the calculation drops.
At \rbig/2, the mean-signal measurement noise is near its minimum, yet this radius is in general outside of the \rsmall~integration radius used to compute \Ysmall.
For the BOXSZ sample, \rbig/2 varies from approximately 1\arcmin~to 4\arcmin, with a median of approximately 2.5\arcmin.
\begin{deluxetable*}{lccccccccc} \tabletypesize{\scriptsize} \tablecaption{Best-Fit gNFW Pressure Parameters for the BOXSZ cluster sample.} \tablewidth{0pt} \tablehead{ Name&~~~$\Delta$RA&~~~$\Delta$DEC&$p_0$&$r_{500}/c_{500}$&$\epsilon$&$\theta$&$\chi^2$/DOF&PTE&MP\\
&~~~(arcmin)&~~~~(arcmin)&($10^{-11}\rm\frac{erg}{cm^3}$)&~(arcmin)&&(deg E of N)&&&} \startdata 
Abell~2204&$-0.13$$\pm 0.07$&\phs{$ 0.00$}$\pm 0.05$&$23.7$$\pm\phn{ 3.7}$&\phn$4.3$$\pm 0.4$&$0.26$$\pm0.06$&\phs{$ 82.6$}$\pm\phn{ 8.1}$&$1197.4/1117$&$0.03$&$ 4$\\[2pt]
Abell~383&$-0.02$$\pm 0.17$&$-0.22$$\pm 0.16$&\phn{$ 4.5$}$\pm\phn{ 0.6}$&$11.6$$\pm 3.1$&\nodata &\phs{\phn{\nodata}} &$1156.2/1118$&$0.19$&$ 2$\\[2pt]
Abell~209&$-0.01$$\pm 0.09$&$-0.02$$\pm 0.11$&\phn{$ 9.2$}$\pm\phn{ 0.8}$&\phn$6.3$ &$0.25$$\pm0.08$&$-18.4$$\pm\phn{ 9.7}$&$1206.8/1118$&$0.03$&$ 3$\\[2pt]
Abell~963&\phs{$ 0.17$}$\pm 0.11$&\phs{$ 0.08$}$\pm 0.12$&$41.8$$\pm22.7$&\phn$1.6$$\pm 0.9$&\nodata &\phs{\phn{\nodata}} &$1179.9/1118$&$0.14$&$ 2$\\[2pt]
Abell~1423&$-0.34$$\pm 0.34$&\phs{$ 0.27$}$\pm 0.19$&\phn{$ 7.2$}$\pm\phn{ 1.4}$&\phn$ 5.5$ &$0.50$$\pm0.15$&\phs{$ 69.8$}$\pm11.2$&$1149.8/1118$&$0.17$&$ 3$\\[2pt]
Abell~2261&$-0.48$$\pm 0.23$&\phs{$ 0.04$}$\pm 0.13$&\phn{$ 3.7$}$\pm\phn{ 0.7}$&\phn$ 6.3$ &$0.42$$\pm0.12$&\phs{$ 82.6$}$\pm\phn{ 9.0}$&$1111.8/1116$&$0.51$&$ 3$\\[2pt]
Abell~2219&$-0.16$$\pm 0.14$&\phs{$ 0.28$}$\pm 0.13$&$13.4$$\pm\phn{ 1.7}$&\phn$ 6.7$ &\nodata &\phs{\phn{\nodata}} &$1084.3/1120$&$0.70$&$ 1$\\[2pt] 
Abell~267&$-0.25$$\pm 0.15$&\phs{$ 0.15$}$\pm 0.15$&\phn{$ 7.9$}$\pm\phn{ 1.3}$&\phn$ 4.7$ &\nodata &\phs{\phn{\nodata}} &$1011.6/1119$&$0.98$&$ 1$\\[2pt]
RX~J2129.6+0005&$-0.19$$\pm 0.14$&\phs{$ 0.28$}$\pm 0.21$&\phn{$ 6.4$}$\pm\phn{ 1.0}$&\phn$ 4.8$ &$0.45$$\pm0.12$&\phs{$ 17.6$}$\pm10.4$&$1182.8/1118$&$0.07$&$ 3$\\[2pt]
Abell~1835&$-0.10$$\pm 0.07$&\phs{$ 0.04$}$\pm 0.11$&\phn{$ 9.3$}$\pm\phn{ 1.1}$&\phn$ 5.4$ &$0.26$$\pm0.07$&$-15.6$$\pm\phn{ 9.5}$&$ 967.1/ 946$&$0.23$&$ 3$\\[2pt]
Abell~697&$-0.07$$\pm 0.05$&$-0.27$$\pm 0.09$&\phn{$ 9.1$}$\pm\phn{ 0.6}$&\phn$ 5.5$ &$0.37$$\pm0.04$&$-21.2$$\pm\phn{ 3.8}$&$1284.2/1118$&$0.00$&$ 3$\\[2pt] 
Abell~611&$-0.08$$\pm 0.15$&$-0.33$$\pm 0.13$&\phn{$ 8.4$}$\pm\phn{ 1.1}$&\phn$ 4.0$ &\nodata &\phs{\phn{\nodata}} &$1120.5/1120$&$0.46$&$ 1$\\[2pt]
MS~2137&\phs{$ 0.03$}$\pm 0.24$&$-0.20$$\pm 0.26$&\phn{$ 5.5$}$\pm\phn{ 1.1}$&\phn$ 3.3$ &\nodata &\phs{\phn{\nodata}} &$1124.8/1120$&$0.42$&$ 1$\\[2pt]
Abell~S1063&\phs{$ 0.20$}$\pm 0.13$&\phs{$ 0.10$}$\pm 0.13$&$15.6$$\pm\phn{ 1.8}$&\phn$ 5.0$ &\nodata &\phs{\phn{\nodata}} &$1113.5/1120$&$0.43$&$ 1$\\[2pt]
MACS~J1931.8-2634&$-0.06$$\pm 0.12$&\phs{$ 0.33$}$\pm 0.16$&\phn{$ 9.9$}$\pm\phn{ 1.2}$&\phn$ 3.8$ &\nodata &\phs{\phn{\nodata}} &$1180.4/1120$&$0.11$&$ 1$\\[2pt]
MACS~J1115.8+0129&$-0.04$$\pm 0.13$&\phs{$ 0.61$}$\pm 0.16$&\phn{$ 4.5$}$\pm\phn{ 0.8}$&\phn$6.6$$\pm 1.6$&$0.30$$\pm0.09$&\phn{$ -0.0$}$\pm10.3$&$1179.2/1117$&$0.07$&$ 4$\\[2pt]
MACS~J1532.8+3021&\phs{$ 0.05$}$\pm 0.15$&\phs{$ 0.04$}$\pm 0.15$&\phn{$ 6.3$}$\pm\phn{ 1.1}$&\phn$ 3.7$ &\nodata &\phs{\phn{\nodata}} &$1204.2/1120$&$0.03$&$ 1$\\[2pt]
Abell~370&$-0.06$$\pm 0.10$&$-0.34$$\pm 0.10$&{$10.0$}$\pm\phn{1.1}$&\phn$3.8$ &\nodata &\phs{\phn{\nodata}} &$1143.2/1120$&$0.29$&$ 1$\\[2pt]
MACS~J1720.2+3536&$-0.37$$\pm 0.24$&\phs{$ 0.14$}$\pm 0.11$&\phn{$ 1.9$}$\pm\phn{ 0.4}$&$21.4$$\pm 5.8$&$0.47$$\pm0.07$&$-83.8\pm6.6$\footnotemark[1]&$1210.7/1117$&$0.02$&$ 4$\\[2pt]
ZWCL~0024+17&\phs{$ 1.04$}$\pm 0.37$&$-0.32$$\pm 0.46$&\phn{$ 4.4$}$\pm\phn{ 1.8}$&\phn$ 2.7$ &\nodata &\phs{\phn{\nodata}} &$1201.4/1120$&$0.04$&$ 1$\\[2pt]
MACS~J2211.7-0349&\phs{$ 0.05$}$\pm 0.09$&\phs{$ 0.15$}$\pm 0.09$&$16.2$$\pm\phn{ 1.8}$&\phn$ 4.2$ &\nodata &\phs{\phn{\nodata}} &$1153.0/1120$&$0.18$&$ 1$\\[2pt]
MACS~J0429.6-0253&$-0.29$$\pm 0.18$&$-0.08$$\pm 0.18$&\phn{$ 3.3$}$\pm\phn{ 0.7}$&\phn$9.7$$\pm 2.9$&\nodata &\phs{\phn{\nodata}} &$1168.4/1119$&$0.12$&$ 2$\\[2pt]
MACS~J0416.1-2403&\phs{$ 0.16$}$\pm 0.14$&\phs{$ 0.19$}$\pm 0.14$&\phn{$ 9.6$}$\pm\phn{1.3}$&\phn$ 3.2$ &\nodata &\phs{\phn{\nodata}} &$ 996.5/ 948$&$0.13$&$ 1$\\[2pt]
MACS~J0451.9+0006&\phs{$ 0.09$}$\pm 0.13$&$-0.02$$\pm 0.14$&\phn{$ 7.7$}$\pm\phn{ 1.1}$&\phn$ 2.8$ &\nodata &\phs{\phn{\nodata}} &$1164.6/1120$&$0.17$&$ 1$\\[2pt]
MACS~J1206.2-0847&\phs{$ 0.09$}$\pm 0.06$&\phs{$ 0.13$}$\pm 0.06$&$12.6$$\pm\phn{ 0.9}$&\phn$ 4.0$ &\nodata &\phs{\phn{\nodata}} &$1102.7/1120$&$0.60$&$ 1$\\[2pt]
MACS~J0417.5-1154&$-0.35$$\pm 0.06$&\phs{$ 0.19$}$\pm 0.06$&\phn{$ 8.3$}$\pm\phn{ 0.9}$&\phn$6.7$$\pm 0.9$&\nodata &\phs{\phn{\nodata}} &$1165.8/1119$&$0.13$&$ 2$\\[2pt]
MACS~J0329.6-0211&$-0.20$$\pm 0.09$&$-0.05$$\pm 0.13$&$11.0$$\pm\phn{ 1.3}$&\phn$ 2.9$ &$0.40$$\pm0.10$&\phn{$ -5.0$}$\pm\phn{ 8.5}$&$1212.5/1118$&$0.02$&$ 3$\\[2pt]
MACS~J1347.5-1144&\phs{$ 0.08$}$\pm 0.03$&$-0.11$$\pm 0.03$&$36.3$$\pm\phn{ 5.4}$&\phn$2.4$$\pm 0.2$&$0.19$$\pm0.05$&$-20.4$$\pm\phn{ 8.4}$&$1073.2/1117$&$0.79$&$ 4$\\[2pt]
MACS~J1311.0-0310&$-0.45$$\pm 0.15$&\phs{$ 0.13$}$\pm 0.15$&\phn{$ 2.9$}$\pm\phn{ 0.7}$&\phn$5.7$$\pm 1.7$&\nodata &\phs{\phn{\nodata}} &$1118.5/1119$&$0.50$&$ 2$\\[2pt]
MACS~J2214.9-1359&$-0.09$$\pm 0.10$&\phs{$ 0.07$}$\pm 0.12$&$12.9$$\pm\phn{ 1.5}$&\phn$ 3.2$ &$0.39$$\pm0.10$&\phs{$ 14.2$}$\pm\phn{ 9.8}$&$1131.5/1115$&$0.20$&$ 3$\\[2pt]
MACS~J0257.1-2325&$-0.14$$\pm 0.15$&$-0.00$$\pm 0.11$&\phn{$ 3.3$}$\pm\phn{ 0.4}$&$14.4$$\pm 3.8$&\nodata &\phs{\phn{\nodata}} &$1062.9/1119$&$0.87$&$ 2$\\[2pt]
MACS~J0911.2+1746&$-0.65$$\pm 0.33$&\phs{$ 0.08$}$\pm 0.11$&\phn{$ 7.1$}$\pm\phn{ 1.5}$&\phn$ 2.8$ &$0.79$$\pm0.11$&$-83.2$$\pm\phn{ 7.1}$&$1127.8/1118$&$0.37$&$ 3$\\[2pt]
MACS~J0454.1-0300&\phs{$ 0.09$}$\pm 0.07$&\phs{$ 0.03$}$\pm 0.05$&\phn{$ 8.0$}$\pm\phn{ 1.1}$&\phn$4.2$$\pm 0.3$&$0.26$$\pm0.06$&\phs{$ 86.4$}$\pm\phn{ 7.2}$&$1188.8/1117$&$0.07$&$ 4$\\[2pt]
MACS~J1423.8+2404&\phs{$ 0.11$}$\pm 0.11$&\phs{$ 0.25$}$\pm 0.11$&\phn{$ 9.5$}$\pm\phn{ 1.4}$&\phn$ 2.4$ &\nodata &\phs{\phn{\nodata}} &$1052.6/1120$&$0.92$&$ 1$\\[2pt]
MACS~J1149.5+2223&\phs{$ 0.00$}$\pm 0.07$&$-0.14$$\pm 0.08$&\phn{$ 5.8$}$\pm\phn{ 0.9}$&\phn$5.9$$\pm 1.4$&$0.24$$\pm0.06$&$-51.3$$\pm\phn{ 9.0}$&$1119.6/1117$&$0.44$&$ 4$\\[2pt]
MACS~J0018.5+1626&\phs{$ 0.29$}$\pm 0.10$&$-0.08$$\pm 0.10$&\phn{$ 5.6$}$\pm\phn{ 0.8}$&\phn$5.5$$\pm 1.1$&\nodata &\phs{\phn{\nodata}} &$1098.9/1119$&$0.62$&$ 2$\\[2pt]
MACS~J0717.5+3745&$-0.03$$\pm 0.04$&$-0.01$$\pm 0.04$&$38.2$$\pm\phn{ 9.9}$&\phn$2.2$$\pm 0.4$&\nodata &\phs{\phn{\nodata}} &$1188.7/1119$&$0.05$&$ 2$\\[2pt]
MS~2053.7-0449&$-0.53$$\pm 0.24$&\phs{$ 0.11$}$\pm 0.21$&\phn{$ 5.1$}$\pm\phn{ 1.4}$&\phn$ 1.8$ &\nodata &\phs{\phn{\nodata}} &$1205.9/1120$&$0.05$&$ 1$\\[2pt]
MACS~J0025.4-1222&$-0.14$$\pm 0.09$&\phs{$ 0.02$}$\pm 0.08$&\phn{$ 9.1$}$\pm\phn{ 1.0}$&\phn$ 2.4$ &\nodata &\phs{\phn{\nodata}} &$1226.7/1120$&$0.01$&$ 1$\\[2pt]
MACS~J2129.4-0741&$-0.05$$\pm 0.08$&$-0.09$$\pm 0.06$&$13.9$$\pm\phn{ 1.5}$&\phn$ 2.7$ &$0.31$$\pm0.09$&\phs{$ 65.8$}$\pm\phn{ 8.8}$&$1124.5/1118$&$0.38$&$ 3$\\[2pt]
MACS~J0647.7+7015&\phs{$ 0.02$}$\pm 0.09$&\phs{$ 0.01$}$\pm 0.09$&\phn{$ 7.1$}$\pm\phn{ 1.4}$&\phn$4.4$$\pm 0.8$&\nodata &\phs{\phn{\nodata}} &$1128.9/1119$&$0.26$&$ 2$\\[2pt]
MACS~J0744.8+3927&\phs{$ 0.10$}$\pm 0.06$&\phs{$ 0.04$}$\pm 0.13$&$10.9$$\pm\phn{ 1.0}$&\phn$ 2.5$ &$0.56$$\pm0.09$&\phn{$ -2.8$}$\pm\phn{ 4.9}$&$1265.2/1118$&$0.00$&$ 3$\\[2pt]
MS~1054.4-0321&\phs{$ 0.23$}$\pm 0.09$&\phs{$ 0.08$}$\pm 0.08$&\phn{$ 5.7$}$\pm\phn{ 1.4}$&\phn$3.7$$\pm 0.7$&\nodata &\phs{\phn{\nodata}} &$1086.1/1119$&$0.77$&$ 2$\\[2pt]
CL~J0152.7&\phs{$ 0.38$}$\pm 0.12$&\phs{$ 0.68$}$\pm 0.21$&\phn{$ 2.0$}$\pm\phn{ 0.4}$&\phn$8.3$$\pm 2.7$&$0.36$$\pm0.09$&\phs{\phn{$  8.4$}}$\pm\phn{ 8.3}$&$1220.3/1117$&$0.01$&$ 4$\\[2pt]
CL~J1226.9+3332&$-0.09$$\pm 0.11$&\phs{$ 0.14$}$\pm 0.06$&$17.8$$\pm\phn{ 1.9}$&\phn$ 1.8$ &$0.54$$\pm0.10$&\phs{$ 71.2$}$\pm\phn{ 7.6}$&$1293.2/1118$&$0.00$&$ 3$
\enddata
\footnotetext[1]{\scriptsize The model+noise fits for the preferred MACS J1720.3 4-MP model do not return a physically reasonable distribution of position angles, and therefore do not provide an accurate characterization of the uncertainty on this parameter. This is because the fits do not fully explore the range of possible position angles, perhaps due to the large value of \rbig/$c_{500}$ for this cluster. As a result, we have estimated the uncertainty on the position angle for MACS~J1720.3 using the distribution of values from the 3-MP model+noise fits.}
\tablecomments{The best-fit pressure profile parameters for the BOXSZ cluster sample. The second and third columns give the shift of the SZE-centroid of the best-fit model with respect to the X-ray centroid given in Table \ref{tab:obs_info}. The fourth, fifth, sixth, and seventh columns give the amplitude, scale radius in terms of \rbig~and $c_{500}$, ellipticity, and position angle of the major elliptical axis, $\theta$ (these parameters are introduced in Section \ref{sec:xfer_model_offset}). The eighth column gives the best-fit \chisqr~followed by the number of degrees of freedom of the gNFW profile fits. 
The ninth column gives the probability for the model+noise-derived \chisqr~values to exceed the measured \chisqr~for the best-fit minimal model. 
Specifically, the model+noise-derived \chisqr~distributions, as introduced in Section \ref{sec:xfer_model_offset}, are for the best-fit minimal model added to each of the noise realizations and fit with the minimal number of model parameters. 
A $0.00$ entry indicates this probability is less than 1\%. The final column gives the number of model parameters of the minimal model as described in Section \ref{sec:xfer_model_offset}. (1) represents a spherical model with a scale radius fixed based on the X-ray-derived \rbig~and the value $c_{500}=1.18$ from Arnaud10, (2) represents a spherical model with a floating scale radius, (3) represents an elliptical model where the principal axes are fixed based on the value from (1) according to the procedure outlined in Section \ref{sec:xfer_model_offset} , and (4) represents an elliptical model with a floating radius.} \label{tab:fit_info}
\end{deluxetable*}

The best-fit pressure profile parameters for the BOXSZ sample are presented in Table~\ref{tab:fit_info}. 
Because the noise covariance matrix is not strictly diagonal, as assumed in the fit, we compute the uncertainties on the fitted parameter values using the distributions of parameter values obtained from fits to the model+noise realization maps described in Appendix \ref{sec:sig_offset}.
The upper (lower) uncertainty of each fit parameter is the distance between the 84.1 (15.9) percentile and the median of the corresponding parameter value distributions.

The Bolocam processed and deconvolved maps, including the 1000 noise realizations, for the clusters in the BOXSZ sample are now available at \url{http://irsa.ipac.caltech.edu/Missions/bolocam.html}. Appendix~\ref{sec:thumbnails} contains thumbnails of the processed and deconvolved SZE maps for our entire data set.
\subsection{\Ysmall~Estimation}
\label{sec:ysz_est}
\begin{figure}
  \centering
  \epsscale{1.17}
  \plotone{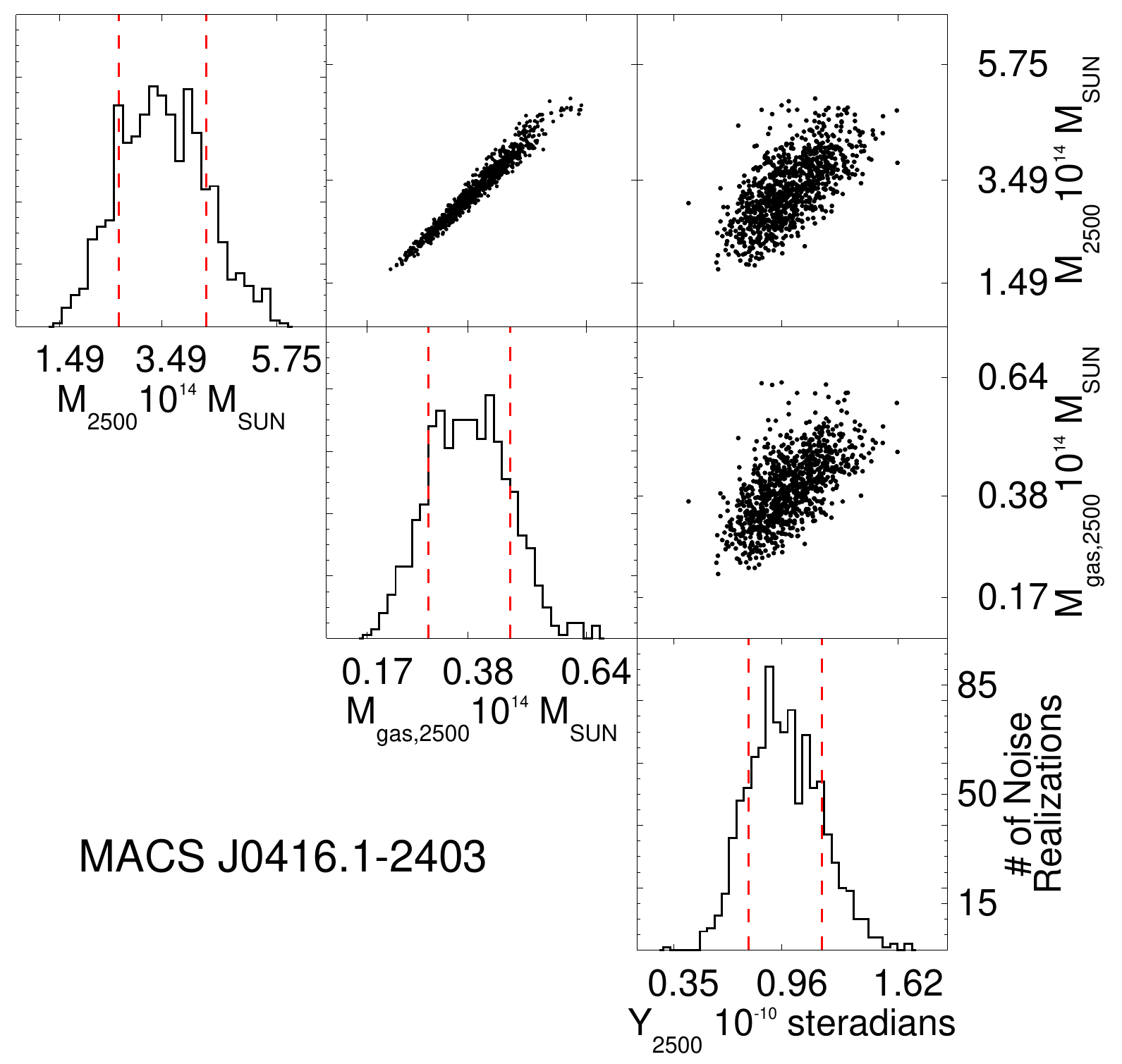}
  \caption{Histograms: one-dimensional marginalized probability distributions for the measured \Msmall, \Mgassmall, and \Ysmall~of MACS~J0416.1-2403, a typical Bolocam cluster at $z=0.42$, with $M_{2500}=3.4 \times 10^{14}$\Msun. 
The dashed red lines are the 68\% confidence regions as determined directly from the noise realizations.
All histograms are normalized to 100 noise realizations and share a common horizontal axis with the scatter plots in the same column.
Scatter plots: two-dimensional distributions for the physical properties given in the corresponding vertical and horizontal histograms.
Note that the uncertainty in \Mgassmall~is predominantly a result of the uncertainty in $f_{\rm{gas},2500}$, which affects the uncertainty in \Msmall~and subsequently the integration aperture, \rsmall. In the plot, \Ysmall~is not corrected for beam smoothing effects (discussed in the text).
}\label{fig:margin}
\end{figure}
The signal-offset-corrected deconvolved SZE images are directly integrated using Equation \ref{eq:big_y} to determine the best-fit value of \Ysmall~for each cluster, with the integration extending over the solid angle within \rsmall.
The motivation for choosing \rsmall~instead of \rbig, which is an oft-adopted mass proxy radius,
is described in Appendix \ref{sec:r500_v_r2500}.
Each of the 1000 signal-offset-corrected deconvolved noise realizations is integrated within \rsmall. 
The integrated value, $\Delta Y_{2500,i}$, for noise realization $i$, contains no cluster signal. We therefore use the quantity $Y_{2500} + \Delta Y_{2500,i}$ to estimate the distribution of $Y_{2500}$ values given the noise properties of the Bolocam data (see Figure \ref{fig:margin}). 
As can be seen from Equation \ref{eq:M10mass}, an uncertainty in \Msmall~translates directly into an uncertainty in the X-ray estimated \rsmall.
To account for the uncertainty in \Ysmall~due to uncertainties in the X-ray-derived value of \rsmall,
the integration radius for each noise realization is randomly drawn from the distribution of \rsmall~values produced by the Monte Carlo chains obtained from the X-ray data.
An example of the final \Msmall, \Mgassmall, and \Ysmall~probability distributions for MACS~J0416.1 is shown in Figure \ref{fig:margin}. 

In contrast to the distribution of \Msmall~values, which is approximately log-normal, the distribution of \Ysmall~values is approximately normal.
Since the scaling relation formalism in Section~\ref{sec:SR} assumes log-Gaussian error, the effects of the Gaussian distribution of \Ysmall~values are accounted for when we implement our default scaling relation fitting procedure as part of our selection bias characterization, and we describe this in detail in Appendix~\ref{sec:sf}.
\begin{figure}
\begin{center}
  \includegraphics[scale=0.525]{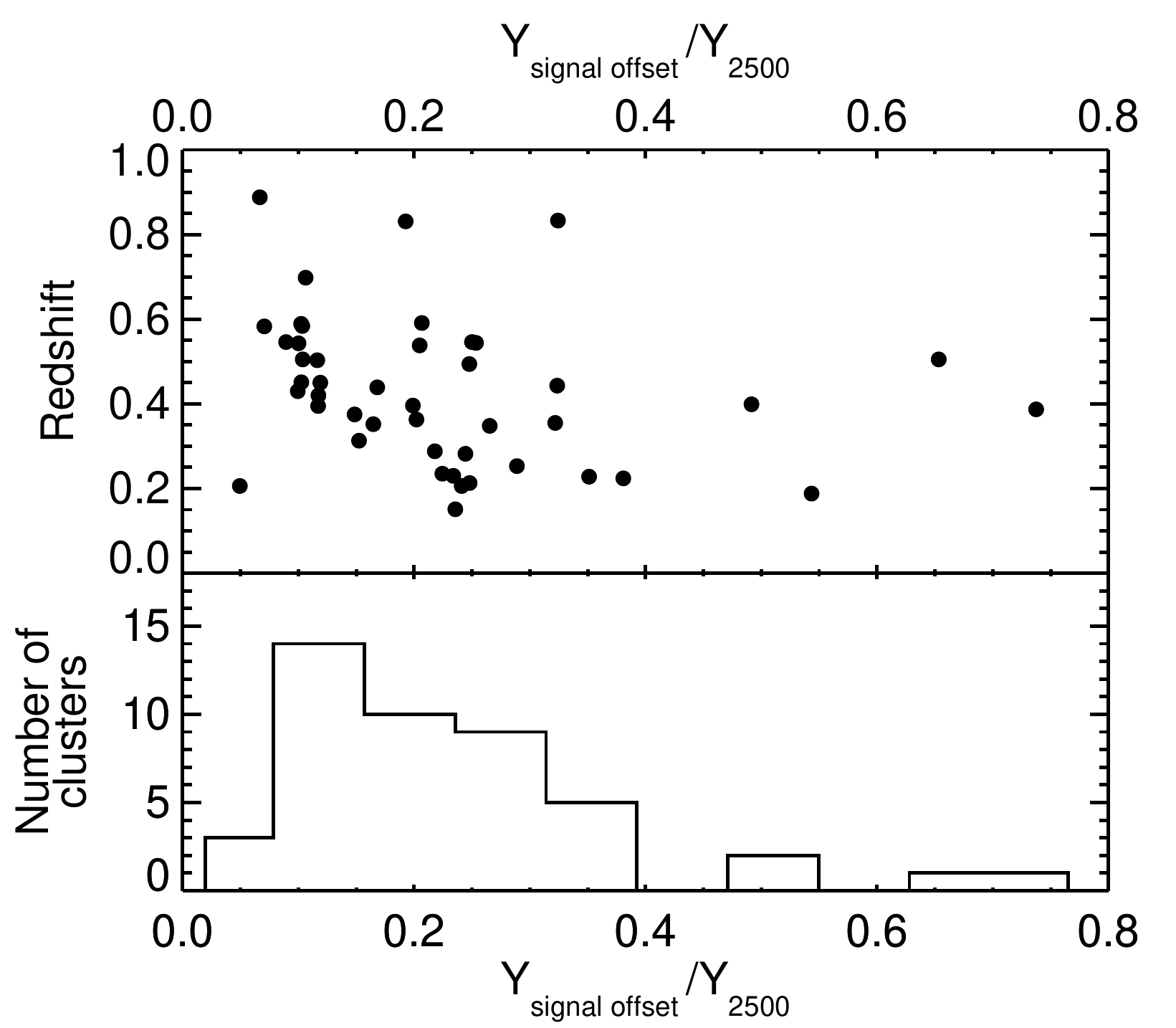}
\end{center}
  \caption{The fractional contribution of the SZE signal offset to \Ysmall.
``$Y_{\rm signal\ offset}$" is depicted schematically as the red box in Figure \ref{fig:schematic} and refers to the integrated SZE signal offset (described in Section \ref{sec:xfer_model_offset}) within \rsmall, constrained using the minimal parametric model outside \rbig/2.
As described in the text, these $Y_{\rm signal\ offset}$ values can be interpreted as upper limits on the model
dependence of our \Ysmall{ }values.
Upper plot: redshift versus the fractional $Y_{\rm signal\ offset}$ for the BOXSZ cluster sample. Lower plot: histogram of the fractional $Y_{\rm signal\ offset}$ for the BOXSZ cluster sample.}\label{fig:offsety}
\end{figure}

The method we have employed to compute \Ysz~differs from the parametric fitting methods used in other scaling relation analyses (e.g., \citealt{Bonamente2008,Marrone2009,Marrone2012,Planck2011SR,Planck2013SZELens}), as we do not parameterize the detected signal.
We use parametric models solely to determine the signal transfer function (which depends very weakly on the cluster shape) 
and to constrain the SZE signal offset (as described in Section \ref{sec:xfer_model_offset}). The fractional contribution of the SZE signal offset to \Ysmall{ }is shown in Figure \ref{fig:offsety}.
In general, this contribution is approximately $20$\%, although it is higher in a set of four clusters with large values of \rbig$/c_{500}$ (Abell~383, MACS~J1720.2+3536, MACS~J0257.1-2325, MACSJ~0429.6-0253). This fraction can be interpreted as an upper limit on the model dependence of our results, as it provides the change in $Y_{2500}$ that would result from making the maximally extreme assumption that the deconvolved map should have zero mean outside of $r_{500}/2$.

\begin{figure}
  \plotone{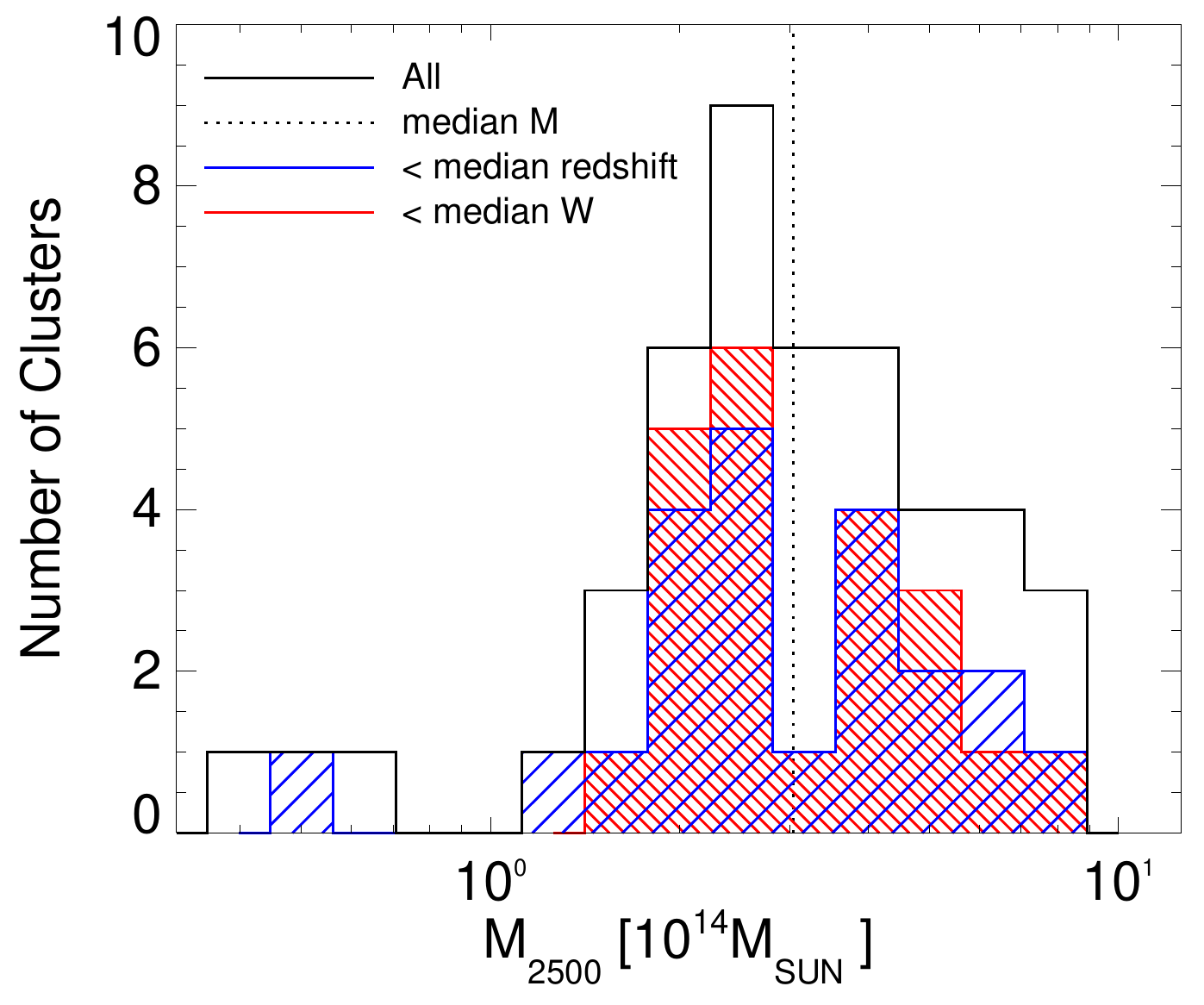}
          \caption{Distribution of \Msmall~values in the BOXSZ sample.
The median value, $\langle$\Msmall$\rangle=3.0\times10^{14}$\Msun, is indicated by the vertical dotted black line.
Blue histogram (hashed from the upper right to the bottom left): the 23 clusters that lie at or below the median sample redshift ($\langle z\rangle =0.42$).
Red histogram (hashed from the upper left to the bottom right): the 23 clusters that lie at or below the median \wbig~parameter ($\langle$\wbig$\rangle=0.7\times10^{-2}$) and are classified as being the least disturbed. Both morphological state and redshift have unbiased distributions with respect to cluster mass.}\label{fig:M}
\end{figure}

\begin{figure}
  \plotone{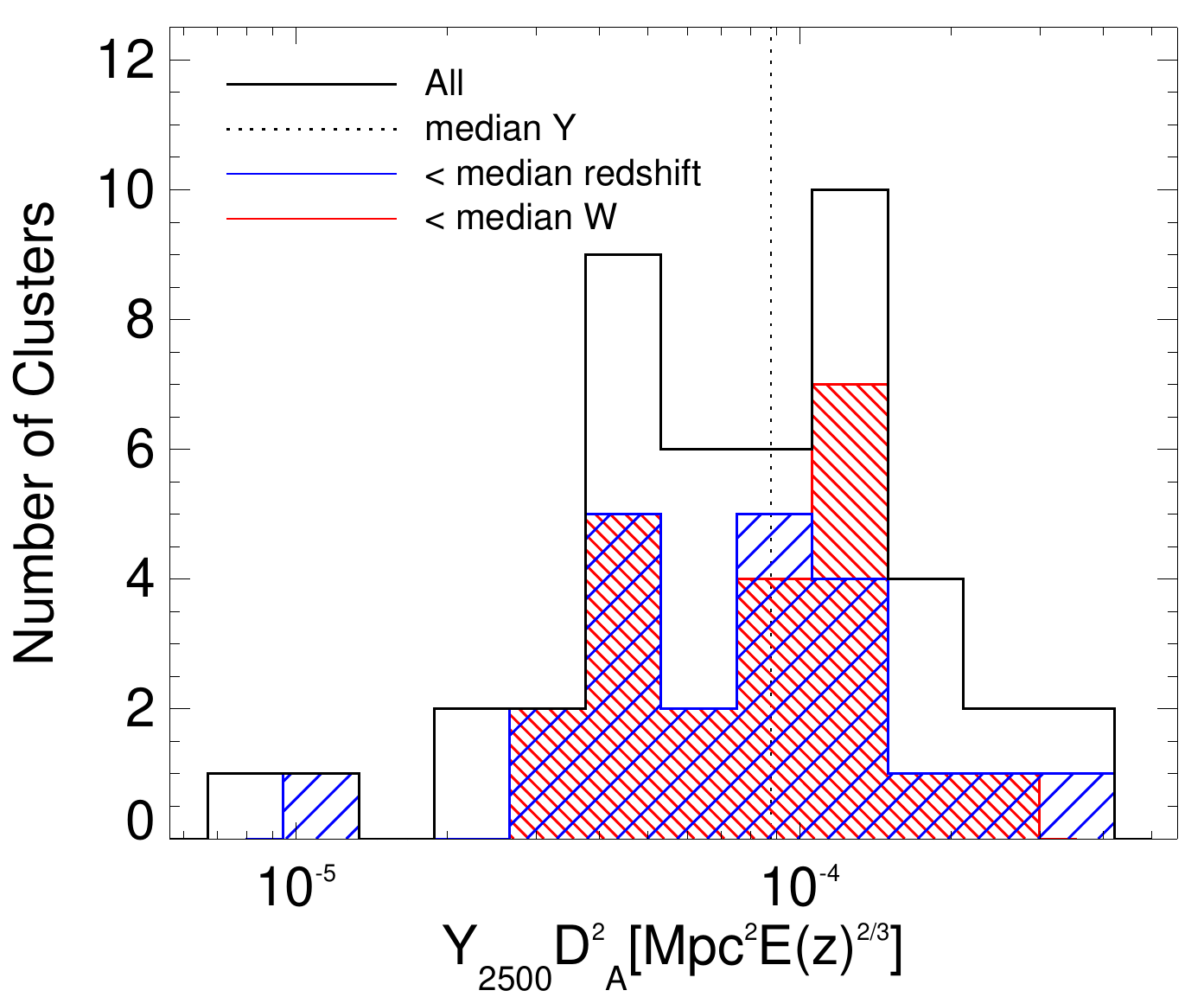}
          \caption{Analogous to Figure \ref{fig:M}, except using \Ysmall~instead of \Msmall. Again, morphological state and redshift have unbiased distributions with respect to \Ysmall.}
          \label{fig:Y}
\end{figure}

Best-fit values for the entire cluster sample are presented in Table~\ref{tab:szx_info}. We derive the upper (lower) uncertainties from the distance between the 84.1 (15.9) percentile and the median of the distribution of $\Delta Y_{2500}$ values.
Due to the way in which $\Delta Y_{2500}$ is constructed, these uncertainties on \Ysmall{ }marginalize over all of our uncertainties on \Msmall{ }(Section \ref{sec:xray} and above) and on the SZE signal offset (Appendix \ref{sec:sig_offset}).
Figures \ref{fig:M} and \ref{fig:Y}, respectively, depict the histograms of the best-fit \Msmall~and \Ysmall~values for the entire cluster sample.\\
\begin{deluxetable*}{lccccrccc} \tabletypesize{\scriptsize} \tablecaption{Physical X-ray and SZE Parameters Measured in this Analysis.} \tablewidth{0pt} \tablehead{ Name&$z$&$r_{2500}$&$M_{\rm gas,2500}$&$M_{\rm tot,2500}$&\myalign{c}{$kT_x$}&\myalign{c}{$Y_{2500}$}&~~$\delta_B$&~~$\delta_R$\\
&&(Mpc)&($10^{14}$\Msun)&($10^{14}$\Msun)&(keV) \ \ \ &($10^{-10}$ ster)&&} \startdata Abell~2204&0.151&0.62$^{+0.03}_{-0.03}$& 0.44$^{+0.07}_{-0.05}$& 4.00$^{+0.68}_{-0.51}$& 8.55$\pm0.58$& 3.58$^{+0.63}_{-0.67}$&  0.00&  0.06\\[2.3pt]
Abell~383&0.188&0.44$^{+0.02}_{-0.03}$& 0.16$^{+0.02}_{-0.02}$& 1.46$^{+0.22}_{-0.24}$& 5.36$\pm0.19$& 1.77$^{+0.49}_{-0.42}$&  0.02&  0.04\\[2.3pt] Abell~209&0.206&0.53$^{+0.03}_{-0.03}$& 0.29$^{+0.04}_{-0.05}$& 2.61$^{+0.41}_{-0.47}$& 8.23$\pm0.66$& 2.47$^{+0.39}_{-0.36}$&  0.01&  0.06\\[2.3pt]
Abell~963&0.206&0.50$^{+0.03}_{-0.02}$& 0.25$^{+0.03}_{-0.03}$& 2.22$^{+0.39}_{-0.30}$& 6.08$\pm0.30$& 0.60$^{+0.26}_{-0.26}$&  0.02&  0.04\\[2.3pt] Abell~1423&0.213&0.42$^{+0.03}_{-0.03}$& 0.14$^{+0.03}_{-0.02}$& 1.31$^{+0.29}_{-0.22}$& 5.75$\pm0.59$& 0.85$^{+0.34}_{-0.33}$&  0.03&  0.04\\[2.3pt]
Abell~2261&0.224&0.60$^{+0.03}_{-0.03}$& 0.43$^{+0.05}_{-0.06}$& 3.87$^{+0.56}_{-0.58}$& 6.10$\pm0.32$& 1.19$^{+0.31}_{-0.28}$&  0.01&  0.04\\[2.3pt] Abell~2219&0.228&0.71$^{+0.04}_{-0.03}$& 0.69$^{+0.10}_{-0.08}$& 6.29$^{+1.08}_{-0.87}$&10.90$\pm0.53$& 3.96$^{+0.90}_{-0.88}$&  0.01&  0.08\\[2.3pt]
Abell~267&0.230&0.48$^{+0.02}_{-0.02}$& 0.21$^{+0.03}_{-0.02}$& 1.93$^{+0.27}_{-0.26}$& 7.13$\pm0.71$& 0.89$^{+0.25}_{-0.21}$&  0.02&  0.05\\[2.3pt] RX~J2129.6+0005&0.235&0.52$^{+0.03}_{-0.02}$& 0.27$^{+0.04}_{-0.03}$& 2.47$^{+0.39}_{-0.33}$& 6.34$\pm0.62$& 0.88$^{+0.23}_{-0.23}$&  0.02&  0.05\\[2.3pt]
Abell~1835&0.253&0.65$^{+0.03}_{-0.03}$& 0.56$^{+0.07}_{-0.05}$& 5.11$^{+0.80}_{-0.57}$& 9.00$\pm0.25$& 1.86$^{+0.33}_{-0.31}$&  0.01&  0.06\\[2.3pt] Abell~697&0.282&0.64$^{+0.04}_{-0.04}$& 0.54$^{+0.09}_{-0.08}$& 4.90$^{+0.96}_{-0.89}$&10.93$\pm1.11$& 2.02$^{+0.30}_{-0.28}$&  0.01&  0.08\\[2.3pt]
Abell~611&0.288&0.49$^{+0.02}_{-0.02}$& 0.24$^{+0.03}_{-0.02}$& 2.21$^{+0.35}_{-0.27}$& 6.85$\pm0.34$& 0.65$^{+0.16}_{-0.15}$&  0.03&  0.05\\[2.3pt]
MS~2137&0.313&0.47$^{+0.02}_{-0.02}$& 0.22$^{+0.02}_{-0.02}$& 1.98$^{+0.27}_{-0.23}$& 4.67$\pm0.43$& 0.41$^{+0.11}_{-0.12}$&  0.04&  0.03\\[2.3pt]
Abell~S1063&0.348&0.75$^{+0.04}_{-0.04}$& 0.94$^{+0.15}_{-0.11}$& 8.57$^{+1.61}_{-1.28}$&10.90$\pm0.50$& 3.54$^{+0.68}_{-0.65}$&  0.01&  0.08\\[2.3pt]
MACS~J1931.8-2634&0.352&0.57$^{+0.02}_{-0.02}$& 0.42$^{+0.05}_{-0.04}$& 3.83$^{+0.51}_{-0.44}$& 7.47$\pm1.40$& 1.33$^{+0.22}_{-0.21}$&  0.03&  0.05\\[2.3pt]
MACS~J1115.8+0129&0.355&0.56$^{+0.02}_{-0.02}$& 0.40$^{+0.04}_{-0.04}$& 3.65$^{+0.44}_{-0.46}$& 9.20$\pm0.98$& 1.13$^{+0.32}_{-0.35}$&  0.03&  0.07\\[2.3pt]
MACS~J1532.8+3021&0.363&0.55$^{+0.03}_{-0.02}$& 0.38$^{+0.05}_{-0.04}$& 3.39$^{+0.55}_{-0.39}$& 6.83$\pm1.00$& 0.46$^{+0.16}_{-0.16}$&  0.04&  0.05\\[2.3pt]
Abell~370&0.375&0.48$^{+0.03}_{-0.03}$& 0.26$^{+0.04}_{-0.04}$& 2.35$^{+0.41}_{-0.47}$& 7.34$\pm0.52$& 0.91$^{+0.16}_{-0.16}$&  0.06&  0.05\\[2.3pt]
MACS~J1720.2+3536&0.387&0.49$^{+0.03}_{-0.02}$& 0.28$^{+0.04}_{-0.03}$& 2.54$^{+0.42}_{-0.33}$& 7.90$\pm0.74$& 1.21$^{+0.66}_{-0.36}$&  0.06&  0.06\\[2.3pt]
ZWCL~0024+17&0.395&0.30$^{+0.02}_{-0.02}$& 0.06$^{+0.01}_{-0.01}$& 0.55$^{+0.13}_{-0.11}$& 5.94$\pm0.87$& 0.13$^{+0.07}_{-0.07}$&  0.22&  0.04\\[2.3pt]
MACS~J2211.7-0349&0.396&0.66$^{+0.03}_{-0.03}$& 0.69$^{+0.10}_{-0.08}$& 6.30$^{+1.01}_{-0.84}$&13.97$\pm2.74$& 2.58$^{+0.36}_{-0.37}$&  0.03&  0.10\\[2.3pt]
MACS~J0429.6-0253&0.399&0.47$^{+0.02}_{-0.02}$& 0.25$^{+0.03}_{-0.03}$& 2.25$^{+0.35}_{-0.30}$& 8.33$\pm1.58$& 0.82$^{+0.25}_{-0.21}$&  0.07&  0.06\\[2.3pt]
MACS~J0416.1-2403&0.420&0.54$^{+0.04}_{-0.05}$& 0.38$^{+0.10}_{-0.10}$& 3.45$^{+0.88}_{-0.94}$& 8.21$\pm0.99$& 1.06$^{+0.25}_{-0.22}$&  0.05&  0.06\\[2.3pt]
MACS~J0451.9+0006&0.430&0.43$^{+0.04}_{-0.03}$& 0.19$^{+0.05}_{-0.04}$& 1.77$^{+0.53}_{-0.37}$& 6.70$\pm0.99$& 0.44$^{+0.11}_{-0.09}$&  0.10&  0.05\\[2.3pt]
MACS~J1206.2-0847&0.439&0.64$^{+0.03}_{-0.03}$& 0.66$^{+0.09}_{-0.07}$& 6.00$^{+0.98}_{-0.83}$&10.71$\pm1.29$& 1.91$^{+0.23}_{-0.22}$&  0.03&  0.08\\[2.3pt]
MACS~J0417.5-1154&0.443&0.70$^{+0.04}_{-0.04}$& 0.88$^{+0.13}_{-0.12}$& 7.96$^{+1.40}_{-1.28}$& 9.49$\pm1.12$& 2.81$^{+0.46}_{-0.47}$&  0.02&  0.07\\[2.3pt]
MACS~J0329.6-0211&0.450&0.49$^{+0.02}_{-0.02}$& 0.30$^{+0.03}_{-0.03}$& 2.71$^{+0.39}_{-0.32}$& 6.34$\pm0.31$& 0.64$^{+0.11}_{-0.10}$&  0.07&  0.05\\[2.3pt]
MACS~J1347.5-1144&0.451&0.71$^{+0.03}_{-0.03}$& 0.92$^{+0.10}_{-0.10}$& 8.37$^{+1.12}_{-1.05}$&10.75$\pm0.83$& 1.89$^{+0.18}_{-0.17}$&  0.02&  0.08\\[2.3pt]
MACS~J1311.0-0310&0.494&0.43$^{+0.02}_{-0.02}$& 0.21$^{+0.02}_{-0.02}$& 1.93$^{+0.28}_{-0.22}$& 6.00$\pm0.32$& 0.48$^{+0.10}_{-0.09}$&  0.11&  0.04\\[2.3pt]
MACS~J2214.9-1359&0.503&0.52$^{+0.03}_{-0.03}$& 0.38$^{+0.06}_{-0.05}$& 3.46$^{+0.70}_{-0.54}$& 9.65$\pm0.78$& 1.13$^{+0.21}_{-0.21}$&  0.07&  0.07\\[2.3pt]
MACS~J0257.1-2325&0.505&0.45$^{+0.03}_{-0.02}$& 0.23$^{+0.04}_{-0.03}$& 2.10$^{+0.40}_{-0.31}$& 9.90$\pm0.90$& 1.02$^{+0.29}_{-0.23}$&  0.11&  0.07\\[2.3pt]
MACS~J0911.2+1746&0.505&0.41$^{+0.02}_{-0.03}$& 0.17$^{+0.03}_{-0.03}$& 1.59$^{+0.29}_{-0.31}$& 6.60$\pm0.60$& 0.20$^{+0.09}_{-0.08}$&  0.15&  0.05\\[2.3pt]
MACS~J0454.1-0300&0.538&0.56$^{+0.03}_{-0.03}$& 0.51$^{+0.07}_{-0.06}$& 4.59$^{+0.79}_{-0.68}$& 9.15$\pm0.49$& 0.92$^{+0.12}_{-0.10}$&  0.06&  0.07\\[2.3pt]
MACS~J1423.8+2404&0.543&0.44$^{+0.02}_{-0.02}$& 0.25$^{+0.04}_{-0.03}$& 2.30$^{+0.39}_{-0.31}$& 6.92$\pm0.32$& 0.35$^{+0.07}_{-0.09}$&  0.12&  0.05\\[2.3pt]
MACS~J1149.5+2223&0.544&0.54$^{+0.03}_{-0.03}$& 0.46$^{+0.07}_{-0.06}$& 4.16$^{+0.78}_{-0.62}$& 8.50$\pm0.57$& 1.16$^{+0.19}_{-0.18}$&  0.07&  0.06\\[2.3pt]
MACS~J0018.5+1626&0.546&0.58$^{+0.03}_{-0.03}$& 0.54$^{+0.08}_{-0.07}$& 4.87$^{+0.82}_{-0.77}$& 9.14$\pm0.43$& 1.06$^{+0.19}_{-0.15}$&  0.06&  0.07\\[2.3pt]
MACS~J0717.5+3745&0.546&0.65$^{+0.03}_{-0.04}$& 0.77$^{+0.11}_{-0.10}$& 7.00$^{+1.14}_{-1.09}$&11.84$\pm0.54$& 1.17$^{+0.24}_{-0.22}$&  0.05&  0.08\\[2.3pt]
MS~2053.7-0449&0.583&0.28$^{+0.02}_{-0.02}$& 0.07$^{+0.02}_{-0.01}$& 0.59$^{+0.16}_{-0.12}$& 4.45$\pm0.58$& 0.06$^{+0.03}_{-0.02}$&  0.36&  0.03\\[2.3pt]
MACS~J0025.4-1222&0.584&0.45$^{+0.04}_{-0.03}$& 0.26$^{+0.06}_{-0.05}$& 2.38$^{+0.66}_{-0.50}$& 6.49$\pm0.50$& 0.29$^{+0.06}_{-0.06}$&  0.13&  0.05\\[2.3pt]
MACS~J2129.4-0741&0.589&0.48$^{+0.03}_{-0.02}$& 0.33$^{+0.05}_{-0.04}$& 3.03$^{+0.54}_{-0.43}$& 8.57$\pm0.74$& 0.73$^{+0.11}_{-0.10}$&  0.11&  0.06\\[2.3pt]
MACS~J0647.7+7015&0.591&0.52$^{+0.02}_{-0.03}$& 0.42$^{+0.05}_{-0.05}$& 3.83$^{+0.51}_{-0.54}$&11.50$\pm1.10$& 0.91$^{+0.15}_{-0.14}$&  0.09&  0.08\\[2.3pt]
MACS~J0744.8+3927&0.698&0.49$^{+0.02}_{-0.02}$& 0.38$^{+0.05}_{-0.04}$& 3.50$^{+0.53}_{-0.46}$& 8.08$\pm0.44$& 0.31$^{+0.06}_{-0.05}$&  0.13&  0.06\\[2.3pt]
MS~1054.4-0321&0.831&0.44$^{+0.03}_{-0.02}$& 0.34$^{+0.07}_{-0.04}$& 3.16$^{+0.71}_{-0.35}$&11.98$\pm1.44$& 0.32$^{+0.06}_{-0.05}$&  0.19&  0.08\\[2.3pt]
CL~J0152.7&0.833&0.22$^{+0.05}_{-0.03}$& 0.04$^{+0.03}_{-0.01}$& 0.37$^{+0.29}_{-0.12}$& 6.48$\pm0.90$& 0.14$^{+0.06}_{-0.04}$&  0.40&  0.05\\[2.3pt] CL~J1226.9+3332&0.888&0.42$^{+0.02}_{-0.02}$& 0.31$^{+0.04}_{-0.04}$& 2.77$^{+0.45}_{-0.36}$&11.97$\pm1.27$& 0.34$^{+0.06}_{-0.06}$&  0.23&  0.08
\enddata \tablecomments{The X-ray and SZE-derived properties used in the BOXSZ scaling relations analysis.The first two columns give the cluster ID and redshift. The references for the individual cluster redshift measurements are given in \citet{SayersPressure}. The third column gives \rsmall~followed by $M_{gas,2500}$, $M_{tot,2500}$ and  $kT$, which are calculated as described in \citet{Mantz2010b}.The seventh column gives \Ysmall~as measured in this work. 
The last two columns give the fractional beam-smoothing and relativistic \Ysmall~corrections.
Both terms are positive and boost the \Ysz~value compared to that obtained from direct integration of the data (see Section~\ref{sec:boloy}).} \label{tab:szx_info} \end{deluxetable*}
\section{Scaling Relations, Fitting Technique, and Bias Corrections}
\label{sec:SR}
The scale-free nature of gravitational collapse leads to the prediction that cluster ICM observables scale in a self-similar fashion with the total cluster mass in the absence of non-gravitational physics.
Cluster observables are converted to logarithmic form and are normalized to the approximate median value for the BOXSZ sample:
\begin{eqnarray}
m_{2500} &\equiv& \log_{10}\left[\frac{E(z)M_{2500}}{10^{14.5}~M_{\odot}}\right] \\
m_{500} &\equiv& \log_{10}\left[\frac{E(z)M_{500}}{10^{15}~M_{\odot}}\right]  \\
l &\equiv& \log_{10}\left[\frac{L_{500}}{E(z)10^{44}~\rm erg\cdot s^{-1}}\right] \\
t &\equiv& \log_{10}\left[\frac{kT_{\rm x}}{\rm keV}\right] \\
y_{x} &\equiv& \log_{10}\left[E(z)10^{4.5}C_X \ kT_{\rm{x}} \ M_{\rm{gas},2500}\right] \\
y &\equiv& \log_{10}\left[E(z)10^4D_{\rm{A}}^2Y_{2500}\right].
\end{eqnarray}
Where the term
\begin{eqnarray}
  C_X = \frac{\sigma_T}{m_ec^2}\frac{1}{\rho_{\rm gas}/n_e} = 1.406\times\frac{10^{-5}\rm Mpc^2}{10^{14}\rm keV \ M_\odot},
\end{eqnarray}
normalizes \Yx~to \Ysmall~with $\sigma_T$, the Thompson cross-section, $m_e$ and $m_p$, the electron and proton rest masses, respectively, and $c$ the speed of light.
For a fully ionized gas with cosmic He abundance, $\rho_{\rm gas}/n_e = 1.149~m_p$.
For this work, the \Tx~utilized to calculate \Yx~is always determined within the region $[0.15,1.0]$~\rbig. We note that this value generally differs by less than a few percent from \Tx~computed within the region $[0.15r_{500},r_{2500}]$, as is demonstrated by both M10 and V09.
Finally, the normalization factors in the definitions of the mass and Compton-$y$ variables have been chosen to force the median of each parameter over the entire sample to be approximately zero.
Effectively, this allows us to decorrelate the uncertainties in the best-fit slopes and intercepts for each scaling relation.

Using the logarithmic representations for the cluster observables, we can formulate linear relations between cluster properties, $u$ and $v$, as:
\begin{eqnarray}
\label{eq:sr}
  u=\beta_0^{uv}+\beta_1^{uv}v.
\end{eqnarray}
We occasionally will refer to the ensemble of fit parameters for a particular scaling relation as $\theta_{u|v}=(\beta^{uv}_0,\beta^{uv}_1,\sigma_{uv}^2)$, where $\sigma_{uv}^2$ is the Gaussian intrinsic scatter of the observable $u\in[l,t,y,y_{\rm x}]$ at a fixed $v$. 
We refer to $\sigma_{uv}$ as ``intrinsic scatter", and we use the term ``fractional intrinsic scatter" when referring to the fractional intrinsic scatter of the non-logarithmic observables (e.g., \Lbig, \Tx, \Ysmall, and \Yx). We calculate the fractional intrinsic scatter by dividing the relevant $\sigma_{uv}$ by $\log_{10}(e)$.

The various factors of $E(z)$ are included to account for the fact that these cluster properties are measured at constant overdensity with respect to an evolving critical density. 
By assuming self-similarity and HSE, cluster temperature should scale with cluster mass according to $\beta_1^{tm}=2/3$.
From Equations \ref{eq:little_y} and \ref{eq:big_y} we see that the \Ysz~observable is a line-of-sight integral of cluster pressure, which under the ideal gas law scales as the product of density and temperature.  In the limit that the electron density scales with total cluster mass and the cluster is in HSE, we expect the scaling between \Ysz~and \Mtot~to be $\beta_1^{ym}=5/3$.
We refer to this type of scaling as self-similar scaling and use it as a general reference point for comparison.
All of our scaling relation fits are performed using the Bayesian fitting code, \ttfamily linmix\_err\normalfont \citep{Kelly2007}, and are corrected for selection- and regression-induced biases using the procedure described below.

All of the clusters in the BOXSZ sample were selected based on the availability of \emph{Chandra} X-ray data.
In addition to this, several other factors affected the selection process.
First, some clusters were chosen to have high X-ray luminosities and spectroscopic temperatures under the expectation that these X-ray observables would correlate with a bright SZE signal. 
Second, moderate redshift clusters were given preference because those clusters were expected to have \rbig~values within the resulting 14\arcmin$\times$14\arcmin~Bolocam image. 
Finally, as there already was a large degree of overlap with the MACS $z>0.5$ \citep{Ebeling2007} and CLASH \citep{Postman2012} samples, a few clusters were chosen so that BOXSZ would have complete observations for these two catalogs.
Out of concern that the \emph{ad hoc} nature of the BOXSZ cluster selection would bias the measured scaling relations, selection effects specific to our cluster sample have been modeled.
This procedure, which includes correlations in the intrinsic scatter of different observables at fixed mass and redshift, is discussed in Appendix \ref{sec:sf}. 
As expected, due to its large intrinsic scatter, the \olm~relation is most influenced by selection effects associated with how the BOXSZ clusters were originally drawn from X-ray flux limited samples.
Due to the weak covariance of \Lx~with \Tx~and \Ysz, the BOXSZ selection has very little impact on the \otm~and \oym~relations, although our underlying fitting procedure does produce small biases in those two relations, which we correct for.
The selection-bias-corrected scaling relations are presented in Table \ref{tab:boxsz_meas}, and the correction factors are given in Table \ref{tab:boxsz_delta} of Appendix~\ref{sec:sf}.
We note that the uncertainties given in Table~\ref{tab:boxsz_meas} do not incorporate the regression- and selection-induced bias correction uncertainties given in Table~\ref{tab:boxsz_delta}, which should be considered to be systematic uncertainties on the best-fit scaling relation parameters.
The recovered \otm~and \olm~relations are consistent within 2$\sigma$ with those presented using a full Bayesian analysis of a sample of 94 clusters in M10. 
The scaling relation results will be discussed in detail in Section \ref{sec:sr_results}.

All of the uncertainties in Table \ref{tab:boxsz_meas} are directly obtained from the standard deviation of the posterior output of the best-fit parameters obtained from \ttfamily linmix\_err\normalfont. While these uncertainties do not account for the covariances and non-gaussianities in the measurement uncertainties of the observables, we have checked that this omission has a small effect. Specifically, our fits to the ensemble of mock cluster realizations in Appendix \ref{sec:sf} fully sample the \Ysmall{ }and \Msmall{ }noise distributions, including their covariance (e.g., see Figure \ref{fig:margin}), and we find that the standard deviations of the best-fit scaling relation parameters from these ensembles of fits for both the \Ysmall--\Msmall{ } and \Tx--\Mbig{ }relations are within $15\%$ of the uncertainties obtained from \ttfamily linmix\_err\normalfont.
One can therefore attribute an additional 15\% systematic uncertaintiy to the uncertainties we have quoted for the best-fit \Ysmall--\Msmall{ }scaling relation parameters.
While we have not performed such checks for the \Yx--\Msmall{ }and \Tx--\Msmall{ }scaling relations, there is no reason to expect that they would show greater inconsistency.
\section{Results and Discussion}
\label{sec:sr_results}
\begin{figure*}
\begin{center}
\begin{tabular}{cc}
\includegraphics[scale=0.53] {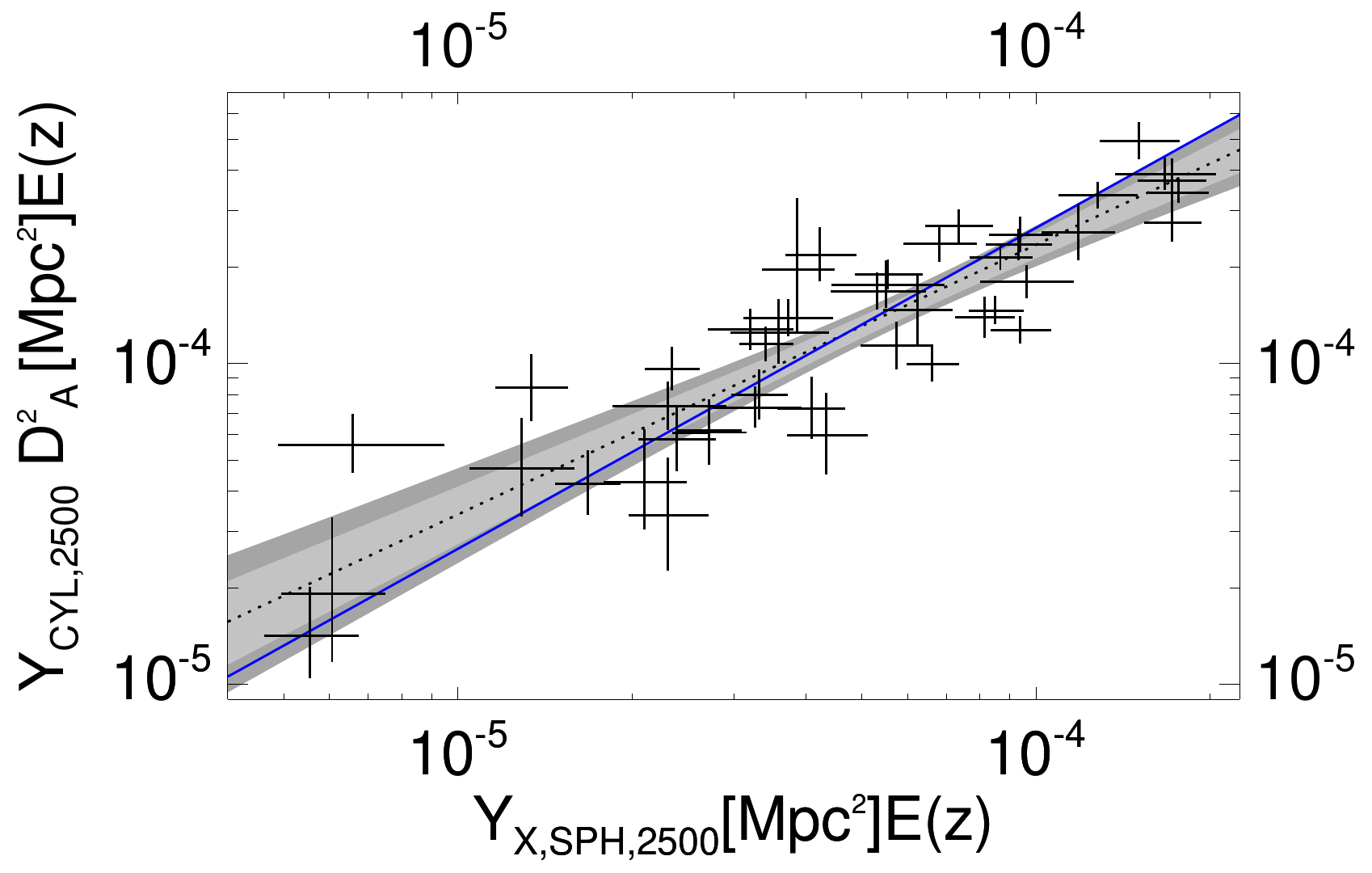} &
\includegraphics[scale=0.53] {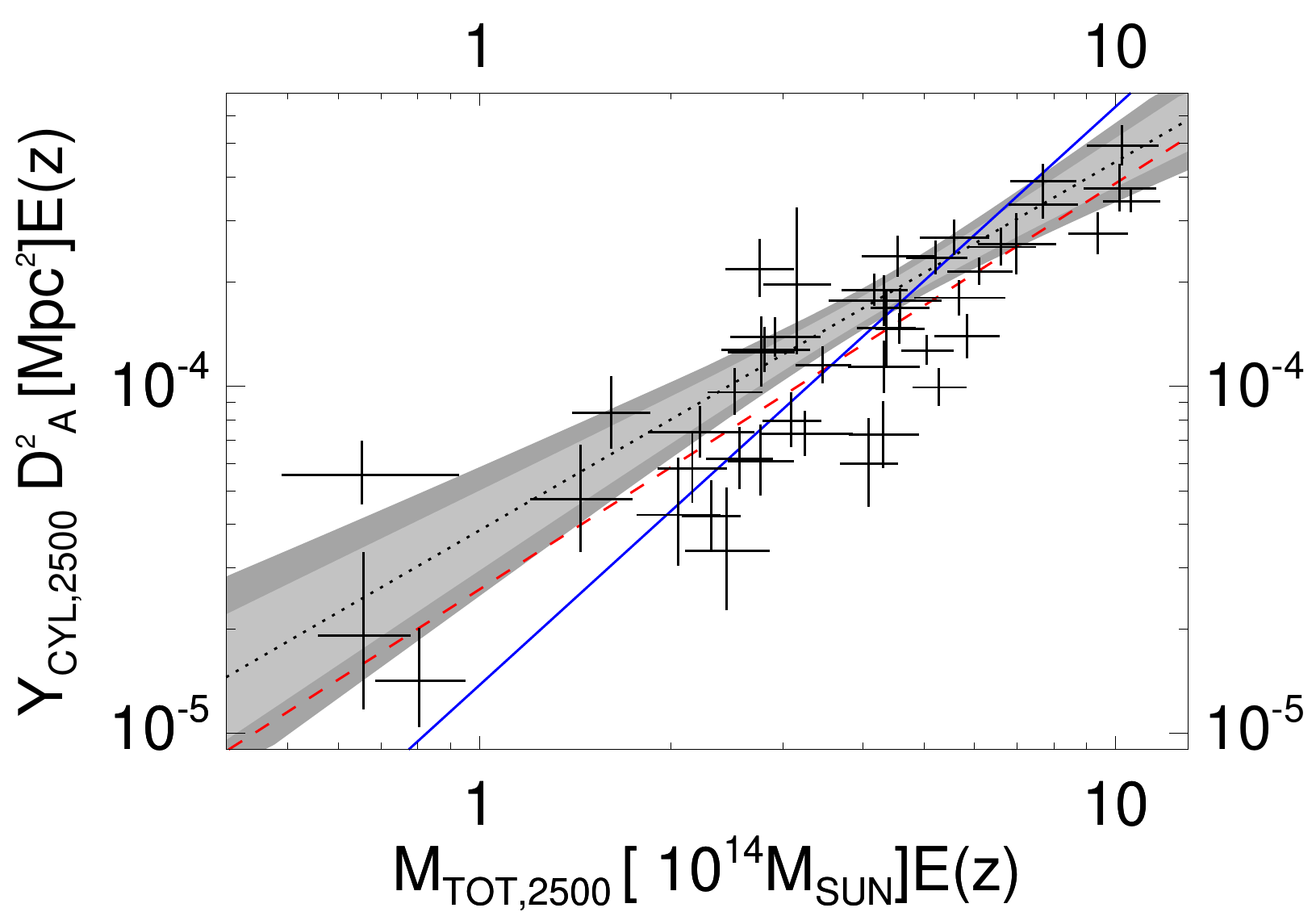}
\end{tabular}
  \caption{Observed \Ysmall--\Yx~(left panel), and \Ysmall--\Msmall~(right panel) scaling relations for the BOXSZ sample.  
The black data points represent the measured parameters and their 1$\sigma$ uncertainties. 
The dashed black line represents the best-fit to the data and the 1$\sigma$ and 2$\sigma$ uncertainties are depicted by the light gray and dark gray shading, respectively. 
These uncertainties correspond to the 68\% and 95\% joint likelihood of the measured slope and intercept for a given scaling relation.
These fits are corrected for selection effects and small biases associated with our fitting procedure. The uncorrected fit for the \Ysmall--\Msmall~relation is given by the dashed red line.
The blue line represents the best fit to the data with a fixed self-similar slope.}
\label{fig:srYvM}
\end{center}
\end{figure*}

All of the measured BOXSZ scaling relations are given in Table \ref{tab:boxsz_meas}, and \Ysmall--\Yx~and \Ysmall--\Msmall~relations are plotted in Figure \ref{fig:srYvM}. Starting with the \Ysmall--\Yx~relation, we measure the slope, \byyx~=~\byyxv, to be approximately $2\sigma$ from unity. For the \Ysmall--\Msmall~relation, plotted in the right-hand panel of Figure \ref{fig:srYvM}, we measure a best-fit slope \bym~=~\bymv, which is approximately 5$\sigma$ away from the self-similar slope of 5/3.
The \Ysmall--\Msmall~slope contrasts with previously published results, which are all consistent with self-similarity. We compare our measurements with these previous results in the following section.

\begin{deluxetable}{crrcr} \tabletypesize{\scriptsize}
\tablecaption{Measured Scaling Relations for BOXSZ Cluster Sample.}
\tablewidth{0pt} \tablehead{
$\theta$ & \multicolumn{1}{c}{~~~$\beta_0$} & \multicolumn{1}{c}{~$\beta_1$}& $\beta_{1,\rm SS}$ & \multicolumn{1}{c}{~$\sigma$}} \startdata 
$Y_{2500}$--$M_{2500}$ &0.12$\pm$0.03 & 1.06$\pm$0.12 & 5/3 & 0.11$\pm$0.04\\
$Y_{X}$--$M_{2500}$ & 0.06$\pm$0.02 & 1.36$\pm$0.06 & 5/3 & 0.03$\pm$0.03\\
$Y_{2500}$--$Y_{X}$ &  $-$0.05$\pm$0.03 & 0.84$\pm$0.07 & 1 & 0.09$\pm$0.03\\
$T_{X}$--$M_{2500}$ & $-$0.13$\pm$0.02 & 0.35$\pm$0.05 & 2/3 & 0.05$\pm$0.02
\enddata\tablecomments{First column, scaling relation; second column, intercept, $\beta_0$; third column, slope, $\beta_1$; fourth column, $\beta_{1,\rm SS}$, predicted slope for the self-similar model; and fifth column, intrinsic scatter $\sigma$. 
Except for $Y_{2500}-Y_{X}$, all relations are corrected for selection effects (see Section \ref{sec:SR} and Appendix~\ref{sec:sf}).}\label{tab:boxsz_meas} 
\end{deluxetable}

For consistency, we check to see whether our X-ray data also exhibit deviations from self-similarity, and we include the best-fit \Tx--\Msmall~and \Yx--\Msmall~scaling relations to the BOXSZ sample in Table \ref{tab:boxsz_meas} as well. 
The best-fit slope for the \Tx--\Msmall~relation, \btm$=$\btmv, is also inconsistent with a self-similar slope of 2/3.
For the \Yx--\Msmall~relation, we measure \byxm$=$\byxmv. 
These measured slopes are 2.5$\sigma$ shallower than the corresponding M10 results based on 94 clusters, which use a similar X-ray analysis (but at \rbig~rather than \rsmall).
Similar to M10 (but with greater significance), our results for the \Ysmall--\Msmall, \Yx--\Msmall~and \Tx--\Msmall{ }scaling relations all have shallower slopes than self-similar predictions.

We measure the fractional intrinsic scatter in \Ysmall~at fixed \Yx~to be \isyyxv~and the fractional intrinsic scatter in \Ysmall~at fixed \Msmall~to be \isymv, both of which are consistent with previous measurements of the intrinsic scatter (see Table \ref{tab:others_sze}).
These measured values of the intrinsic scatter, however, are larger than the 10--15\% scatter predicted by simulations\citep[e.g.,][]{Nagai2006,Fabjan2011,Battaglia2012,Sembolini2013}.
The difference between the predicted and measured scatters may be due to additional sources of measurement uncertainty, projection effects, and/or astrophysics not yet accounted for in the simulations.
In our particular case, some of the additional scatter may also be due to our use of a cylindrical \Ysmall, as described in Section \ref{sec:ysz}, but we expect this difference to be small based on recent simulations \citet{Battaglia2012} and because our intrinsic scatter is consistent with other measurements based on spherical \Ysz. 
\subsection{Physically Motivated \oym~Consistency Checks}
\label{sec:splittests}
\begin{figure*}
\centering
\includegraphics[scale=0.5]{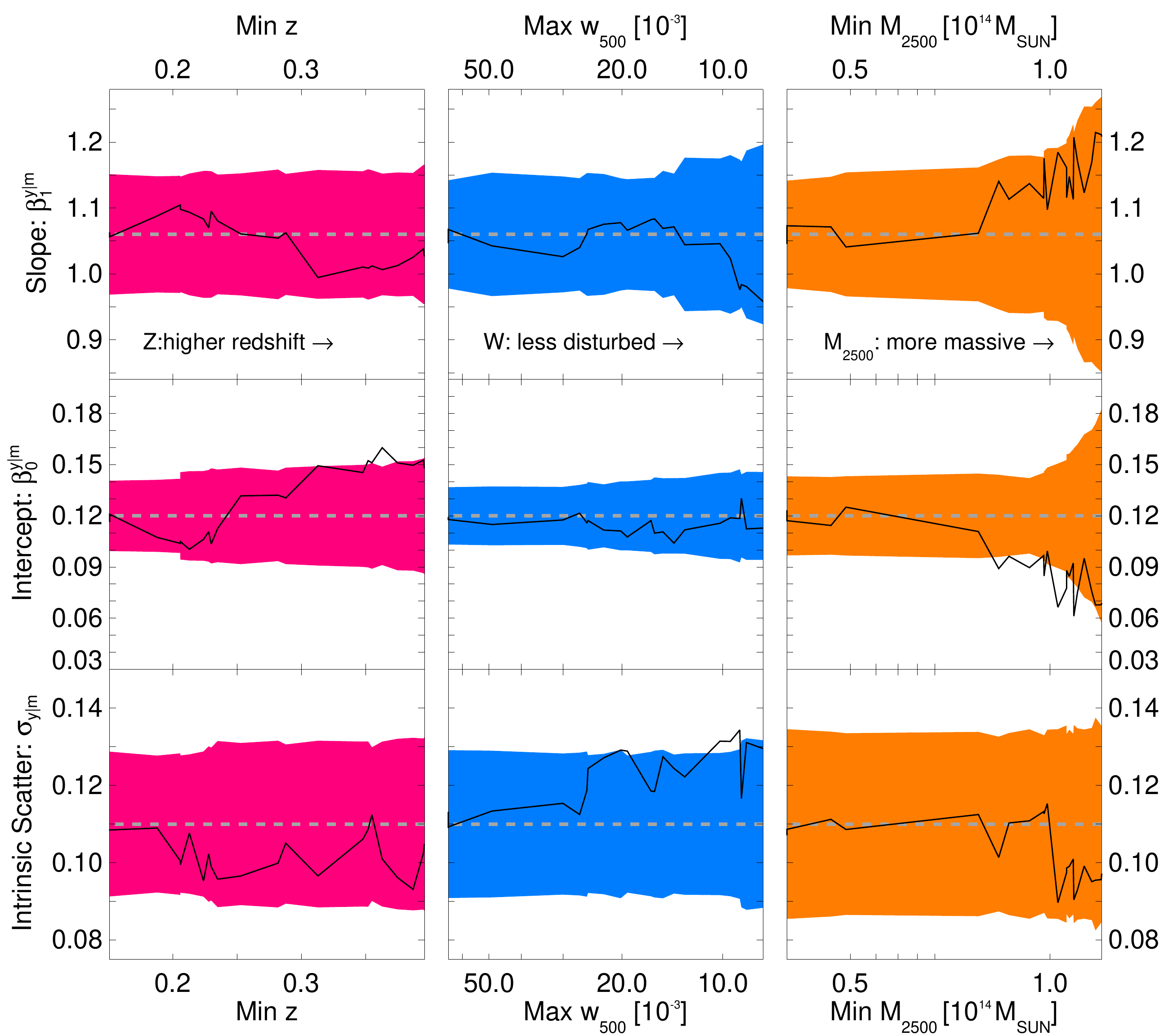} 
\caption{Measured \Ysmall--\Msmall~scaling relation parameters for various subsamples of BOXSZ clusters.
Fit parameters (from top to bottom): slope (\bym), intercept (\aym), and scatter (\oym).  
Selection parameters (from left to right): redshift ($z$), degree of disturbance (\wbig), and mass (\Msmall).
Subsamples are selected to include only those clusters with measured parameters to the right of the particular data point's position on the x-axis. 
Starting with the full sample of 45 clusters, the sample size decreases by one cluster per data point going from left to right and ends at the sample's median value. The values in the middle column are descending instead of ascending. 
Black curves represent the fits to the selected subsamples of BOXSZ clusters.
The dashed gray lines are the fiducial measured scaling relation parameters for the full cluster sample (see Table \ref{tab:boxsz_meas}).
We estimate additional regression biases by repeating the procedure on 100 mock cluster samples, and we correct the BOXSZ and mock cluster sample scaling relations using the median best-fit scaling relation parameters of these samples.
The colored bands indicate the 68\% fluctuation range of the measured mock cluster scaling relation parameters, centered on the fiducial values. The behavior of the data is consistent with this fluctuation range.}
\label{fig:split}
\end{figure*}

A range of consistency checks have been performed on the data not only to  test the robustness of the results but also to search for possible physical effects that are not described by the parameterization chosen for the scaling relations.
First, we perform a series of split tests, fitting scaling relations to subsamples of the BOXSZ sample selected on redshift, \wbig~(our chosen proxy for a cluster's dynamical state, introduced in Section \ref{sec:xray}), and \Msmall, to test if our scaling relations have any dependence on these parameters.
We correct all of these measurements for selection and regression biases using the values in Table \ref{tab:boxsz_delta}, which are calculated for the full BOXSZ sample.
Additional regression biases might arise as the sample size decreases, and the samples selected on \Msmall~will be particularly affected due to the decreased dynamic range of the fits.
We measure this additional bias by repeating our split-test procedure on 100 mock BOXSZ cluster samples, which we generate starting from the 45 measured \Msmall{ }values of the BOXSZ sample, applying the best-fit scaling relations, and adding unique Gaussian realizations of intrinsic scatter and measurement noise.\footnote{These mock samples are created in a less sophisticated manner than those generated to characterize the selection- and regression-induced biases discussed in Section \ref{sec:SR} (and fully described in Appendix \ref{sec:sf}). Specifically, we did not account for any covariance in the \Ysmall{ }and \Msmall{ }measurement uncertainties when characterizing the regression bias of our split test measurements. Since the correlation coefficient, $r$, in the measurement noise is small ($r<0.2$ for most clusters), and we do not expect the covariance to scale with redshift, \Msmall, or cluster morphology, we do not expect that this will significantly affect our results. We then correct the mock samples and the BOXSZ sample for these biases.
In Figure \ref{fig:split}, we plot the results, and in Table \ref{tab:2500_jk}, we give the measured parameters for subsamples of 23 clusters.}
In addition, we have fit subsets of clusters split into cool-core and non-cool-core samples as defined in \citet{SayersPressure}.
These split tests show no evidence of larger-than-expected deviations from the sample-to-sample variation of the best-fit scaling relation parameters of the mock samples.
We therefore conclude that our \Ysmall--\Msmall~scaling relations show no evidence of redshift, morphology, or mass dependence.

Since the value of \rsmall~(in Mpc) is relatively constant over the sample, splitting the sample based on redshift is approximately equivalent to splitting based on angular size.
Therefore, there is no evidence, given our measurement uncertainties, that the scaling relation results depend on cluster angular size, indicating that the high-pass filtering (and consequent deconvolution, including the signal offset estimation) 
has been properly accounted for.
\setlength{\tabcolsep}{4.5pt}
\begin{deluxetable}{cccc}
	\tablecaption{\Ysmall--\Msmall~scaling relations for subsamples of 23 clusters selected on redshift, \wbig, and \Msmall.}
	\tablehead{
Sample & $\beta_1^{y|m}$ & $\beta_0^{y|m}$ & $\sigma^{y|m}$
}
	\startdata
	$z\leq0.42$	& 1.08$\pm$ 0.19 &  0.11$\pm$ 0.04 & 0.13$\pm$ 0.05\\
	$z\geq 0.42$ &  1.02$\pm$ 0.16 &  0.15$\pm$ 0.07 & 0.12$\pm$ 0.05\\ \\
	\wbig~$\leq 0.7\times10^{-2}$  	& 0.96$\pm$ 0.18 &  0.11$\pm$ 0.07 &0.13$\pm$ 0.05\\
	\wbig~$\geq 0.7\times10^{-2}$  	& 1.08$\pm$ 0.13 &  0.14$\pm$ 0.04 &0.09$\pm$ 0.04\\ \\
	\Msmall~$\leq 3.0\times10^{14}$\Msun & 1.10$\pm$ 0.31 &  0.14$\pm$ 0.05 &0.13$\pm$ 0.05\\
	\Msmall~$\geq3.0\times10^{14}$\Msun & 1.22$\pm$ 0.31 &  0.07$\pm$ 0.13 &0.10$\pm$ 0.04
	\enddata
	\tablecomments{The slope ($\beta_1^{y|m}$), intercept ($\beta_0^{y|m}$), and intrinsic scatter ($\sigma^{y|m}$), are measured using the formalism described in Section \ref{sec:SR} and corrected for selection effects using the values given in Table \ref{tab:boxsz_delta}.
} 
	\label{tab:2500_jk}
\end{deluxetable}

We next explore the model dependence of our results by measuring the scaling relations using \Ysmall~values obtained using the 1-MP model rather than the minimal model given in Table \ref{tab:fit_info}.
Recall that the 1-MP model is preferred for only 16 of the BOXSZ clusters.
While all of the scaling relations derived using \Ysmall\ values based on the 1-MP model are consistent with those derived using the selected minimal model, 
we note that the slope of the \Ysmall--\Msmall~scaling relation thus obtained is 1.5$\sigma$ steeper than our fiducial fit.

We further examine how our results depend on the exact shape of the pressure profile model.
The first test that we perform is to use the morphology-dependent pressure profile parameters given in Arnaud10 for those clusters which we classify as relaxed or disturbed in Section \ref{sec:xray}.
The results are indistinguishable from our adopted method, further indicating that the results do not depend strongly on the parametric model adopted to constrain the signal offset. 

The second test that we perform is to see how our results change when using the Bolocam pressure profile as presented in \citet{SayersPressure}. 
The median of the best-fit \chisqr~values, ellipticities ($\epsilon$), and scale radii ($r_{500}/c_{500}$) remain approximately the same, although with some scatter. 
The median value of $p_0$ is lower by approximately 1$\sigma$, likely due to the fact that the Bolocam pressure profile has a lower normalization than the Arnaud10 pressure profile (4.3$P_{500}$ compared to 8.4$P_{500}$). 
In addition, the \Ysmall~values are higher by approximately 1$\sigma$ for all clusters. 
This is caused by the Bolocam pressure profile being shallower at large radii, resulting in a higher value of the mean signal level for the deconvolved maps.
When we fix the number of model parameters, the measured scaling relations are negligibly different for the two pressure profiles. 
However, when we perform the scaling relation fits using the minimal model determined for each cluster, use of the Bolocam pressure profile results in a \Ysmall--\Msmall~slope that is steeper by 0.8$\sigma$. 
This is due to the fact that the 1-MP fit is favored more often with the Bolocam pressure profile than the Arnaud10 profile, and, as described above, the slope from the 1-MP fits tends to be slightly steeper.

In conclusion, after repeating our analysis using different pressure models and different degrees of freedom in our models, none of the alternative fits are significantly different from our fiducial results.
\subsection{Comparison With Previous Studies}
\label{sec:others}
{\setlength{\tabcolsep}{.35em}
\begin{deluxetable*}{ccccccccccc}
\tabletypesize{\scriptsize} 
\tablecaption{Overview of Various SZE-X-ray Scaling Relation Studies.}
\tablewidth{0pt} 
\tablehead{
\colhead{Name} & \colhead{SZE data} & \colhead{X-ray data} & \colhead{Proxy} & \colhead{~$\Delta$} &
  \colhead{$\beta^{y|m}_1$} & \colhead{$\sigma_{y|m}$} & 
  \colhead{$\beta^{y|y_x}_1$} & \colhead{$\sigma_{y|y_x}$} & 
\colhead{$N_{\rm lz}:N_{\rm hz}$} & \colhead{\Mbig$[10^{14}$\Msun$]$}
}
\startdata
this work & Bolocam & CXO & \Mgas & 2500 & \bymv\smallskip & \isymlogv & \byyxv\smallskip & \isyyxlogv & $22:23$ & $\left[3.9,24.9\right]$\vspace{-.15in}\\ 
B08 & OVRO/BIMA & CXO & HSE & 2500 & $1.66\pm0.20$ & \nodata &\nodata & \nodata & $22:16$ & $\left[2.0,16.2\right]$\\
A11 & SPT & CXO/XMM & \Yx & 500 & $1.67\pm0.29$ & $0.09\pm0.05$ & $0.90\pm0.17$ & $0.07\pm0.05$ & $\phn{3}:12$ & $\left[3.5,11.8\right]$\\
P11 & \emph{Planck} &XMM & \Yx & 500 & $1.74\pm0.08$ & $0.10\pm0.01$& $0.95\pm0.04$ & $0.10\pm0.01$& $59:\phn{3}$ & $\left[2.4,19.7\right]$
\enddata
\tablecomments{\small First column: SZE scaling relation study under consideration, including, \citet[B08]{Bonamente2008}, \citet[A11]{Andersson2011}, and \citet[P11]{Planck2011SR}.
Second and third columns: the SZE and X-ray instruments with which the data were taken for each particular study. CXO stands for \emph{Chandra} X-Ray Observatory.
Fourth column: the particular X-ray mass proxy implemented, which is discussed in Section \ref{sec:others}.
Fifth column: the critical overdensity out to which \Ysz~and \Mtot~are integrated.
The sixth through ninth columns, from left to right, give the measured slopes and intrinsic scatters for the \Ysz--\Mtot~and \Ysz--\Yx~scaling relations for the given study.
Tenth column: the number of clusters below and above the BOXSZ median redshift of $\langle z\rangle=0.42$.
For A11, the \bym~values are given for $Y_{spher}$ and the \byyx~values are given for $Y_{cyl}$.
The final column gives the range of \Mbig~masses used in each particular study.
The B08  \Mbig~values are approximated from the measured \Msmall~values by multiplying them by a factor of 2.
Despite the variety in \Ysz--\Mtot~relations, the \Ysz--\Yx~relations are consistent between the various scaling relation studies.}
\label{tab:others_sze}
\end{deluxetable*}
}
Table \ref{tab:others_sze} lists some of the relevant characteristics of the three main studies to which we compare this study. These studies measure SZE-X-ray scaling relations using OVRO/BIMA/\emph{Chandra} \citep[hereafter B08]{Bonamente2008}, \emph{Planck}/\emph{XMM} \citep[hereafter P11]{Planck2011SR}, and SPT/\emph{Chandra}/\emph{XMM} \citep[hereafter A11]{Andersson2011} data.
A direct comparison, however, is made challenging because of differences between the X-ray mass proxies, selection criteria, and analysis methodologies adopted in each study. 
To avoid systematic differences associated with the different mass proxies used for each study, it is helpful to consider the \Ysz--\Mgas{ }and the \Ysz--\Yx{ }relations as well.
We explore the key similarities and differences between our results and these particular scaling relation studies below.

B08 present the first observed \Ysz--\Mtot~scaling relations for a sizeable cluster sample using OVRO/BIMA SZE measurements and \emph{Chandra} X-ray data.
The sample consists of 38 clusters, with a median redshift of $\langle z \rangle=0.30$, and all parameters are derived within \rsmall.
\Msmall, \Mgassmall, and \Ysmall~values are obtained by spherically integrating joint SZE/X-ray fits to spherical isothermal $\beta$-models, and clusters are assumed to be in HSE. 
The \Msmall~values in the B08 sample span from $1.0\times10^{14}$ \Msun~to $8.1\times 10^{14}$ \Msun. 
Of the three cluster samples that are considered in this section, the B08 sample is most similar to the BOXSZ one in terms of redshift, mass, and cluster selection. In fact, the two samples share 21 clusters in common.
B08 measure a \Ysmall--\Mgassmall~slope of $1.41\pm0.13$ and \Ysmall--\Msmall~slope of $1.66\pm0.20$.

An important item to consider when comparing our study to B08 is that while we, together with the other analyses considered in this work, explicitly fit for intrinsic scatter in \Ysz~at fixed \Mtot, B08 quantify the individual sources of scatter as part of their systematic and statistical measurement uncertainty. 
These sources of scatter are calculated in \citet{Laroque2006} and include: kinetic SZE, radio point source contamination, asphericity, hydrostatic equilibrium, and isothermality. 
Consequently, in addition to measurement error, B08 include a 20\% and a 10\% fractional uncertainty in their \Msmall~and \Ysmall~measurements, respectively. Including additional uncertainty in this way, however, is not equivalent to simultaneously fitting for the intercept, slope and intrinsic scatter of the scaling relation.

Interestingly, when we fit the B08 data using the \ttfamily linmix\_err \normalfont method (and without including the additional systematic component to the individual uncertainties),
we measure \bym$=1.15\pm0.15$, \aym$=-0.14\pm0.03$, and \sigym$=0.12\pm0.02$, which are similar to the BOXSZ results.
This exercise demonstrates the complexity of comparing scaling relations parameters calculated using different methodologies.
While a rigorous comparison of the error budgets between our study and that of B08 is beyond the scope of this paper, this result suggests that at least part of the discrepancy between our work and the B08 results is due to a fundamental difference in how each study models intrinsic scatter versus measurement uncertainty.

The sample for the second study under consideration in this section, A11, consists of 15 SZE-significance selected clusters, with 0.29~$<z<$~1.08, within the SPT 178 $\mathrm{deg}^2$ survey. 
The nature of an SZE significance-limited selection of clusters from a relatively small survey results in a less massive cluster selection than the BOXSZ sample---all but one of the A11 clusters lie below the BOXSZ median $\langle$\Mbig$\rangle=9.1\times10^{14}$\Msun. A further difference is that they use \rbig~as their integration radius.
A11 calculate both spherical and cylindrical \Ybig~values by integrating cluster-specific pressure models derived from X-ray-constrained $n_e$ and $T_x$ parametric models, allowing the SZE data to constrain only the overall normalization.
A11 measure \byyx$=0.90 \pm 0.17$ using cylindrical \Ybig. 
A11 derive \Mbig~values from the \Mbig--\Yx~relation of V09, with \bmyx$=0.57\pm0.03$, and measure \bym$=1.67\pm0.29$,  using spherical \Ybig.
A11 characterize their selection bias using simulated SZE sky maps derived directly from N-body simulations (including semi-analytic distributions of cluster gas) to estimate how their detection significance depends on \Ybig.

The next sample that we compare our results with, P11, contains 62 clusters and is the largest sample considered in this work.
This sample was constructed primarily based on membership in both the \emph{Planck} Early Release Compact Source Catalog (\citealt{Planck2011ESZ}) and the Meta Catalog of X-ray Clusters (\citealt{Piffaretti2011}). It shares a similar mass range ($2\times10^{14}$\Msun$<M_{500}<2\times10^{15}$\Msun) but covers lower redshifts than the BOXSZ cluster sample. 
Of the 62 clusters in the P11 sample, 59 lie below the median BOXSZ redshift.
Similar to A11, P11 use \rbig~as an integration radius and \Yx~as a mass proxy. 
\Ybig~is calculated by assuming the universal pressure model given in Arnaud10, allowing the SZE data to constrain the overall normalization of the cylindrically projected model out to $5$\rbig, which is then converted to a spherically integrated \Ybig.
They measure \bymg$=1.39\pm0.06$ and \byyx$=0.95\pm0.04$.\footnote{The recent scaling relations derived in \cite{Planck2013SZEcosmo} contain an additional 9 confirmed clusters with respect to the P11 sample. As the results from this slightly expanded sample are very similar to P11, they are not explicitly examined in this analysis.}
P11 derive \Mbig~from the \Mbig--\Yx~relation of Arnaud10, with \bmyx$=0.548\pm0.027$, and measure \bym$=1.74\pm0.08$.
While there are similarities between the P11 and BOXSZ calculation of selection effects (both sample simulation-derived mass functions to construct mock cluster catalogs with scaling-relation-derived observables), there are key differences in our methodologies.
First, P11 do not allow their assumed X-ray scaling relations to float and they do not include covariance in intrinsic scatter. 
Also, since P11 use a different regression method, the level of regression bias may differ.
Another difference is that their sample is partially SZE-selected. 
Interestingly, while P11 estimate their selection bias for the \Ybig--\Mbig~power-law index to be negligible, their estimated selection bias for the \Ybig--\Mgas~relation is not negligible, necessitating a correction from \bymg$=1.39$ to \bymg$=1.48$. This difference in bias correction is in contrast with our method, where, given the similar treatment of \Mgas{ }and \Mtot{ }in our selection function characterization, our corrections to \bym{ }and \bymg{ }would be approximately equal.

Even more recently, \citet[hereafter B14]{Bender2014} have presented \Ysz--\Mgas, \Ysz--\Tx, and \Ysz--\Yx~scaling relations for 35 clusters observed with APEX-SZ.
They derive \Ysz~cluster observables by spherically integrating the best-fit Arnaud10 pressure profile out to \rbig, where \rbig~is derived from the \citet{Vikhlinin2006} X-ray  based \rbig--\Tx~scaling relation.
As their sample contains some non-detections, they have decided to use a modified version of \ttfamily linmix\_err\normalfont{ }and perform a linear, instead of a logarithmic, regression analysis. They still model and constrain intrinsic scatter in a fashion identical to our analysis, as a Gaussian variance on the logarithmic scaling.
B14 measure the \Ysz--\Yx~slope to be consistent with unity, \byyx$=0.98^{+0.07}_{-0.12}$.
Although B14 do not measure \Ysz--\Mtot~scaling relations, we can compare our result for this relation to their \Ysz--\Mgas~result because we assume constant \fgas.  They measure a power-law index of $1.16^{+0.10}_{-0.17}$, which is within 1$\sigma$ of our result.
Despite the agreement, B14 measure a fractional intrinsic scatter in \Ybig~at fixed mass over twice as large (55$\pm$7)\% as the BOXSZ results. 
As with all of the previously discussed \Ysz--\Mtot~analyses, the extent to which we can compare our results to B14 is limited. The B14 \Mgas{ }measurements are not derived in a uniform fashion, and they specifically note that their intrinsic scatter measurement is considerably reduced (down to 12\% in one instance) when using subsets of data with uniformly analyzed X-ray data.

Our measured \Ysz--\Mtot~power-law index is in some tension with current state-of-the-art simulations, such as those by \citet{Fabjan2011}, \citet{Battaglia2012}, and \citet{Sembolini2013}.
Under a variety of physically motivated scenarios, with sample redshifts ranging from $z=0$ to $z=1$, these simulations give values for the power-law index of \Ybig--\Mbig~between $1.60$ and $1.75$.
Some of these differences might be due to the low mass range of the particular simulations or the use of $\Delta=500$ instead of $\Delta=2500$ (see e.g., \citealt{Fabjan2011,Battaglia2012}).
This, however, is not the case for \citet{Sembolini2013}, who make measurements at both $\Delta=500$ and $\Delta=2500$ and specifically limit their sample to high cluster masses.
They measure the \Ysz--\Mtot~power-law index to be consistent with the self-similar prediction at both overdensities, and they measure the \Ysz--\Mgas~slope to become shallower at higher overdensities: from 1.61 at $\Delta=500$ to 1.48 at $\Delta=2500$ (including CSF but not AGN feedback).
One possible explanation for the discrepancy between our results and simulations is that the \Ysz--\Mtot{ }relation is not a single power law, although it is generally modeled as such. Consideration of such a deviation is motivated by the results of \citet{Stanek2010} and by our analysis of the Sembolini et al. (2013) simulation (Appendix \ref{sec:fgas} and Figure \ref{fig:fgas_all2500}), which suggest a flattening of the power-law relationship between \fgas{ }and \Mtot{ }at high mass.
\subsection{Discussion}
\label{sec:discussion}
Part of the discrepancy between the A11, P11, and BOXSZ results might be a result of physical differences between the cluster samples themselves.
The A11 sample, for example, spans a similar redshift range but a lower mass range than the BOXSZ sample.
In contrast, the P11 sample covers a lower redshift range but a nearly identical mass range.
Based on these samples, it seems unlikely that either a mass or redshift dependence alone can explain the incompatibility of the present results with these other analyses.
Furthermore, in Section~\ref{sec:splittests}, when we fit subsamples selected on redshift, \wbig, and \Mtot, we find that there is no evidence in our data that the \Ysmall--\Mtot~scaling relations depend on these parameters.
Another possibility is that the differences in our results arise due to our choice of \rsmall~as the radius of integration.
Again, this hypothesis alone is not sufficient to explain all of the discrepancies, as B08 also use \rsmall~as an integration radius.\footnote{Note, however, that we obtain good agreement with B08 when the same regression algorithm is employed (Section \ref{sec:others}).}

The discrepancies might also be explained by differences in \Ysz~estimation and scaling relation fitting methodologies between the different groups.
The largely model-independent method by which we estimate \Ysz~does differ from these previous studies, which have relied on parameterized models with shapes constrained using X-ray data.
A bias induced by the highly X-ray-constrained models employed in the B08, A11, P11 results could therefore potentially explain the difference between their results and ours.
When we naively refit the B08 sample including intrinsic scatter, however, we find a result similar to the BOXSZ scaling relations, suggesting that, in this case, the discrepancy with the BOXSZ results is more likely due to differences in fitting method and error estimation.

We conclude that, if the differences between the various \Ysz~scaling relations are primarily due to systematic differences in the estimation of the \Ysz~values, their uncertainties, and/or the fitting methodologies themselves, these differences are not easily teased apart and require a systematic cross-calibration between the different groups, which is beyond the scope of the current analysis.

Another difference between the \Ysz--\Mtot~scaling relation analyses is the method by which they correct for selection effects, if at all.
Differences in the adopted mass function and differences in the treatment of the covariance of the intrinsic scatter between different observables could bias these results. 
Since \Ysz~is a low-scatter observable at fixed \Mtot, P11 and BOXSZ estimate that selection effects require a $\lesssim 0.1$ correction in the slope of the \Ysz--\Mtot~scaling relation. 
The BOXSZ selection bias estimates are further sensitive to the assumptions of log-normal intrinsic scatter and the covariance of the \Ysz~and \Mgas~intrinsic scatter, both of which are not sufficiently constrained using current observations. We estimate that our lack of information about this covariance might add a systematic uncertainty of approximately $\pm$0.1 to the slope of the BOXSZ \Ysmall--\Msmall~scaling relations.

If the source of the deviation of the \Ysz--\Mtot~scaling relations from self-similar predictions is not due to systematics in our \Ysz~analysis, then the \Tx--\Msmall{ }scaling relation should also be affected. This is indeed the case. M10 measure a \Tx--\Msmall~power-law index at \rbig{ }of \btm$=0.48\pm0.04$, over 4$\sigma${ }shallower than the self-similar prediction of $2/3$.
M10 explore potential reasons for a \Tx--\Mtot{ }slope that is shallower than self-similar predictions, such as an excess heating mechanism in the cluster core, and we refer the reader to that work for more details.
The BOXSZ power-law index is even shallower than that measured in M10 and is 6$\sigma$ shallower than self-similar predictions. Since the BOXSZ and M10 samples have similar mass ranges, use the same mass function to account for selection effects, and none of our scaling relation results indicate any redshift dependence (see Section~\ref{sec:splittests}), we do not believe that the difference between our results is due to any selection-dependent, mass-dependent, or redshift-dependent effect. However, it is possible that the inconsistencies between the two analyses are due to the different overdensity radii employed (\rsmall{ }for BOXSZ versus \rbig{ }for M10), potentially enhanced by statistical fluctuations.

As for the discrepancies between the BOXSZ, A11, and P11 results, possible systematic differences in the \Mtot{ }estimates would directly propagate to differences in the measured scaling relation slopes. Given the relatively high mass range of our sample, we have estimated masses by adopting a constant-\fgas{} model, a choice which is widely supported in the literature. A related issue is that of calibration of X-ray temperature (and hence HSE mass) measurements, which potentially affects all scaling relations that rely on HSE masses. We address these questions in more detail in Appendix~\ref{sec:fgas}.
There, we demonstrate that the BOXSZ \Ysmall--\Msmall~power-law index can be made consistent with the A11 and P11 results if we assume \fgas~to have a similar power-law scaling ($f_{\rm gas}\propto M_{2500}^{0.2}$) as that of their adopted mass proxies: V09\footnote{V09 formulate their \fgas--\Mtot~relation slightly differently than we do in this analysis, but, as we show in Figure \ref{fig:fgas_all2500} of Appendix \ref{sec:fgas}, their results are approximately consistent with $f_{\rm gas,500}\propto M_{500}^{0.2}$.} and Arnaud10\footnote{Although Arnaud2010 do not specifically measure the \fgas--\Mtot~relation, their analysis and results are fully consistent with \citet{Pratt2009}, who measure $f_{\rm gas,500}\propto M_{500}^{0.21\pm0.03}$.}, respectively. 
Conversely, similar consistency could have been obtained had we scaled the A11 and P11 results to a constant-\fgas~model. 
We conclude that much of the discrepancy between the BOXSZ \Ysz--\Mtot~power-law index and those of P11 and A11 is driven by differences in \Mtot{ }calibration.

Note that the \Ysz--\Yx{ }relation is largely immune to systematics related to \Mtot, making it the most straightforward of our results to compare to the literature. Here, the agreement in the measurements summarized in Table~\ref{tab:others_sze} is encouraging. As for the disagreements among measurements of the \Ysz--\Mtot{ }relation, we can only conclude at this point that there must be systematic differences associated with the SZE and/or X-ray data analysis or the methods used to fit the scaling relations and to account for selection biases. If the \Ysz--\Mtot{ }slope is significantly shallower than the value found in hydrodynamic simulations, as our analysis concludes, this is an indication that some important astrophysical processes have yet to be accounted for in the simulations.
\section{Summary and Directions of Future Work}
We present SZE measurements for the BOXSZ sample of 45 galaxy clusters collected with Bolocam at 140 GHz.
Relative to most cluster catalogs, BOXSZ is a distant (median redshift of $\langle z \rangle=0.42$), massive (median $\langle$\Msmall$\rangle=3.0\times10^{14}$\Msun), and hot (\Tx$\gtrsim5$ keV) sample of galaxy clusters.
Using the SZE data, we determine scaling relations between our measured \Ysmall~and \emph{Chandra} X-ray measurements of \Msmall~and \Yx. 
We account for various sources of systematic biases in our noise characterization, including contamination from other astronomical sources, and loss of SZE signal from noise-filtering and beam-smoothing effects. 
We find that the modeled uncertainties are minimized at \rsmall~and we therefore present all results using this aperture radius.
We characterize the selection effects due to our \emph{ad hoc} cluster sample by simulating and analyzing mock data sets (Appendix \ref{sec:sf}).
We find that such selection effects create biases that are smaller than or comparable to our measurement uncertainties, and we fully account for these biases in our analysis. We measure a slope of \byyxv~for the \Ysmall--\Yx~relation, consistent with previously published results.
Furthermore, we measure a slope of of \bymv~for the \Ysmall--\Msmall~relation, which is approximately 5$\sigma$ shallower than predicted by self-similarity and inconsistent with previously measured \Ysz--\Mtot~results.
We have also fit scaling relations to subsamples of clusters based on cuts in redshift and morphology, and we find results that are consistent with those obtained from the full BOXSZ sample.

To reconcile the differences between the various \Ysz--\Mtot~scaling results in the literature, one must ensure that both the data and the analysis techniques employed are consistent.     
The values of \Ysz~obtained using different instruments and different analysis techniques should be compared using as large a set of common clusters from the available samples as possible. 
Multi-probe data sets, particularly those with both strong- and weak-lensing constraints, will allow for robustly estimated masses at a range of overdensity radii, including our choice of \rsmall.

With self-consistent \Ysz~and \Mtot~measurements, one can then explore the consistency between the different scaling relation measurement techniques, such as the dependence of these scaling relations on the choice of integration radius and the corrections of selection biases. 
Ultimately, a fully consistent analysis of all of these cluster samples should be able to resolve these discrepancies and give a unified treatment across a larger range of mass and redshift than any of the individual analyses alone.
\section{Acknowledgments}
This material is based upon work at the Caltech Submillimeter Observatory, which, when the data used in this analysis were taken, was operated by the California Institute of Technology under cooperative agreement with the National Science Foundation.
Bolocam was constructed and commissioned using funds from NSF/AST-9618798, NSF/AST-0098737, NSF/AST-9980846, NSF/AST-0229008, and NSF/AST-0206158. Bolocam observations were partially supported by the Gordon and Betty Moore Foundation, the Jet Propulsion Laboratory Research and Technology Development Program, as well as the National Science Council of Taiwan grant NSC100-2112-M-001-008-MY3.  
We acknowledge the assistance of: 
the Bolocam instrument team:
P.~A.~R. Ade, J.~E. Aguirre, J.~J. Bock, S.~F. Edgington, J. Glenn, A. Goldin, S.~R. Golwala, 
D. Haig, A.~E. Lange, G.~T. Laurent, P.~D. Mauskopf, H.~T. Nguyen, P. Rossinot, and J. Sayers;
Matt Ferry, who helped collect the data for Abell 1835 and MS 1054.4-0321;
the day crew and Hilo staff of the Caltech Submillimeter Observatory, who provided invaluable assistance during commissioning and data-taking for this survey data set; 
and Kathy Deniston, Barbara Wertz, and Diana Bisel, who provided effective administrative support at Caltech and in Hilo.
This research made extensive use of high performance computing resources at the U.S. Planck Data Center, which is a part of the Infrared Processing and Analysis Center at Caltech.
We also thank an anonymous referee for a detailed report that substantially improved the quality of this paper.
NC  was partially supported by a NASA Graduate Student Research Fellowship;
JS was partially supported by a NASA Graduate Student Research Fellowship, a NASA Postdoctoral Program fellowship, and a Norris Foundation CCAT Postdoctoral Fellowship;
AM was funded by NSF AST-0838187 and AST-1140019;
The work of LAM was carried out at Jet Propulsion Laboratory, California Institute of Technology, under a contract with NASA;
support for TM was provided by NASA through Einstein Fellowship Program grant No. PF0-110077 awarded by the \emph{Chandra} X-ray Center, which is operated by the Smithsonian Astrophysical Observatory for NASA under contract NAS8-03060, and by a National Research Council Fellowship;
EP and JAS were partially supported by NASA/NNX07AH59G;
KU acknowledges partial support from the Academia Sinica Career Development Award. 
{\it Facilities:} \facility{CSO,\emph{Chandra}}.
\bibliography{allref}   
\bibliographystyle{apj}   
\appendix
\setcounter{figure}{0}
\renewcommand{\thefigure}{\Alph{section}.\arabic{figure}}
\section{Mean Signal Offset Determination for the BOXSZ Noise Realizations}
\label{sec:sig_offset}
In Section \ref{sec:xfer_model_offset}, we explained how we constrain the mean signal level of the deconvolved images using the minimal model fit to the data.
Here, we describe an analogous procedure to constrain the mean signal level of the deconvolved noise realizations, which we use to characterize our overall \Ysmall~uncertainty.
First, we define the following types of two-dimensional images:

\begin{itemize}
  \item[$d$:]{image of the processed Bolocam data}
  \item[$m$:]{best-fit minimal model to $d$, convolved with the Bolocam transfer function (i.e., valid for comparison with $d$, the image of the processed data)}
  \item[$M$:]{best-fit minimal model to $d$, not convolved with the Bolocam transfer function (i.e., 
    the unfiltered version of the model)}
  \item[$n_i$:]{$i^{\textrm{th}}$ noise realization of the processed Bolocam data}
\end{itemize}
We also define a deconvolution operator, $\mathcal{D}$, which 
transforms an image of the processed Bolocam data into an image free from the filtering effects of the Bolocam processing.
The exact details of the deconvolution operator are described in Section 5 of \citet{SayersMorphology}. Briefly, we define the transfer function of the Bolocam data
processing in the two-dimensional Fourier space of the image. Therefore, the deconvolution is performed by Fourier transforming the processed image $d$,
dividing the result by the two-dimensional transfer function, and then Fourier transforming back to physical image space.

Following the procedure detailed in Section \ref{sec:xfer_model_offset}, we then produce deconvolved images with a constrained mean signal level according to
\begin{equation}
\mathcal{D}(d)^{\prime}= \mathcal{D}(d) - \langle\mathcal{D}(d) \rangle_{j \epsilon A} +  \langle M\rangle_{j \epsilon A}
\end{equation}
where the angle brackets represent a noise-weighted mean computed from all map pixels $j$ contained within the region outside of $r_{500}/2$ (denoted by $A$ and shown in blue in Figure 3).
As a result, the deconvolved image has the same weighted mean signal level as the model in the region outside of $r_{500}/2$.

To constrain the mean signal level of the noise realizations in an analogous way, we first add the best-fit minimal model to each noise realization so that it contains an SZE signal similar to the real image. Specifically, we form the image
\begin{equation}
  \delta_i = n_i + m.
\end{equation}
Next, we determine the best-fit model to $\delta_i$, which we call $\mathfrak{M}_i$, and we
note that in general $\mathfrak{M}_i$ is not equal to $M$ due to the influence of the noise.
The deconvolved noise realization with a constrained mean signal level is then computed according to
\begin{equation}
  \mathcal{D}(n_i)^{\prime} = \mathcal{D}(n_i)  - \langle\mathcal{D}(\delta_i) \rangle_{j \epsilon A} + \langle \mathfrak{M}_i \rangle_{j \epsilon A}.
\end{equation}
Therefore, the mean signal level of each noise realization depends on the model-fitting procedure in the same way as the real data, and as a result the set of noise realizations $\mathcal{D}(n_i)^\prime$ fully describe the noise properties of the deconvolved image $\mathcal{D}(d)^\prime$.
\section{Choice of Integration Aperture}
\label{sec:r500_v_r2500}
\begin{figure}
\begin{center}
\begin{tabular}{cc}
  \includegraphics[scale=0.5]{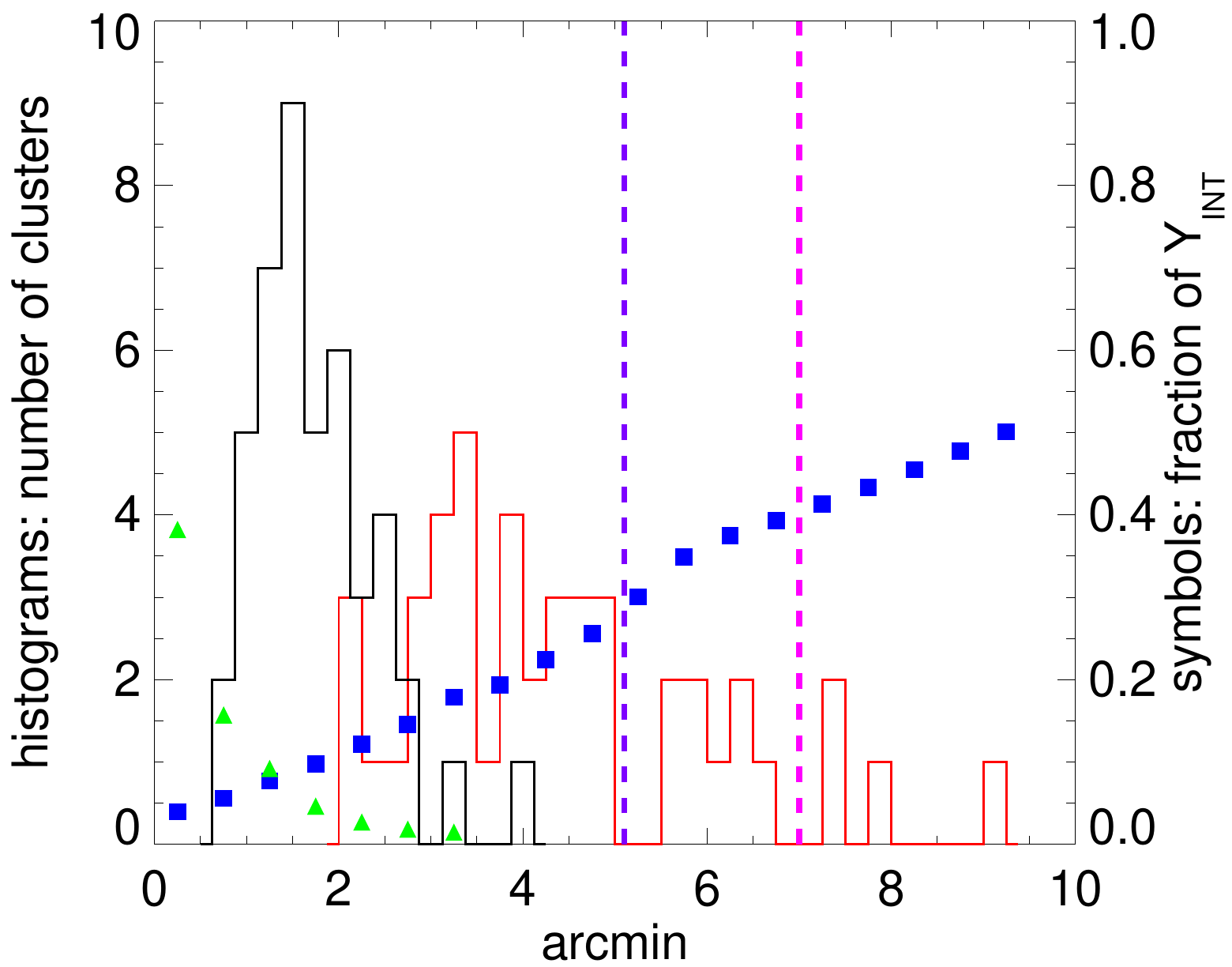} &
  \includegraphics[scale=0.525]{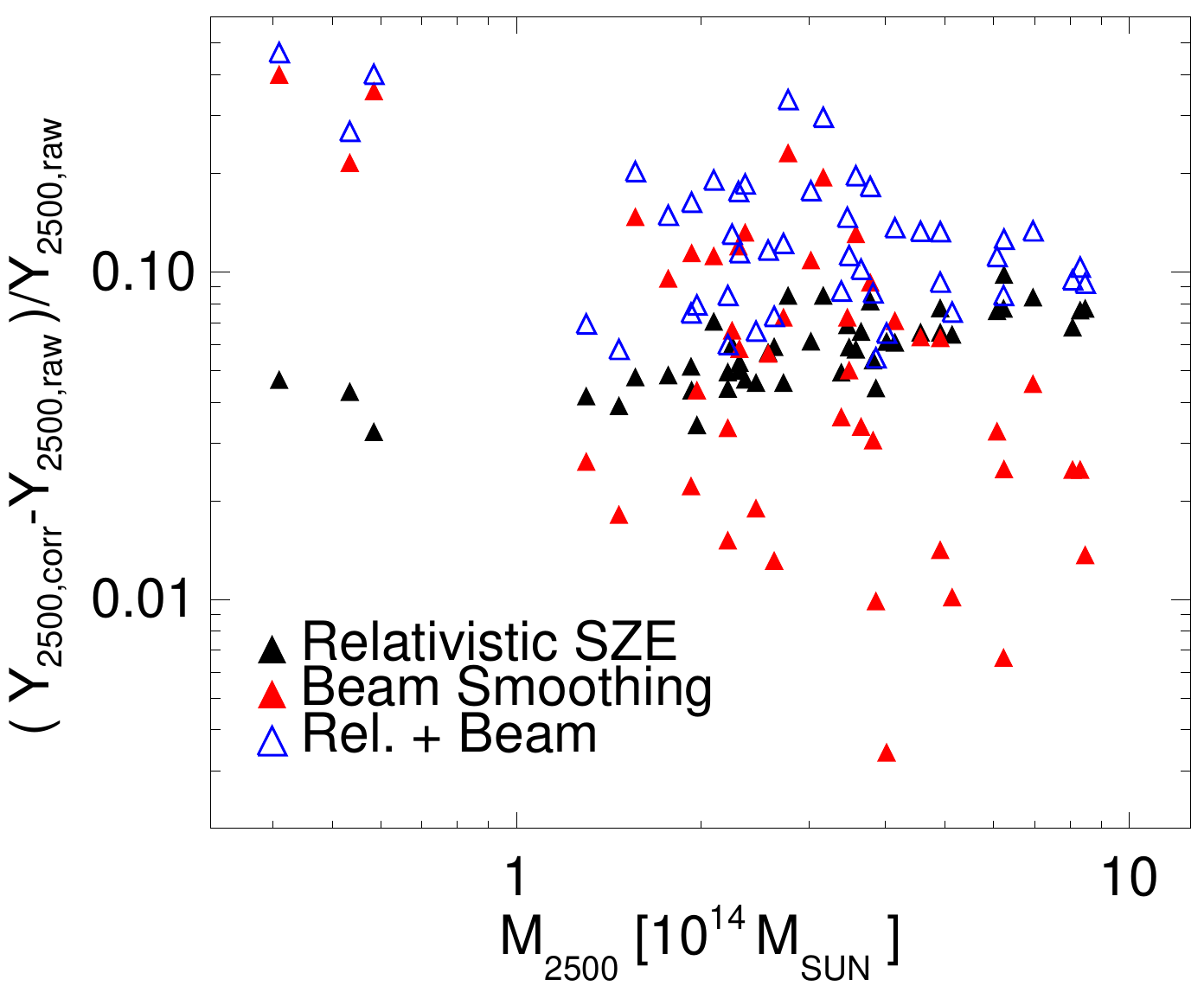}
\end{tabular}
\end{center}
  \caption{Left panel: distribution of the \rbig~(red) and \rsmall~(black) values for the BOXSZ sample. 
    Magenta dashed: 7 arcminute radial extent of the processed BOXSZ maps.
    Violet dashed: 5 arcminute radial extent of the deconvolved BOXSZ maps.
    Green triangles: systematic reduction in $Y_{\Delta}$ as a function of integration radius due to the finite size of the Bolocam PSF. 
    Blue squares: uncertainty in $Y_{\Delta}$ as a function of integration radius only due to uncertainties in the signal offset of the deconvolved SZE image. 
    Right panel: individual \Ysmall~correction factors for the 45 BOXSZ clusters (blue open triangles).
These corrections account for the loss of SZE signal inside \rsmall~due to beam-smoothing with the Bolocam PSF (red filled triangles) and relativistic effects (black filled triangles).
    The anti-correlation between the beam-smoothing correction factor and cluster mass shows that neglecting the beam-smoothing effect would bias our derived slope to higher values. Nevertheless, this effect is not strong and only affects the slope at approximately the 5\% level.}
  \label{fig:whatY}
\end{figure}

The fractional uncertainty on the integrated value of \Ysz~tends to increase relatively sharply with increasing radius (as demonstrated in Figure~\ref{fig:whatY}, by approximately a factor of two between \rsmall~and \rbig).
This is because the noise spectrum of the deconvolved Bolocam images increases at large angular scales, while the SZE signal drops with radius. 
We have therefore chosen to use \rsmall~as the integration radius for our analysis as it is the smallest commonly used overdensity radius which is large enough to approximately capture the global properties of the cluster.
An additional motivation for using this smaller radius is that it mitigates the impact of the deconvolved image signal offset that must be constrained using a parametric model (see Section \ref{sec:xfer_model_offset} and Appendix \ref{sec:sig_offset}).
Furthermore, a few clusters in the BOXSZ sample have large values of \rbig~which do not lie within the 10\arcmin$\times$10\arcmin~deconvolved images, reinforcing the preference for the use of \rsmall.

One consequence of this choice of integration radius is that it is not significantly larger than the Bolocam PSF, and therefore \Ysmall~estimates obtained from directly integrating the images are biased low. 
Effectively, some of the SZE emission within \rsmall~appears in the Bolocam image outside of \rsmall~due to beam smearing.
To estimate this bias, \Ysmall~is computed using the minimal parametric model determined in Section~\ref{sec:xfer_model_offset}, both before and after convolution with the Bolocam PSF.
The Bolocam-measured \Ysmall~value is then corrected by the ratio of \Ysmall~values determined from the un-smoothed and beam-smoothed model for each cluster.
This beam-smoothing correction is generally $\lesssim 10$\% and anti-correlated with mass due to the fact that more massive clusters tend to have larger \rsmall~(see Figure~\ref{fig:whatY}).
Therefore, although this beam-smoothing bias is relatively minor, its mass dependence can bias our scaling relations and thus we correct for it.
Figure \ref{fig:whatY} shows the fractional bias due to beam smoothing as well as the fractional uncertainty on \Ysz~due to the uncertainty of the mean signal offset. 
In contrast, relativistic corrections, discussed in Section \ref{sec:BOXSZ}, tend to have the opposite mass scaling due to the tight correlation between mass and temperature.
These corrections are plotted together with the beam-smoothing corrections in the right-hand panel of Figure~\ref{fig:whatY}.

This choice of integration radius stands in contrast with several observational analyses that adopt \rbig~as their integration radius, primarily motivated by simulations that indicate that this region is relatively unaffected by the non-thermal activity of the cluster core and additional massive structure in the cluster outskirts \citep{Evrard2008}.
Due mainly to observational considerations, many analyses involving X-ray data choose to use \rsmall~(e.g., \citealt{Bonamente2008}).
Observationally, the optimal radius is a function of the resolution and sensitivity limit for a particular telescope.
Although the choice of radius is not driven by considerations related to the X-ray analysis, we note that the use of \rsmall~is advantageous for the X-ray measurements in two ways. First, for the redshift range of the BOXSZ cluster sample, reliable X-ray measurements out to \rbig~using XMM-\emph{Newton} and \emph{Chandra} are often difficult to obtain due to the significant background dominating the dim cluster emissiom; and second, the noise in the \emph{Chandra} X-ray measurements is lower at \rsmall~than \rbig.
\section{Minimal Model Selection}
	\label{sec:ftest}
\begin{figure}
  \epsscale{0.6}
  \plotone{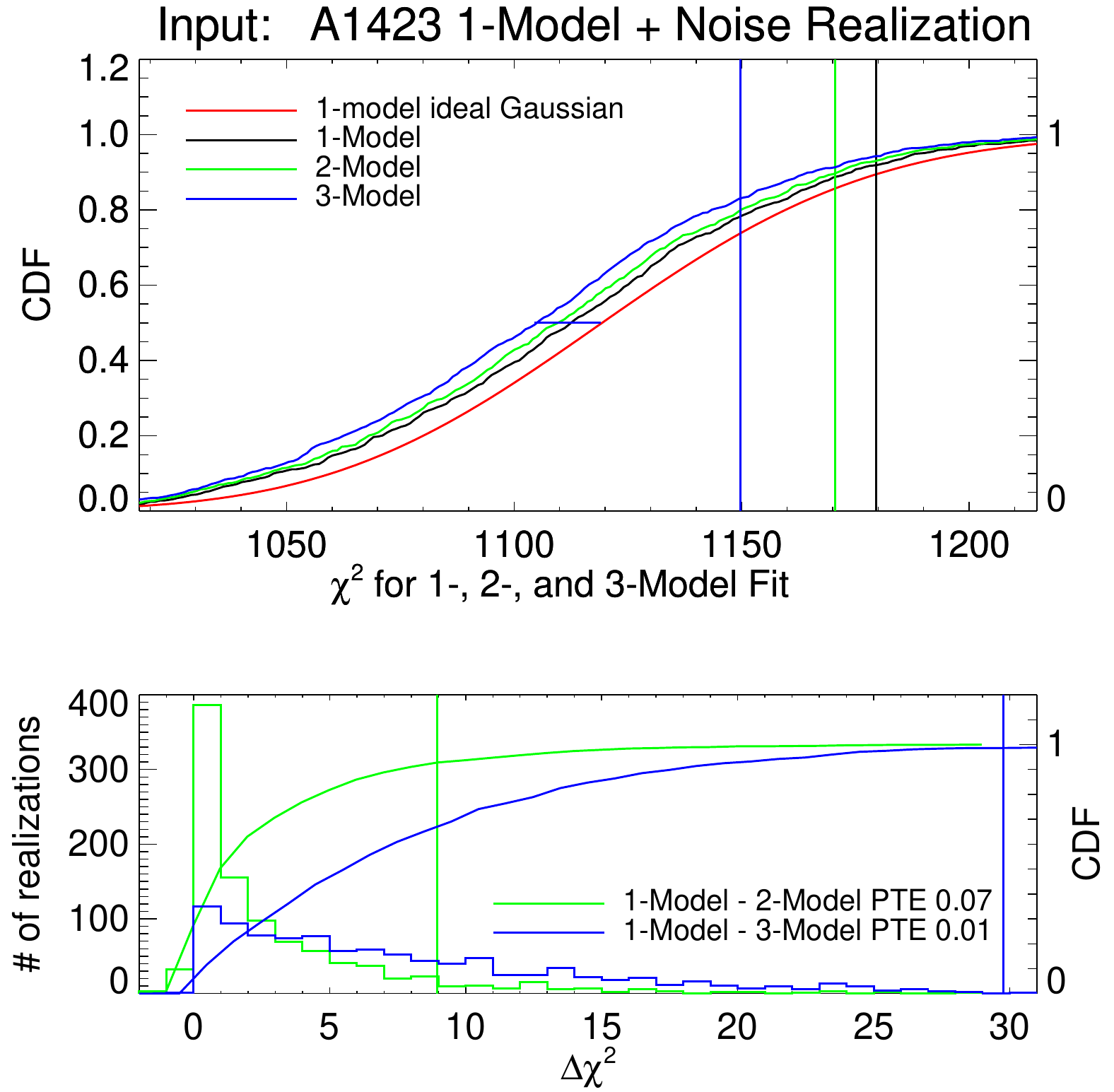}
  \caption{A demonstration of the first step of our model selection procedure for the cluster Abell~1423 (see text).
This procedure is used to determine the minimal number of model parameters (MPs) to include in the gNFW fits.
Upper panel:
(Solid colored curves) The CDF of $\chi^2$ values measured using an input 1-MP model (i.e. a spherical gNFW model with the $c_{500}$ parameter from Equation~\ref{eq:GNFW} fixed to the Arnaud10 value) added to each of the 1000 noise realizations described in Section~\ref{sec:noise} and fit with various numbers of MPs;
(vertical lines) the measured $\chi^2$ values for the observed Bolocam SZE data fit with various numbers of MP.
Both for the measured and simulated $\chi^2$ values, the black, green, and blue coloring represent the 1-, 2-, and 3-MP model fits respectively.
The red curve represents the expected $\chi^2$ CDF for the data being fit with 1-MP under the assumption of ideal Gaussian noise and is included as a visual aid.
The difference between the red and the black CDFs depicts how an ideal Gaussian $\chi^2$-distribution differs from the $\chi^2$-distributions derived from our noise realizations when fit with the same number of MPs.
The horizontal line in the center of the CDF curves is drawn to help the reader observe how the median values for the various noise models differ.
Lower panel: 
(Solid colored curves) For the same model+noise realizations as in the upper panel, the $\Delta\chi^2$ distributions for the 2-MP (green) and 3-MP (blue) models relative to the 1-MP model ($\chi^2_{11}-\chi^2_{12}$ and $\chi^2_{11}-\chi^2_{13}$);
(vertical lines) the corresponding $\Delta\chi^2$ values for the Bolocam SZE data ($\chi^2_{B1}-\chi^2_{B2}$ and $\chi^2_{B1}-\chi^2_{B3}$).
The PTE values indicate the probability that the differenced $\chi^2$ values for an input 1-MP model fit with both a 1-MP model and the indicated higher level model exceed the analogous differenced $\chi^2$ values from the observed BOXSZ data.
The PTE for the 3-MP model fit is less than 2\%, and therefore the model selection procedure indicates that the 3-MP model is a statistically preferred description of the data. 
From here, we continue the model selection at the third step of the procedure, as outlined in the text.}
\label{fig:JK}
\end{figure}
We have implemented a modified version of the F-test to determine the necessity of adding additional parameters to our pressure profile fit \citep{Bevington1992}.
The standard F-test statistic calculates the difference between the $\chi^2$-distributions for fits to models with differing numbers of free parameters, normalized by the reduced \chisqr~of the original model, and is given by:
\begin{equation}
  F_\chi = \frac{\chi^2(m)-\chi^2(m+1)}{\chi^2(m+1)/(N-m-1)} = \frac{\Delta\chi^2}{\chi^2_\nu}.
\end{equation}
The \chisqr~statistic is computed for both the candidate model with $m$ free parameters and the proposed model with $m+1$ free parameters, fit to the $N$ data points of the SZE map.

While our assumption that pixel-to-pixel noise is uncorrelated is sufficient for fitting pressure profiles to our SZE maps, this is not the case when testing for additional model parameters.
We therefore model the \chisqr~and $\Delta$\chisqr~distribution for each gNFW fit to each BOXSZ cluster using the 1000 noise realizations generated for each cluster and described in Section~\ref{sec:noise}, as they contain the full statistical information for the noise. 
We then select the best-fit cluster model using these $\Delta$\chisqr~model distributions using the procedure described below.

First we calculate $\chi_{Bq}^2$, which is the $\chi^2$-value for the Bolocam maps, $B$, fit to each of the four models under consideration, $q$.
The subscript $q\in[1,2,3,4]$ indicates the number of MPs for the model fit.
The \emph{modeled} $\chi^2$-distributions have a slightly different naming convention, $ \chi_{pq}^2$, and are calculated by adding a representative input model, with $p$ MPs, to each of the 1000 noise realizations and fitting each one of these model$+$noise realizations with a model with $q$ MPs.
As can be seen in the upper panel of Figure \ref{fig:JK}, each $\chi^2_{Bq}$ represents a single data point, while each $\chi^2_{pq}$ represents an
entire \emph{modeled} $\chi^2$-distribution.
As an example, $\chi^2_{13}$ is the 1000-realization $\chi^2$-distribution for an input 1-MP model$+$noise realization fit with a 3-MP model, while $\chi^2_{B3}$ represents the observed Bolocam data fit with a 3-MP model.

We then test for the necessity of additional parameters by comparing the difference of $\chi_{Bq}^2$ values for two different $q$'s with the distribution of the difference of the $\chi_{pq}^2$ of these same $q$'s.
The lower plot of Figure \ref{fig:JK} depicts $\chi_{11}^2-\chi_{12}^2$ and $\chi_{11}^2-\chi_{13}^2$ distributions for Abell~1423. The corresponding $\chi_{B1}^2-\chi_{B2}^2$ and $\chi_{B1}^2-\chi_{B3}^2$ values are represented by vertical lines.
Since the additional MP represented by the concentration parameter, $c_{500}$, is independent of the two additional MPs describing the ellipticity, $\epsilon$ and $\theta$, the 2-MP model and the 3-MP model represent two independent branches of comparison in our model selection procedure.
We implement a hierarchical decision tree to choose the minimal model for each particular cluster. Starting with the 1-MP model and progressing towards the 4-MP model:
\\
\begin{enumerate}[a.]
\item 
First we quantify how our fits improve by allowing $c_{500}$ to float. 
Starting with an input 1-MP model, we generate the $\chi^2_{11}-\chi^2_{12}$ distribution for a particular cluster.
If $\chi^2_{B1}-\chi^2_{B2}$~is greater than 98\% of the model distribution,
the 1-MP model is ruled out and the process proceeds to step b.\footnote{The choice of 98\% as a cutoff for the model selection procedure is necessarily arbitrary, but it is motivated by a desire to have on average no more than one cluster from our sample of 45 fall above the cutoff due to a noise fluctuation rather than a true need for an additional MP in the model fit.}
Otherwise, the 1-MP model is determined to be a sufficient model for this branch and the process continues with step c, which tests the justification for adding elliptical degrees of freedom to the fits.
\item  
This step is analogous to step a, but using a 2-MP input model fit with a 4-MP model.
The $\chi^2_{B2}-\chi^2_{B4}$~values are compared with the $\chi^2_{22}-\chi^2_{24}$ distributions. If the $\chi^2_{B2}-\chi^2_{B4}$ value is greater than 98\% of the differenced \chisqr~model distribution, the 4-MP model is chosen as the minimal model and the model selection procedure is finished.
Otherwise, the 2-MP model is determined to be a sufficient model along this branch, and the process proceeds to step c.
\item  This is the second branch of the model selection procedure, this time replacing the 2-MP model of steps a and b with the 3-MP model. Again, if the 4-MP model is selected along this branch, it represents the minimal model and the process is finished.
\item If both branches select the 1-MP model, then the 1-MP model is chosen.
\item If only one branch selects a 2-MP or a 3-MP model, then the model selected along that branch is chosen.
\item If both the 3-MP model and the 2-MP model are selected, the 2-MP model is chosen, as it has fewer MPs.
\end{enumerate}
\section{Scaling Relation Bias Due to Selection Effects}
\label{sec:sf}
We now assess biases in our measured scaling relations specific to the the \emph{ad hoc} method by which we chose clusters for the BOXSZ sample.
This procedure also accounts for other biases associated with our fitting procedure, such as non-Gaussian confidence intervals for our measured \Msmall~and \Ysmall~values.
Our results, which are presented below, indicate that selection effects influence the BOXSZ \Ysmall--\Msmall~scaling relation by less than the 1$\sigma$ measurement uncertainty of the best-fit parameters, and all of the results presented in the body of the manuscript have been corrected for these effects. This methodology should be generally applicable for other cluster samples that have non-analytic selection functions.

We now briefly review our mass function formulation. Readers who are already familiar with this formalism and who are not interested in the specifics of our implementation may skip to the paragraph following Equation \ref{eq:stanek_sigma}. The mass function characterizes the number of clusters per unit volume with masses between $[M,M+dM]$ and redshifts, $[z,z+dz]$: 
\begin{equation}
\label{eq:Tinker}
  \frac{dn(M,z)}{dM}=f(\sigma)\frac{\overline{\rho}_{\rm{m}}(z)}{M}\frac{d\ln \sigma^{-1}}{dM},
\end{equation}
where
\begin{equation}
  f(\sigma) = A\left[\left(\frac{\sigma}{b}\right)^{-a}+1\right]\textnormal{e}^{-c/\sigma^2}.
\end{equation}
The $\sigma$ and $\overline{\rho}_m(z)$ terms represent the variance in the matter power spectrum and the mean matter density at the redshift of the cluster, respectively.
We employ the measured values given in \citet{Tinker2008} for $\Delta=300\Omega_m(z)$, $[A,a,b,c]=[0.200,1.52,2.25,1.27]$, as these are the same parameters used in the M10 analysis.
Multiplying equation~\ref{eq:Tinker} by $dV/dz$ yields the total predicted number of clusters per unit redshift at the redshift of interest:
\begin{equation}
\label{eq:dNdMdz}
  \frac{d^2N}{dMdz}=f(\sigma)\frac{\overline{\rho}_m(z)}{M}\frac{d\ln\sigma^{-1}}{dM}\frac{dV}{dz}.
\end{equation}
The variance in the matter power spectrum, $\sigma$, is a monotonic function of mass, which evolves with redshift as:
\begin{equation}
\sigma(M,z)=\sigma(M,z_{init})\frac{G(z)}{G(z_{init})}\frac{(1+z_{init})}{(1+z)}.
\label{eq:sigma_m_z}
\end{equation}
Here $G(z)$ is the growth function, which we obtain by numerically integrating the following second-order differential equation derived from the Einstein equations (see \citealt{Mortonson2009} for a particularly well-presented introduction to the topic):
\begin{equation}
\frac{d^2G}{d\ln a}+\left(4+\frac{d\ln H}{d \ln a}\right)\frac{dG}{d\ln a}+\left[3+\frac{d\ln H}{d\ln a}-\frac{3}{2}\Omega_m(z)\right]G=0.
\label{eq:einstein_growth}
\end{equation}
\citet{Evrard2002} demonstrate using N-body simulations that $\sigma(M_{200},z=0)$ is well fit using a log quadratic relation of the form:
\begin{equation}
\ln \sigma^{-1}(M_{200},z=0)=s_0+s_1\ln M_{200}+s_2\left(\ln M_{200}\right)^2.
\label{eq:stanek_sigma}
\end{equation}
We have chosen to use a $z_{init}=0$ calibration of $\sigma(M,z)$ derived from these simulations and given in \citet{Stanek2006} for $\Omega_M=0.3$: $s_0=0.667$, $s_1=0.281$, and $s_2=0.0123$. Here we have renormalized the relation from $\sigma_8=0.9$ to $\sigma_8=0.8$ by adding $\ln(0.9/0.8)$ to $s_0$.

We sample the Tinker mass function, Equation \ref{eq:dNdMdz}, defined at $\Delta=300\Omega_m$, using a grid of halo masses, $[M_{300\Omega_m},M_{300\Omega_m}+\Delta M_{300\Omega_m}]$, and redshifts, $[z,z+\Delta z]$.
Instead of generating a mock sky realization within the specific solid angle observable with Bolocam, the sky is over-populated with enough clusters as not to introduce numerical selection effects. 
This corresponds to about eighty times the solid angle observable with Bolocam.
The exact justification for this approach is discussed in more detail later in this Appendix, when we describe specifically how the candidate mock clusters are selected.

Our procedure to generate masses at different overdensities is designed to be as consistent with the analysis of M10 as possible.
Following the procedure given in \citet{Hu2003}, we use the obtained $M_{300\Omega_m}$ values to generate \Mbig{ }values assuming an NFW concentration parameter of $c_{200}=4$.\footnote{This concentration parameter is consistent with the M10 analysis and is close to the empirically derived mean concentration of the CLASH sample obtained from both a joint weak- and strong-lensing analysis \citep[$c_{200}=3.65$ with a standard deviation of $0.65$]{Merten2014} as well as from a stacked weak-lensing-only analysis \citep[$c_{200}=4.01$ with a standard error of approximately $0.3$]{Umetsu2014}.}
The sum of all the grid points represents the total mean number of clusters in the parameter range of interest.
We have chosen an \Mbig~range from approximately $1.5\times10^{14}$ to $4\times10^{15}$\Msun, and have confirmed that the measured scaling relation bias is insensitive to any extension of mass range or increase in cluster density in the mock sky realizations.
We randomize the process by assigning each $[M_{500},z]$ grid point to a specific segment of probability space (weighted by the mean number of clusters for that grid point) and then sampling this space until the total number of clusters corresponds to the total mean number of clusters within the chosen solid angle---approximately $10^7$ clusters, dominated by the least massive halos.
While the mass ratio between $\Delta=300\Omega_m$ and $\Delta=500$ is not sensitive to the precise value of $c_{200}$ for typical massive clusters, the mass ratio between $\Delta=500$ and $\Delta=2500$ is sensitive to $c_{200}$ because the NFW scale radius typically falls near \rsmall. We therefore use the measured \Msmall/\Mbig~ratios from Table~\ref{tab:szx_info}~to assign \Msmall~values to each mock cluster, effectively accounting for the concentration parameter specific to each cluster in our data set.

For each of the 45 clusters in the BOXSZ sample, we generate observables for all mock clusters within $\Delta z=\pm 0.02$ of the given BOXSZ cluster redshift.
We apply a complete set of observable-mass scaling relations, $\Theta$ (introduced in Section \ref{sec:SR}), including a proposed correlation matrix, $\rho_{lty}$, for the intrinsic scatter, to the sampled mock cluster masses to generate mock $l$,$t$, and $y$ observables.
Initially, we construct $\Theta$ using the X-ray-only scaling relations from M10 and the uncorrected best-fit \Ysmall--\Msmall~scaling relation.
We determine $\rho_{lty}$ using a combination of observational and simulation results. 
We adopt the $l$-$t$~correlation coefficient from the measured value of M10: $\rho_{lt}=0.1$.
However, as observed measurements of $\rho_{ly}$ and $\rho_{ty}$ are limited, we use the simulated results of \citet[their ``pre-heating'' scenario; hereafter, S10]{Stanek2010} as a starting point from which we estimate our fiducial X-ray observable-\Ysz~correlation. 
We set $\rho_{ty}$ to $0.6$, which is the S10-simulated correlation coefficient between $y$ and a spectroscopic-like temperature.
For $\rho_{ly}$, we first note that the M10-observed $\rho_{lt}=0.1$ is lower than the S10 value of $0.7$.
This is a result of the use of bolometric luminosity in S10 as opposed to the use of soft-band $0.1-2.4$~keV luminosity in M10.
As the values of $l$ we use to characterize our scaling relation bias are calculated in a fashion identical to M10, we set $\rho_{ly}$ to $0.1$ under the assumption that $\rho_{ly}$ will be similar to $\rho_{lt}$.
Using $\rho_{lty}$, we generate a covariance matrix using the proposed marginalized intrinsic scatter for the individual observables in $\Theta$, and, to each mock cluster observable ($l$,$t$, and $y$), we add a random intrinsic scatter realization using the \ttfamily mrandomn\footnote{http://idlastro.gsfc.nasa.gov/ftp/pro/math/mrandomn.pro}\normalfont~function in IDL.
We add log-normal measurement noise realizations to the mock $l$ and $t$ values based on the 1$\sigma$ measurement uncertainties given in Table \ref{tab:szx_info}.

We formulate the selection function process to mimic our \emph{ad hoc} selection of galaxy clusters that, to a large extent, we chose to have \Tx$\gtrsim5$keV from X-ray luminosity-selected studies.
With a full set of observables assigned to each mock halo, denoted with the subscript $m$ below, we select the mock halo that best matches the measured $l$ and $t$ values (measured in Section \ref{sec:xray}) within $\Delta z=\pm 0.02$ of each BOXSZ cluster, $i$, using:
\begin{eqnarray}
\begin{array}{c}\\[-2pt]
\displaystyle \operatorname{argmax}\\[-3pt]
\scriptstyle m
\end{array}
~\mathcal{L}(m|i) =
\begin{array}{c}\\[-2pt]
\displaystyle \operatorname{argmax}\\[-3pt]
\scriptstyle m
\end{array}
\exp\left(-\frac{(l_i - l_{m})^2}{s_{l}^2} - \frac{(t_i - t_{m})^2}{s_{t}^2}\right),
\label{eq:sf}
\end{eqnarray}
where $s_l$ and $s_t$ are the measurement errors on $l$ and $t$ for a given BOXSZ cluster.\footnote{The $\operatorname{argmax}$ operator returns the argument (here the mock galaxy cluster, $m$) that maximizes $\mathcal{L}(m|i)$ for a given cluster, $i$.}
The underlying mass function is thereby indirectly sampled, providing the associated distribution of $y$ values for a given observed BOXSZ cluster.
For the $y$ and mass values, we logarithmically add the noise from a single noise realization described in Section~\ref{sec:ysz_est}. 
Our choice to add $y$ measurement noise logarithmically is not completely correct, and this is most apparent with 6 of our clusters where a few ($<10\%$) of their noise realizations generate negative $y$ values.
For this application, we drop those noise realizations that generate negative $y$ values, and we determine later in this appendix that this treatment is adequate.
We call the 45 selected mock clusters a ``simBOXSZ" sample. Each mock cluster is assigned its BOXSZ analogue cluster's measured noise model. 
We repeat the above process and generate a total of 1000 simBOXSZ sample realizations.

The discrete nature of this selection process motivates our dense sampling of the mass function.
Given the rarity of the high-mass BOXSZ sample clusters in the observable universe, a realistically populated mock sky produces a very small number of clusters to select as possible counterparts to the true BOXSZ clusters.
The nature of an Eddington bias is such that, for an observable like luminosity that has a large scatter with respect to mass, and given a steeply falling mass function, that observable is more likely to be obtained from a lower-mass cluster that has an upwards scattered observable signal than from a higher-mass cluster with the observable matching the value expected from the scaling relation.
Without populating the sky densely enough ($\sim$80 skies), Equation \ref{eq:sf} is more likely to choose clusters with systematically low mock luminosities, which would introduce an additional, unwanted selection effect.
\begin{figure*}
  \centering
  {\setlength{\tabcolsep}{1.5em}
  \begin{tabular}{cc}
    \includegraphics[scale=0.555]{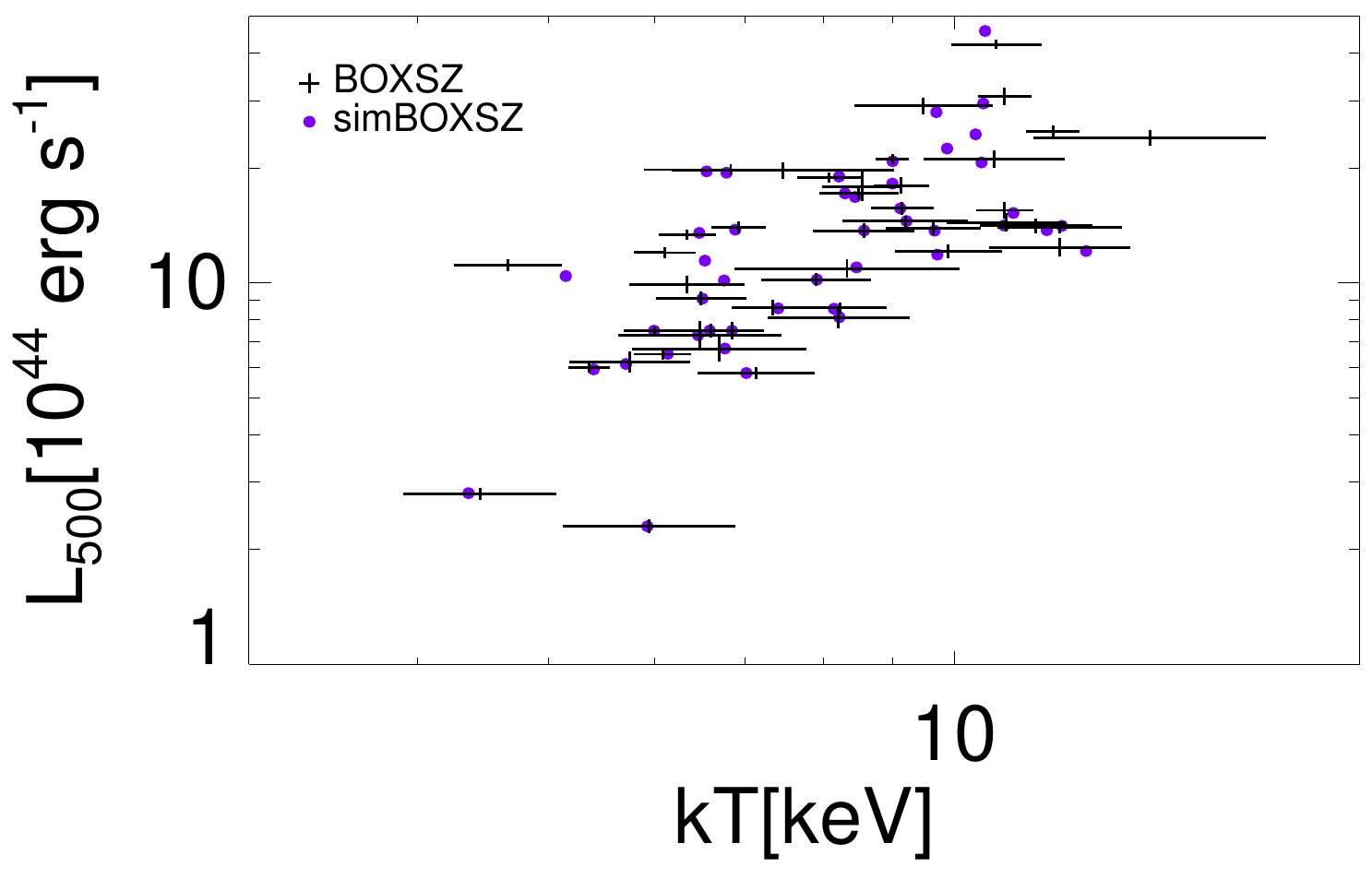} &
    \includegraphics[scale=0.53]{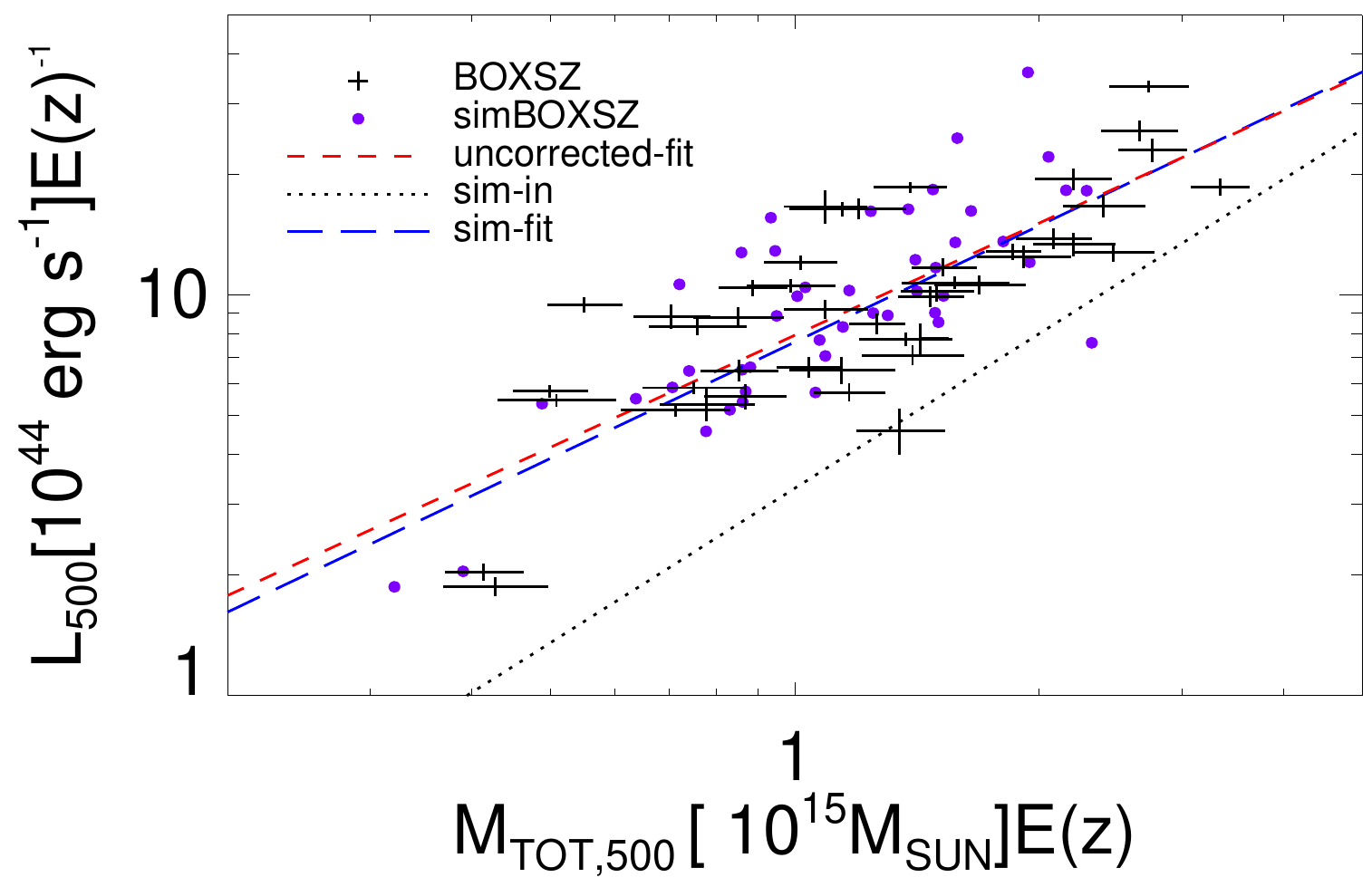} \\ \\ \\
    \includegraphics[scale=0.555]{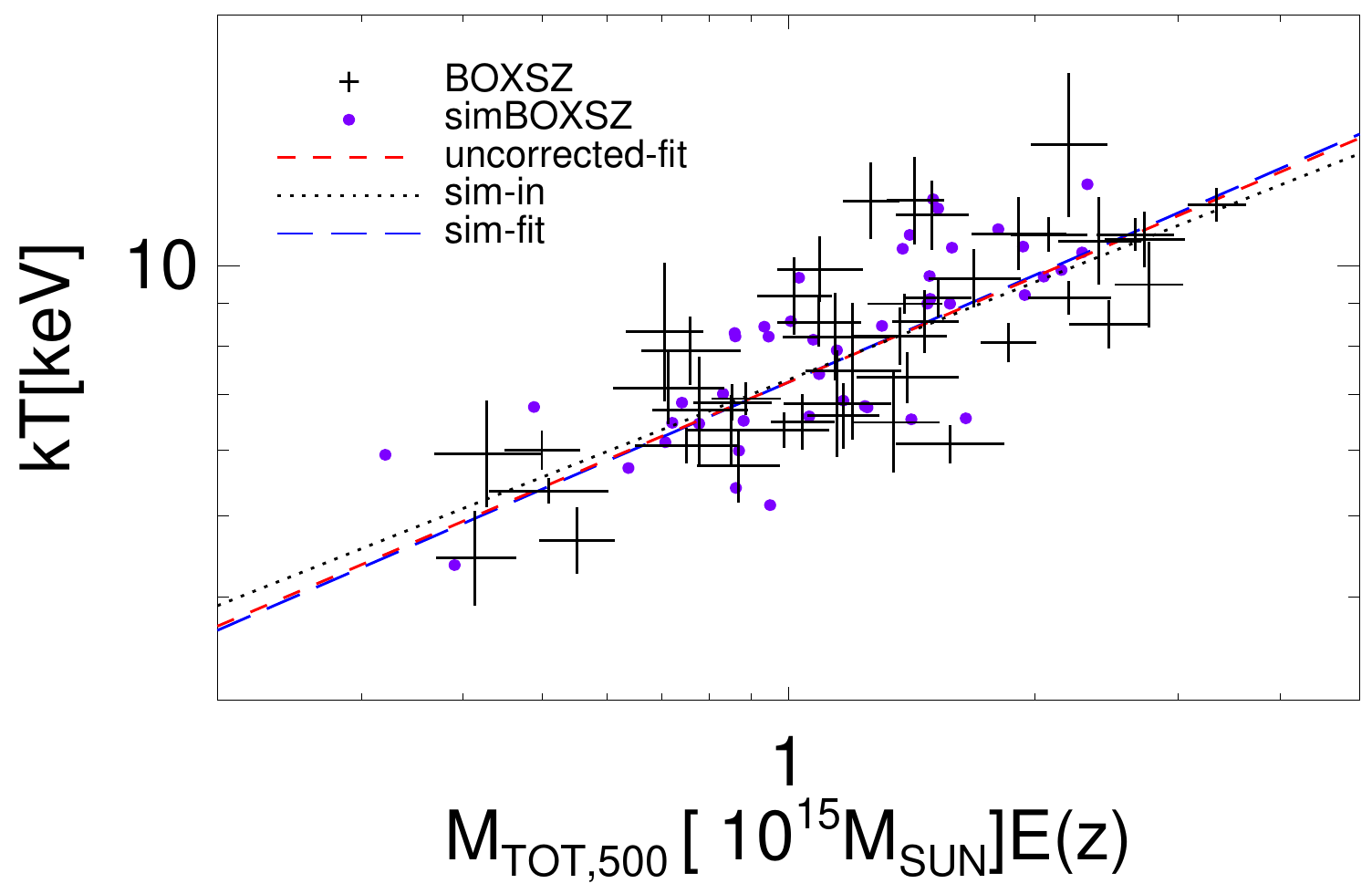} &
    \includegraphics[scale=0.53]{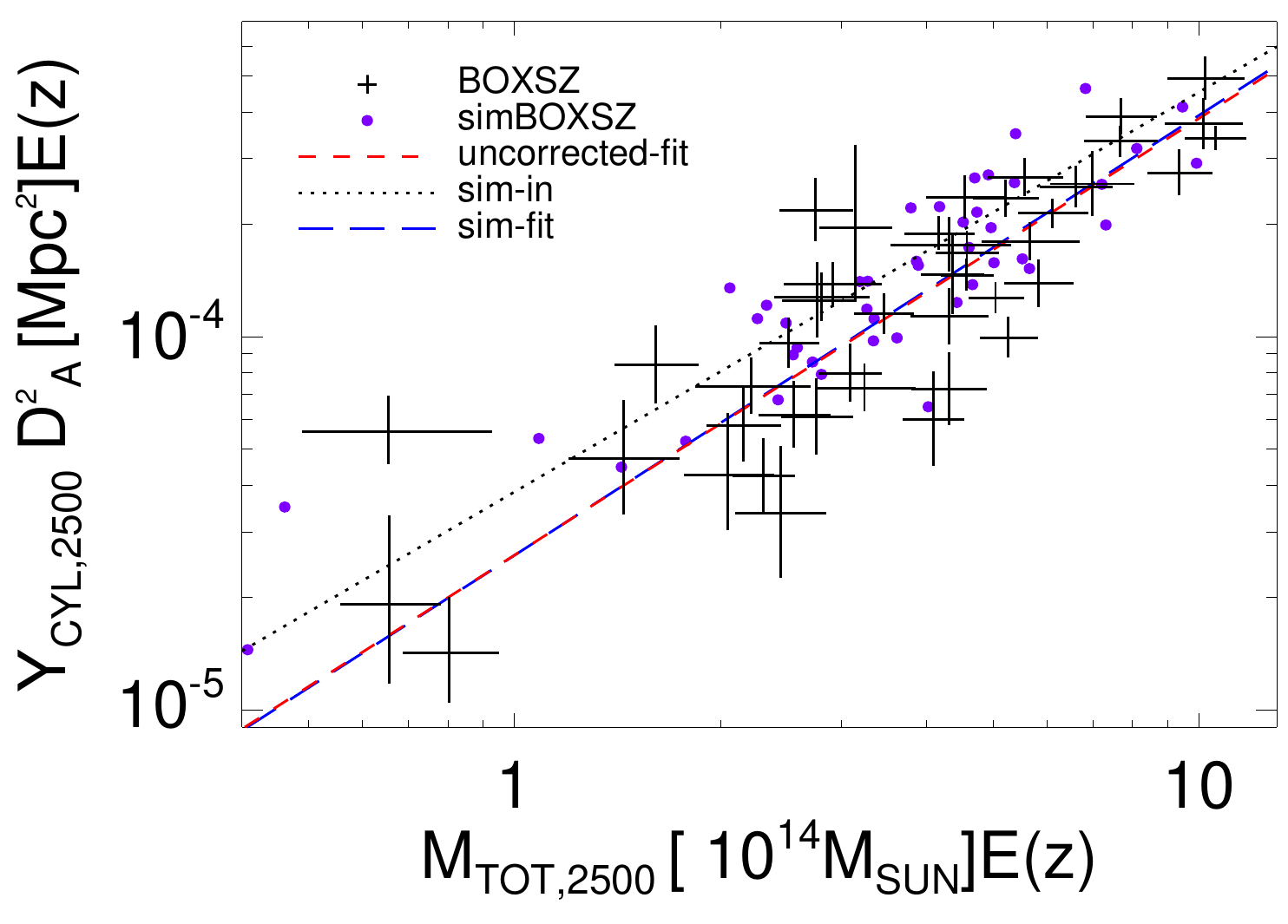}
  \end{tabular}}
  \caption{Clockwise starting in the top-left panel: the \Lbig--\Tbig, \Lbig--\Mbig, \Tbig--\Mbig, and \Ysmall--\Msmall~relations for one realization of the selection bias simulation.
Black data points represent the observed parameter pairs together with their measured 1$\sigma$ uncertainties. Purple dots mark the positions of one simBOXSZ sample. 
The input and median output scaling relations (taken over all simBOXSZ samples) are depicted by the black dotted and the blue dashed lines, respectively. 
The uncorrected best-fit scaling relations for the BOXSZ data are given by the red dashed lines, which by design, closely match the output scaling relations from the simulation.
The \Lbig--\Mbig~relation is most affected in this process, as it the sole observable in the X-ray selection process for most of the clusters.}
  \label{fig:simBOXSZ}
\end{figure*}

An example of the process by which we choose these mock clusters is given in the top-left panel of Figure \ref{fig:simBOXSZ}.
The figure depicts the distribution of $L_{500}$ versus \Tx~for one of the simBOXSZ samples. 
Due to the increased density at the low-mass end of the mass function, it is extremely likely to find a cluster in the simulation with almost exactly the same luminosity-temperature properties as the cluster under consideration.
At the more massive end, the difference between the measured and mock parameters is larger but generally within the measurement error.
To maximize computational speed, the final cluster density was chosen such that a further increase was not observed to significantly change the results.

We fit scaling relations to each of the simBOXSZ samples and find the median fit parameters over all 1000 simBOXSZ samples.
We then compare the median simBOXSZ scaling relation parameters to the uncorrected scaling relation fit to the true BOXSZ data.
We iterate this entire process, perturbing the input $\Theta$ until the median best-fit parameters of the simBOXSZ samples match the best-fit parameters of the true BOXSZ data. We find that the \oym~bias is unchanged when we match the output simBOXSZ scaling relations over the range of scaling relation parameters consistent with the BOXSZ scaling relation measurement uncertainties.
In Figure \ref{fig:simBOXSZ}, we compare the BOXSZ and the simBOXSZ-measured $\theta_{l|m_{500}}$, $\theta_{t|m_{500}}$, and $\theta_{y|m_{2500}}$.
\begin{deluxetable}{cccccl} \tabletypesize{}
\tablecaption{Measured biases in the scaling relation parameters for the BOXSZ cluster sample due to selection effects.}
\tablewidth{0pt} \tablehead{$\theta$ & &Sample&Output & Input & Out-In} 
\startdata 
$Y_{2500}-M_{2500}$ & $\Delta\beta_1$ &BOXSZ&1.179&1.071&\phs0.108$\pm$0.05\\ \vspace{2 mm}
					   & 		  &\Ysmall$>0$&1.125&1.038&\phs0.087\\
					   & $\Delta\beta_0$ &BOXSZ&0.004&0.120&$-$0.116$\pm$0.05\\ \vspace{2 mm}
					   & 		  &\Ysmall$>0$&-0.144&-0.050&$-$0.094\\
					   & $\Delta\sigma$ &BOXSZ&0.103&0.109&$-$0.005$\pm$0.005\\ \vspace{2 mm}
					  &			&\Ysmall$>0$&0.098&0.104&$-$0.006\\
$T_{X}-M_{500}$ & $\Delta\beta_1$ &BOXSZ&0.427	&0.389  &\phs0.038$\pm0.01$\\ \vspace{2 mm}
					&	&\Ysmall$>0$&0.406&0.369&\phs0.037\\
					   & $\Delta\beta_0$ &BOXSZ&0.860&0.863&$-$0.003$\pm0.03$\\ \vspace{2 mm}
					&			&\Ysmall$>0$&0.871&0.863&\phs0.009\\
					&$\Delta\sigma$		&BOXSZ&0.053&0.055&$-$0.002$\pm0.005$\\ \vspace{2 mm}
					&			&\Ysmall$>0$&0.056&0.053&\phs0.003\\
$L_{500}-M_{500}$ & $\Delta\beta_1$ &BOXSZ& 0.965&1.282&$-$0.316$\pm0.07$\\ \vspace{2 mm}
					  &			&\Ysmall$>0$&0.765&1.231&$-$0.475\\
					   & $\Delta\beta_0$ &BOXSZ&0.883&0.518&\phs0.365$\pm0.01$\\ \vspace{2 mm}
					&			&\Ysmall$>0$&0.929	&0.584  &\phs0.3338\\
					&$\Delta\sigma$		&BOXSZ&0.152&0.175&$-$0.023$\pm0.005$\\ \vspace{2 mm}
					&			&\Ysmall$>0$&0.145&0.168&$-$0.023
\enddata
\tablecomments{First column: the scaling relation under investigation. Second column: the measured scaling parameter. 
Third column: the sample under investigation. BOXSZ indicates the full sample of 45 clusters and \Ysmall$>0$ indicates the sample of 39 clusters whose \Ysmall~noise realizations exlusively generate positive \Ysmall~values.
Fourth, fifth and sixth columns indicate the best-fit scaling relation of the simBOXSZ sample, the input into the simulation after matching the output scaling relations to the best-fit values obtained from the real data, and the difference of these two values, where the input is subtracted from the output.
We indicate the $Y_{2500}-M_{2500}$ selection bias corrections in bold font as these are the main results of this section and the other entries have only been included as consistency checks. 
The uncertainties indicate the range of scaling relations biases (``Out-In") obtained when varying the intrinsic scatter correlation values between \Lx, \Tx, and \Ysmall~over the uncertainty ranges discussed in the text.}
\label{tab:boxsz_delta} \end{deluxetable}

We repeat this process, perturbing all of the initial scaling relation parameters by amounts greater than or equal to their measurement uncertainties. When using a fixed covariance matrix, we find that our results do not depend on these initial conditions.
We further explore how our chosen value for the correlation of the intrinsic scatter of the various observables affects the bias in the scaling relations due to selection effects. Since we adopted $\rho_{lt}=0.1$ from the M10 measured value, we perturb this by the M10 measurement error $\pm0.2$ and measure how this changes the results. Due to limited observational constraints on the covariance of \Ysz~with X-ray parameters, we treat $\rho_{ly}$ and $\rho_{ty}$ in an analogous manner and also perturb them by $\pm0.2$. Our results indicate that only $\rho_{ly}$ introduces a noticeable change in our scaling relation bias, on the order of the bias itself, and we add this as an additional systematic uncertainty to our results.

We present the final corrections to the scaling relations for the effects described in this appendix in Table \ref{tab:boxsz_delta}.
These results indicate that the measured departure of the \Ysmall--\Msmall~relation from self-similarity is not due to selection effects for clusters following a conventional mass function predicted by the standard cosmological model.
The change in the \Tx--\Mbig~relation is also negligible. We measure a large selection bias in the \Lx--\Mbig~relation, as expected, since most of the clusters in our sample were initially discovered based on luminosity measurements and the true \Lx--\Mbig~relation has a large amount of intrinsic scatter.

Our results indicate $\beta_1^{y|m}$ to be biased steeper by a little less than $1\sigma$ of the statistical uncertainty of the measured uncorrected relation parameter. Furthermore, the scaling relation bias due to selection effects is more noticeable for the less massive clusters than for the more massive clusters. The selection bias for the $\beta_0^{y|m}$ value is almost entirely due to our choice of normalization and is inversely correlated with the $\beta_1^{y|m}$ value.

The small selection bias for the \Ysmall--\Msmall~relation arises for two main reasons. 
First, the low intrinsic scatter of the \Ysz~signal with fixed cluster mass reduces the overall level of Eddington bias. 
Second, although the BOXSZ \Lx-\Mbig~scaling relation is significantly affected by selection bias, the small expected correlation in the intrinsic scatter between the luminosity and the SZE, at fixed mass, ensures very small cross-over selection effects from luminosity to SZE \citep{Allen2011}.
Finally, lower mass clusters generally received longer integration times, so the introduction of a Malmquist bias due to a hard flux cut-off (such as with a survey of uniform depth) does not apply for the BOXSZ scaling relations. 

Since we directly sample our \Ysmall~and \Msmall~noise realizations, any non-Gaussianities and correlations in the measurement noise are included in the mock samples, and any regression biases thus induced are accounted for in the measurements given in Table \ref{tab:boxsz_delta}.
We end by considering possible systematic biases due to subtleties associated with how we apply our cluster-specific noise realizations to mock clusters as part of our assessment of the bias of the BOXSZ scaling relations. The assumption that noise adds logarithmically is the least valid for the lowest signal-to-noise clusters because noise realizations that generate negative $y$ values must be dropped and small $y$ values in a linear-normal distribution generate a long tail in the logarithmic distribution.
To assess this systematic, we drop all clusters that have individual noise realizations that produce negative \Ysmall~values.
The dropped clusters are: Abell~963, Abell~1423, MACS~J1720.2+3536, ZWCL~0024+17, MACS~J0911.2+1746, and MS~2053.7-0449. Four of these clusters have the lowest peak SZE S/N and are natural candidates to be dropped. 
MACS~J1720.2+3526 and Abell~963 are less obvious candidates to be dropped. In the case of the former, the poorness of the model fit described in the
footnote of Table \ref{tab:fit_info} may be the reason that individual simBOXSZ model+noise realizations can yield a negative \Ysmall.
In the case of the latter, the problem may be associated with noise due to the subtraction of a bright radio galaxy near the cluster.
We repeat the selection bias analysis after dropping these six clusters, the results of which are also given in Table \ref{tab:boxsz_delta}.
Going from the full sample to the reduced sample, the measured selection biases in the parameters of the \Ysmall--\Msmall~scaling relation change by less than the systematic uncertainty of the correction factors themselves. Therefore, we conclude that our treatment of the \Ysz~noise realizations to characterize the scaling relation parameter bias is adequate.
\section{Total Mass Estimates}
\label{sec:fgas}
The method used to estimate total cluster masses has a direct impact on the measured scaling relations. Here we review the motivation for our approach and comment on the impact of potential systematics.

\begin{figure}	
  \begin{center}	
    \includegraphics[scale=0.6]{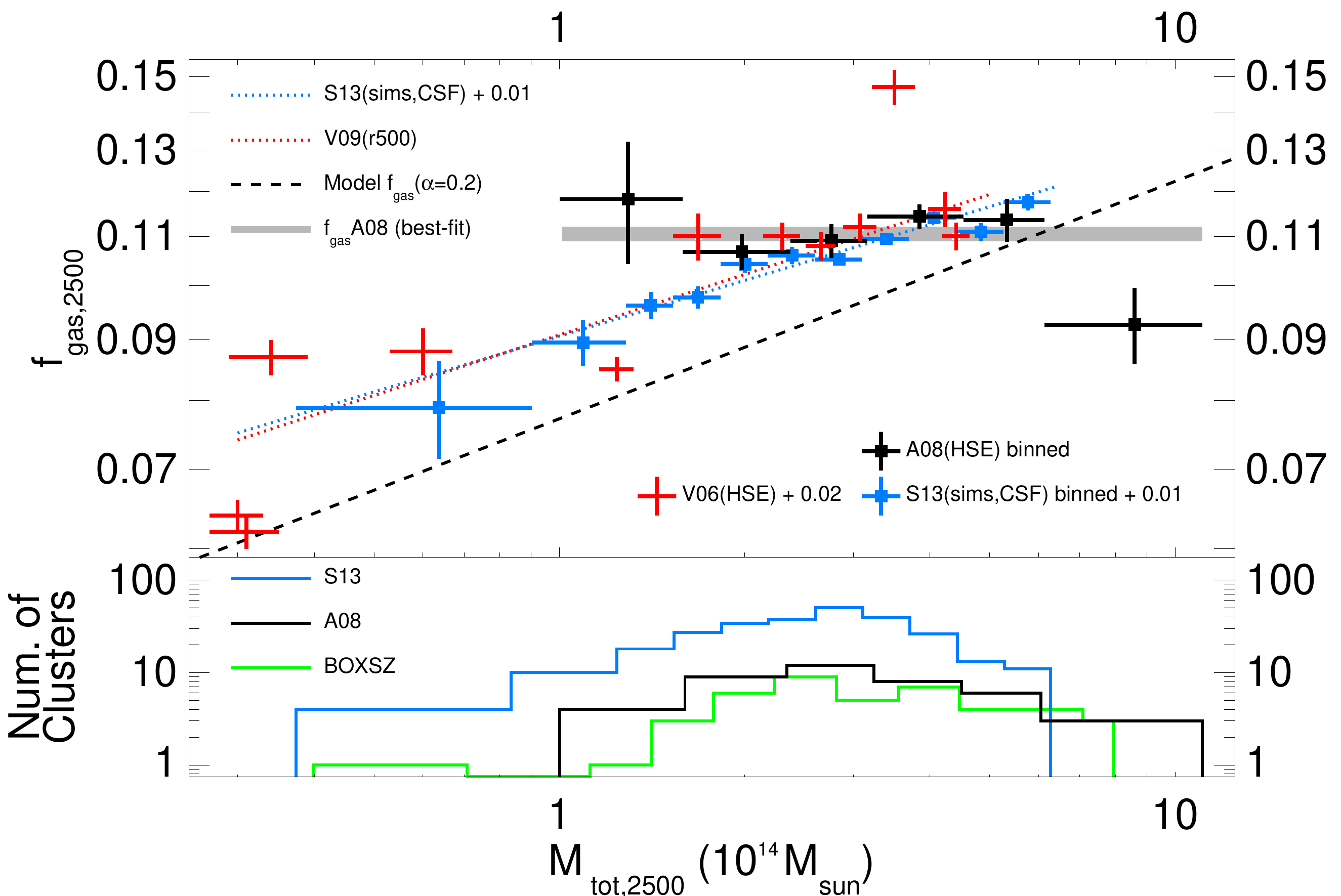}	
    \caption{Measured and simulated values of \fgas{ }as a function of \Msmall.
        Histogram: \Msmall{} distribution of the BOXSZ (green), Allen08 (black, labeled A08), and S13 (blue) clusters.
        Black points: binned Allen08 \fsmall{ }measurements.
        The gray region is the best-fit constant \fsmall~value from Allen08 with $\pm$1$\sigma$ measurement uncertainty.
        Red points: individual \fbig~measurements from V06,\footnote{Regarding the outlier at \Msmall$=3.35\times10^{14}$\Msun, V06 point out that the \fgas{} value for this cluster (Abell~2390) should be treated with caution.
This is due to the presence of large cavities in the intracluster medium near the cluster center, which could result in an underestimate of the mass (overestimate of \fgas). This disturbed region is down-weighted in the Allen08 analysis, and their measured \fgas{} value for Abell~2390 is in agreement with other clusters of similar mass.} shifted upward by 0.02 to account for an overall calibration shift between V06 and Allen08 (see V06).
        Red dotted line: V09 best-fit scaling relation, based primarily on the V06 data.
        Blue dotted line:  \citet[S13]{Sembolini2013} best-fit scaling relation to their CSF simulations.
        Blue points: binned simulated clusters from S13. The S13 best-fit scaling relation and binned data points are shifted upwards by 0.01 to provide consistency with the other data sets. The mass binning for the Allen08 data is the same in both plots, and similarly for the S13 simulations.
	        The dashed black line is the steepest \fgas{} model (Equation \ref{eq:fgas}) considered in Appendix \ref{sec:fgas}.}
    \label{fig:fgas_all2500}
  \end{center}
\end{figure}

As described in Section \ref{sec:xray}, we determine \Msmall{} from gas mass profiles measured with {\it Chandra} and a model for the gas mass fraction. The \fgas{} model that we adopt is based on the measurements of Allen08, who provide \fsmall~measurements for a large sample of 42 clusters that is well matched in mass and redshift to our own. 
Figure \ref{fig:fgas_all2500} shows \fgas{} values for these clusters, binned in mass (black points), as a function of \Msmall. The Allen08 data are consistent with a constant value (gray horizontal band) and span a range in mass that contains all but two of the BOXSZ clusters; as demonstrated in Section \ref{sec:splittests}, these two clusters do not strongly influence our scaling relations.
Also shown in Figure~\ref{fig:fgas_all2500} are \fsmall~measurements for individual clusters from V06 (red points). These data too are consistent with a constant value of \fgas{ }over the mass range most relevant for the BOXSZ analysis, although they show a lower value of \fgas{ }at smaller masses.
\citet{Sun2009}, primarily using galaxy groups at lower masses than our sample, also observe an increasing trend of \fgas{ }with \Mtot{ }(see also \citealt{Pratt2009}).
These results can be reconciled with the Allen08 results by an \fgas--\Mtot{ }relation which generally increases with mass but flattens at the high masses probed by the bulk of the Allen08 and BOXSZ samples.
\citet{Stanek2010} report such a flattening in the Millennium Gas simulations, and a flattening may also be present in the simulations of \citet[][blue points in Figure \ref{fig:fgas_all2500}]{Sembolini2013}.
Note that both of these simulations include CSF but not AGN feedback, which is potentially important within \rsmall.

While there are still significant uncertainties in modeling the gas physics of galaxy clusters (as is evident from the diversity of simulation predictions), differences between the observational studies discussed above may also have contributions from systematic errors in calibration.
In particular, a temperature-dependent disagreement between temperatures measured with the {\it Chandra} ACIS, XMM PN, and XMM MOS detectors is now well documented (e.g.,\ \citealt{Nevalainen2010, Tsujimoto2011, Mahdavi2013, Schellenberger2014}). At least two of these instruments must be systematically biased as a function of temperature, which would result in a mass-dependent bias in HSE mass estimates. Such a bias would straightforwardly alter the slope measured in any scaling relation analysis that ultimately relies on HSE masses.
While masses derived from gravitational lensing would circumvent this issue, current samples either exclude the region $\lesssim$~\rsmall~from their analysis (e.g., the WtG analysis presented in \citealt{Applegate2014}), or are not yet large enough to precisely constrain scaling relation slopes (e.g., the CLASH WL+strong-lensing samples; \citealt{Merten2014}).
 
We explore how a dependence of \fgas{ }on mass would affect our \Ysz--\Mtot{ }scaling relation measurements by adopting a power-law relation between \fgas{} and mass,
\begin{equation}
  f_{\rm gas,2500} = f_{0}~\left(\frac{M_{2500}}{6\times10^{14}M_{\odot}}\right)^\alpha,
  \label{eq:fgas}
\end{equation}
where $f_0$ is a constant. Values of $\alpha$ in the literature range from approximately $0.0$ to $0.2$, with the larger values measured from samples extending to significantly lower masses than BOXSZ, as noted above and plotted in Figure \ref{fig:fgas_all2500}. This range is also commensurate with the correction required to forge agreement between temperatures measured by XMM and {\it Chandra} \citep{Rozo2014a,Schellenberger2014}.
\begin{figure}
  \begin{center}
    \includegraphics[scale=0.7]{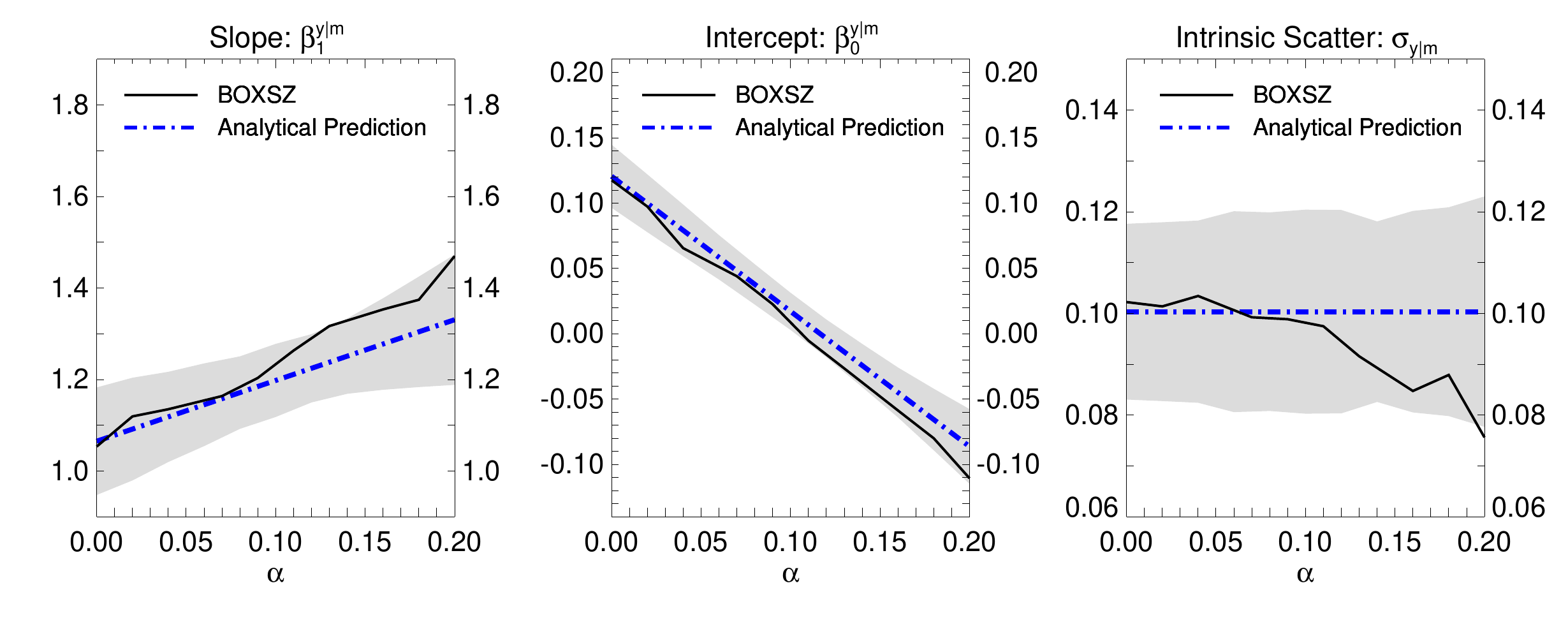}
    \caption{The dependence of our \Ysmall--\Msmall~scaling relation parameters on the assumed power-law scaling of \fgas~with \Msmall, with \fgas$\propto M_{2500}^\alpha$ (see Equation \ref{eq:fgas}).
The black solid curves indicate the best-fit parameters for the BOXSZ sample data, the blue dot-dashed lines indicate our analytical predictions, and the grey bands indicate the 68\% fluctuation band as determined from the mock samples. The method by which we calculate all of these values is described in the text.
The scaling relations for the BOXSZ sample data are corrected for biases induced by selection effects and regression biases.}
\label{fig:ymbias}
  \end{center}
\end{figure}
We would like to use Equation \ref{eq:fgas} to generate new \Msmall{ }values for various values of $\alpha$. However, since our aperture of integration, \rsmall, is a function of \Msmall, ideally, we would repeat our analysis and measure \Mgassmall{ }and \Ysmall{ }directly from the observed X-ray and SZE maps, respectively. Such a process would include using the proposed \fsmall{ }to generate new \rsmall, \Mgassmall, and \Msmall{ }values from the X-ray data.
We would then integrate our SZE maps using these new \rsmall{ }values to calculate new \Ysmall~values and uncertainties. Different \rsmall{ }and \Mtot{ }values would further affect our selection bias estimates and the SZE signal offsets calculated for individual clusters.

For the purposes of this analysis, we approximate the above process as follows. Using the method described in \citet{Rozo2014a} that accounts for the role of \fgas{} in determining \rsmall, we scale our fiducial \Ysmall{} and \Msmall{ }values using the \fgas{ }model of Equation \ref{eq:fgas}, fixing $f_{0}=0.1104$, and varying $\alpha$ between 0.0 and 0.2:
\begin{eqnarray}
\label{eq:F2}
  \frac{M_{\rm 2500,\alpha\ne 0}}{M_{\rm 2500,\alpha=0}}&=& \left(\frac{f_{\rm gas,2500,\alpha=0}}{f_{\rm gas,2500,\alpha\ne 0}}\right)^{1.67}, 
\end{eqnarray}
\begin{eqnarray}
\label{eq:F3}
	\frac{Y_{\rm 2500,\alpha\ne 0}}{Y_{\rm 2500,\alpha=0}}&=& \left(\frac{M_{\rm 2500,\alpha\ne 0}}{M_{\rm 2500,\alpha=0}}\right)^{0.27}=\left(\frac{f_{\rm gas,2500,\alpha=0}}{f_{\rm gas,2500,\alpha\ne 0}}\right)^{0.45}.
\end{eqnarray}
We fit new \Ysmall--\Msmall{ }scaling relations using the newly obtained values, adopting the same \Ysmall{ }and \Msmall{ }logarithmic noise estimates and the selection bias corrections of our fiducial analysis (Table \ref{tab:boxsz_delta}).
Using our fiducial scaling relation, we obtain an analytical prediction for the alternative \fgas~model scaling relations by inserting Equations \ref{eq:F2} and \ref{eq:F3} into Equation \ref{eq:sr}{ }(shown in blue in Figure \ref{fig:ymbias}).
Since \fgas{ }models with $\alpha\ne0$ alter the range of both the \Msmall~and \Ysmall~values, the regression bias varies as a function of $\alpha$, causing our results to depart from the simple analytical prediction.
To correct for these new regression biases, we create mock samples in a manner identical to that used in Section~\ref{sec:splittests}.
As with the true BOXSZ sample data, we scale the mock \Ysmall~and \Msmall~values to the alternative \fgas~model using Equations \ref{eq:F2} and \ref{eq:F3}. We then measure the median best-fitting \Ysmall--\Msmall~scaling relations from the scaled mock samples  and subtract from it the analytical prediction to obtain the additional $\alpha \ne 0$ regression bias, which we then use to correct the $\alpha \ne 0$ BOXSZ scaling relations.
The corrected BOXSZ scaling relation parameters are given by the solid black lines in Figure \ref{fig:ymbias}.
Finally, we use the 68\% fluctuation region about the median of the scaling relation fits to the scaled mock cluster samples to define a band indicating the expected fluctuations around the analytical prediction (shown in gray in Figure \ref{fig:ymbias}). The behavior of the data is consistent with the analytical prediction given these expected fluctuations.
Figure \ref{fig:ymbias} shows that, as expected, $\alpha>0$ makes the slope of the measured \Ysmall--\Msmall{ }scaling relation steeper, reduces the value of the intercept, and has no effect on the intrinsic scatter.
Values of $\alpha \gtrsim 0.2$ would need to be invoked to obtain consistency with typical values of the \Ysz--\Mtot{ }slope found in simulations 
\citep{Fabjan2011,Sembolini2013} and other observations (P11, A11).  
Such a high value of $\alpha$ would, however, be in disagreement with the \fgas{ }results in our mass range from Allen08 ($\alpha=0.005 \pm 0.058$).  Therefore, invoking a non-constant \fsmall--\Msmall{ }relation in our mass range does not fully resolve these discrepancies.

Regarding the observational studies that obtain significantly steeper \Ysz--\Mtot{ }slopes than our analysis, it is worth noting that the X-ray scaling relations used to provide mass proxies in each case nominally include a mass dependence of \fgas, $\alpha>0$. As demonstrated above, some disagreement in the observed \Ysz--\Mtot~slope is therefore expected. Following our discussion above, the value of $\alpha$  implied by an X-ray study is likely to depend on the distribution of masses in the data set relative to the flattening in Figure \ref{fig:fgas_all2500} as well as the telescope employed (XMM vs. \emph{Chandra}) and perhaps finer details of the data reduction and fitting procedure. While we cannot fully resolve these questions here, they certainly motivate a detailed and careful study of the cluster scaling relations, beyond the simple power-law form, over a wide mass range.
\section{Thumbnails}
\label{sec:thumbnails}
This section includes thumbnails of the 14\arcmin$\times$14\arcmin~S/N images and the 10\arcmin$\times$10\arcmin~deconvolved images for all 45 clusters in the BOXSZ sample.
\begin{figure}
  \epsscale{0.95}
  \plotone{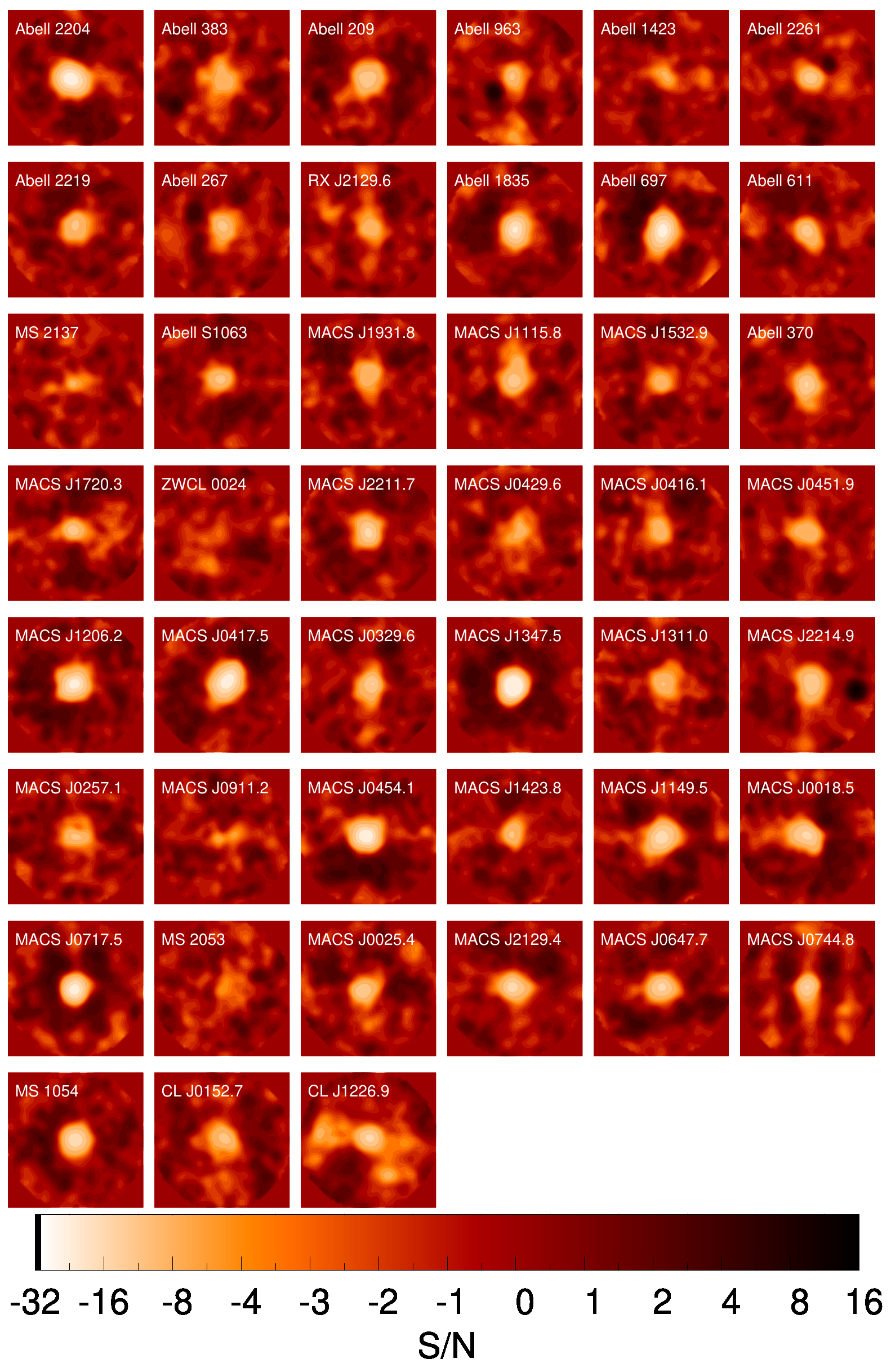}
  \caption{
Thumbnails showing the S/N per beam in the processed SZE images for all 45 BOXSZ clusters.
The images are 14\arcmin$\times$14\arcmin~in size.
The color scale is linear from S/N of $-4$ to S/N of $+2$ to allow an accurate visualization of the noise and low S/N SZE decrements, and the color scale is quasi-logarithmic at lower and higher S/N values.
This logarithmic scale is required due to the large dynamic range of some images due to significant SZE decrements and/or bright point sources.
Note that the point sources are subtracted from the data prior to any estimation of \Ysmall. In this figure, we mask regions beyond 7\arcmin~in radius due to low integration times at the corners of our maps.}
  \label{fig:thumbnails1}
\end{figure}
\begin{figure}
  \epsscale{0.95}
  \plotone{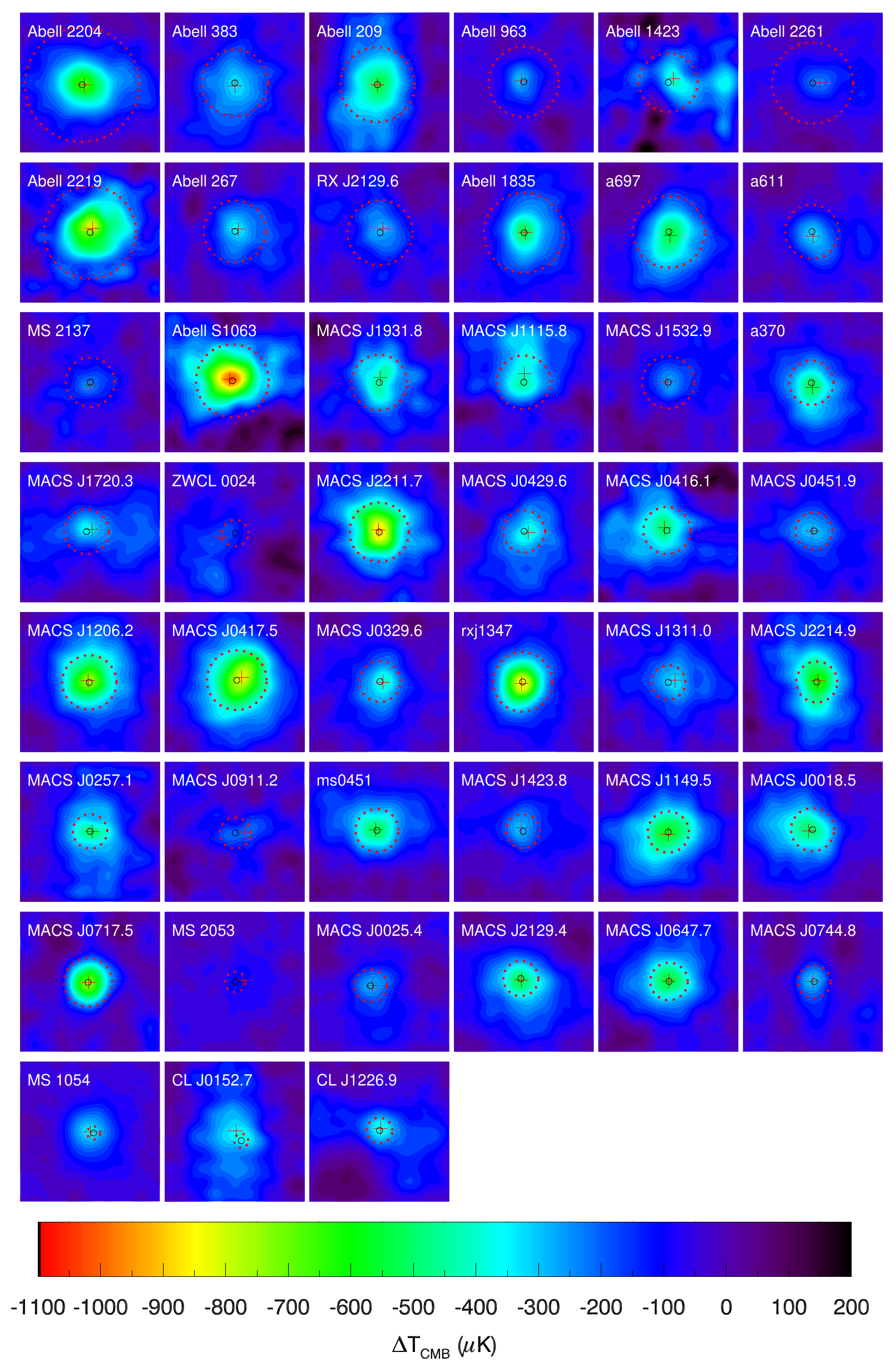}
  \caption{10\arcmin$\times$10\arcmin~deconvolved SZE images for all 45 BOXSZ clusters.
We obtain \Ysmall~from these images by integrating within the region enclosed by the dashed red line (\rsmall) centered on the X-ray centroid (small black circle).
The best-fit SZE centroid is indicated with a 1\arcmin-wide red cross.
Due to the linear color scale, which extends to include the brightest clusters in the sample, the contrast for some clusters appears low even though they are detected at high significance.}
  \label{fig:thumbnails2}
\end{figure}
\end{document}